\documentclass[12pt]{article}
\usepackage[T1]{fontenc}
\usepackage{amsmath}
\usepackage{amssymb}
\usepackage{times}
\usepackage{txfonts}
\DeclareMathAlphabet{\mathbold}{OML}{txr}{b}{it}
\usepackage[font={small}]{caption}

\usepackage{multirow}
\usepackage{units}
\usepackage{epsfig}
\usepackage{hhline}
\usepackage{dsfont}
\usepackage{color}

\usepackage{cite} 

\usepackage{longtable}
\usepackage{xspace}
\usepackage{bm} 
\usepackage{acronym}

\usepackage{graphicx}
\usepackage{rotating}

\usepackage{hyperref} 
\hypersetup{colorlinks=true, urlcolor=blue}
\usepackage{cite} 
\hypersetup{
  colorlinks,
  citecolor=blue,
  linkcolor=red,
  urlcolor=blue
  }
  
\renewcommand{\arraystretch}{1.25} 
\newlength{\dinwidth}
\newlength{\dinmargin}
\usepackage{parskip}

\begin{document}  

\newcommand{\delrel}{\ensuremath{\delta}}
\newcommand{\delabs}{\ensuremath{\Delta}}
\newcommand{\dabs}[2][1]{\ensuremath{\delabs^{{\rm{#1}}}_{{#2}}}}
\newcommand{\drel}[2][1]{\ensuremath{\delrel^{{\rm{#1}}}_{{#2}}}}
\newcommand{\Nsys}{{N_{\rm sys}}}
\newcommand{\stat}{{\rm stat}}
\newcommand{\sys}{{\rm sys}}
\newcommand{\TODO}{{\color{red}TODO}\xspace}

\newcommand{\muf}{\ensuremath{\mu_{f}}\xspace}
\newcommand{\mur}{\ensuremath{\mu_{r}}\xspace}
\newcommand{\as}{\ensuremath{\alpha_s}\xspace}
\newcommand{\asmz}{\ensuremath{\alpha_s(M_Z)}\xspace}
\newcommand{\asmur}{\ensuremath{\alpha_s(\mur)}\xspace}
\newcommand{\aem}{\ensuremath{\alpha_{\mathrm{em}}}\xspace}
\newcommand{\Lumi}{\ensuremath{\mathcal{L}}}
\newcommand{\pb}{\rm pb}
\newcommand{\invpb}{\ensuremath{\rm{pb}^{-1}}}
\newcommand{\PDF}{\ensuremath{{\rm PDF}}\xspace}
\renewcommand{\deg}{\ensuremath{^\circ}\xspace}
\newcommand{\unitmatrix}{1\!\!1}
\newcommand{\fC}{\ensuremath{f^{\rm C}}\xspace}
\newcommand{\fU}{\ensuremath{f^{\rm U}}\xspace}
\newcommand{\bas}{\boldsymbol{{\alpha_s}}} 
\newcommand{\basmz}{\boldsymbol{\alpha_s(M_Z)}} 
\newcommand{\bmur}{\boldsymbol{\mu_{r}}}

\newcommand{\chisq}{\ensuremath{\chi^{2}}}
\newcommand{\chisqA}{\ensuremath{\chi_{\rm A}^{2}}}
\newcommand{\chisqL}{\ensuremath{\chi_{\rm L}^{2}}}
\newcommand{\ndf}{\ensuremath{n_{\rm dof}}}
\newcommand{\A}{\ensuremath{\bm{A}}}
\newcommand{\M}{\ensuremath{\bm{M}}}
\newcommand{\V}{\ensuremath{\bm{V}}}
\newcommand{\B}{\ensuremath{\bm{B}}}
\newcommand{\J}{\ensuremath{\bm{J}}}
\newcommand{\N}{\ensuremath{\bm{N}}}
\newcommand{\LL}{\ensuremath{\bm{L}}}

\newcommand{\femjet}{\ensuremath{f_{\mathrm{em,jet}}}\xspace}
\newcommand{\femjetgen}{\ensuremath{f_{\mathrm{em,jet}}^{\mathrm{gen}}}\xspace}
\newcommand{\femjetrec}{\ensuremath{f_{\mathrm{em,jet}}^{\mathrm{rec}}}\xspace}
\newcommand{\Pem}{\ensuremath{P_{\mathrm{em}}}\xspace}
\newcommand{\Eem}{\ensuremath{E_{\mathrm{em}}}\xspace}
\newcommand{\Ehad}{\ensuremath{E_{\mathrm{had}}}\xspace}
\newcommand{\Ptbal}{\ensuremath{P_{\mathrm{T}}}--balance\ }
\newcommand{\PTbal}{\ensuremath{P_{\mathrm{T,bal}}}\xspace}
\newcommand{\Ptgen}{\ensuremath{P_{\mathrm{T}}^\mathrm{gen}}\xspace}
\newcommand{\Ptda}{\ensuremath{P_{\mathrm{T}}^{\mathrm{da}}}\xspace}
\newcommand{\thetajet}{\ensuremath{\theta_{\mathrm{jet}}}\xspace}
\newcommand{\etajet}{\ensuremath{\eta_{\mathrm{jet}}}\xspace}
\newcommand{\Ejet}{\ensuremath{E_{\mathrm{jet}}}\xspace}
\newcommand{\Ejetda}{\ensuremath{E^{\mathrm{da}}_{\mathrm{jet}}}\xspace}
\newcommand{\relres}{\ensuremath{\sigma(\Ejet)/\Ejet}\xspace}

\newcommand{\Empz}{\ensuremath{E-p_z}\xspace}
\newcommand{\gammah}{\ensuremath{\gamma_{\mathrm{h}}}\xspace}
\newcommand{\Pth}{\ensuremath{P_{\mathrm{T}}^{\mathrm{h}}}\xspace}
\newcommand{\Pzh}{\ensuremath{p_z^{\mathrm{h}}}\xspace}
\newcommand{\h}{\ensuremath{\mathrm{h}}}

\newcommand{\e}{\ensuremath{\mathrm{e}}}
\newcommand{\Ee}{\ensuremath{E_\mathrm{e}^\prime}\xspace}
\newcommand{\Pte}{\ensuremath{P_{\mathrm{T}}^{\mathrm{e}}}\xspace}
\newcommand{\thetae}{\ensuremath{\theta_\mathrm{e^\prime}}\xspace}
\newcommand{\phie}{\ensuremath{\phi_\mathrm{e}}\xspace}
\newcommand{\Eda}{\ensuremath{E^{\mathrm{da}}}\xspace}

\newcommand{\Ebeam}{\ensuremath{E_{\mathrm{e}}}\xspace}
\newcommand{\Pbeam}{\ensuremath{E_{\mathrm{p}}}\xspace}

\newcommand{\xbj}{\ensuremath{x}\xspace}
\newcommand{\Qsq}{\ensuremath{Q^2}\xspace}

\newcommand{\ptjet}{\ensuremath{P_{\rm T}^{\rm jet}}\xspace}
\newcommand{\ptjetmin}{\ensuremath{P_{\rm T,min}^{\rm jet}}\xspace}
\newcommand{\meanpt}{\ensuremath{\langle P_{\rm T} \rangle}}
\newcommand{\meanptdi}{\ensuremath{\langle P_{\mathrm{T}} \rangle_{2}}\xspace}
\newcommand{\meanpttri}{\ensuremath{\langle P_{\mathrm{T}} \rangle_{3}}\xspace}
\newcommand{\pt}{\ensuremath{P_{\rm T}}\xspace}
\newcommand{\Pt}{\ensuremath{P_{\mathrm{T}}}\xspace}
\newcommand{\Ptone}{\ensuremath{P_{\mathrm{T,1}}}\xspace}
\newcommand{\Pttwo}{\ensuremath{P_{\mathrm{T,2}}}\xspace}

\newcommand{\kt}{\ensuremath{{k_{\mathrm{T}}}}\xspace}
\newcommand{\antikt}{\ensuremath{{\mathrm{anti-}k_{\mathrm{T}}}}\xspace}
\newcommand{\bkt}{\ensuremath{\bm{k}_{\boldsymbol{\mathrm{T}}} }\xspace}
\newcommand{\bantikt}{\ensuremath{\boldsymbol{\mathrm{anti-}}\bm{k}_{\boldsymbol{\mathrm{T}}}}\xspace}

\newcommand{\etal}{{\it et al.}}
\newcommand{\Hone}{[H1 Collaboration]}

\newcommand{\Et}{\ensuremath{E_\mathrm{T}}\xspace}
\newcommand{\etalab}{\ensuremath{\eta^{\mathrm{jet}}_{\mathrm{lab}}}\xspace}
\newcommand{\ptlab}{\ensuremath{P^{\mathrm{jet}}_{\mathrm{T,lab}}}\xspace}
\newcommand{\Mjj}{\ensuremath{M_{\mathrm{12}}}\xspace}
\newcommand{\Mjjj}{\ensuremath{M_{\rm 123}}}
\newcommand{\xij}{\ensuremath{\xi}\xspace}
\newcommand{\xidi}{\ensuremath{\xi_2}\xspace}
\newcommand{\xitri}{\ensuremath{\xi_3}\xspace}

\newcommand{\ud}{\ensuremath{\mathrm{d}}\xspace}
\newcommand{\LO}{\ensuremath{\mathcal{O}(\alpha_s^0)}\xspave}
\newcommand{\Oa}{\ensuremath{\mathcal{O}(\alpha_s)}\xspace}
\newcommand{\Oaa}{\ensuremath{\mathcal{O}(\alpha_s^2)}\xspace}
\newcommand{\Oaaa}{\ensuremath{\mathcal{O}(\alpha_s^3)}\xspace}

\newcommand{\eq}{equation}
\newcommand{\fig}{figure}
\newcommand{\tab}{table}

\newcommand{\GeV}{\ensuremath{\mathrm{GeV}}\xspace}
\newcommand{\GeVsq}{\ensuremath{\mathrm{GeV}^2}\xspace}

\newcommand{\fifteen}{\ensuremath{15\,000}}
\newcommand{\dd}{\mathrm{d}}
\newcommand{\Ord}{\ensuremath{\mathcal{O}}}

\newcommand{\dHFS}[2][] {\dabs[RCES]{#2}\xspace}
\newcommand{\dJES}[2][] {\dabs[JES]{#2}\xspace}
\newcommand{\dLAr}[2][] {\dabs[LAr Noise]{#2}\xspace}
\newcommand{\dEe}[2][]  {\dabs[E_{e}]{#2}\xspace}
\newcommand{\dThe}[2][] {\dabs[\theta_{e}]{#2}\xspace}
\newcommand{\dID}[2][]  {\dabs[ID(e)]{#2}\xspace}
\newcommand{\dMod}[2][] {\dabs[Model]{#2}\xspace}
\newcommand{\dModRW}[2][] {\dabs[ModelRW]{#2}\xspace}
\newcommand{\dLumi}[2][]{\dabs[Lumi]{#2}\xspace}
\newcommand{\dTrig}[2][]{\dabs[Trig]{#2}\xspace}
\newcommand{\dTCl}[2][] {\dabs[TrkCl]{#2}\xspace}
\newcommand{\dNorm}[2][]{\dabs[Norm]{#2}\xspace}
\newcommand{\dMCSt}[2][] {\dabs[MCstat]{#2}\xspace}
        
\newcommand{\csdsub}{\ensuremath{}}
\newcommand{\csdsubn}{\ensuremath{}}
\newcommand{\CS}{\ensuremath{\sigma}}
\newcommand{\CSN}{\ensuremath{\sigma/\sigma_{\rm NC}}}
\newcommand{\trenn}{\cdot}
\newcommand{\bQsq}{\bm{Q^2}}
\newcommand{\bxidi}{\bm{\xi_2}}
\newcommand{\bxitri}{\bm{\xi_3}}
\newcommand{\bmeanptdi}{\bm{\langle P_{\mathrm{T}} \rangle_{2}}}
\newcommand{\bmeanpttri}{\bm{\langle P_{\mathrm{T}} \rangle_{3}}}
\newcommand{\bptjet}{\bm{P_{\rm T}^{\rm jet}}}
\newcommand{\bpt}{\bm{P_{\mathrm{T}}}}

\newcommand{\DStat}[1]{\drel[\stat]{#1}\xspace}
\newcommand{\DSys}[1] {\drel[\sys]{#1}\xspace}
\newcommand{\DHFS}[1] {\drel[RCES]{#1}\xspace}
\newcommand{\DJES}[1] {\drel[JES]{#1}\xspace}
\newcommand{\DLAr}[1] {\drel[LAr Noise]{#1}\xspace}
\newcommand{\DEe}[1]  {\drel[E_{e^\prime}]{#1}\xspace}
\newcommand{\DThe}[1] {\drel[\theta_{e^\prime}]{#1}\xspace}
\newcommand{\DID}[1]  {\drel[ID(e)]{#1}\xspace}
\newcommand{\DLumi}[1]{\drel[Lumi]{#1}\xspace}
\newcommand{\DTrig}[1]{\drel[Trig]{#1}\xspace}
\newcommand{\DTCL}[1] {\drel[TrkCl]{#1}\xspace}
\newcommand{\DMod}[1] {\drel[Model]{#1}\xspace}
\newcommand{\DModRW}[1] {\drel[ModelRW]{#1}\xspace}
\newcommand{\DNorm}[1]{\drel[Norm]{#1}\xspace}
\newcommand{\DMCSt}[1][] {\drel[MCstat]{#1}\xspace}

\newcommand{\cHad} {\ensuremath{c^{\rm had}}    }
\newcommand{\cEW}  {\ensuremath{c^{\rm ew}}}
\newcommand{\cRad} {\ensuremath{c^{\rm rad}}    }
\newcommand{\cHadi} {\ensuremath{c_i^{\rm had}}}
\newcommand{\cRadi} {\ensuremath{c_i^{\rm rad}}}
\newcommand{\cEWi}  {\ensuremath{c_i^{\rm ew}}}
\newcommand{\DHad} {\drel[had]{}\xspace}
\newcommand{\DRad} {\drel[rad]{}\xspace}

\newcommand{\sI} {\ensuremath{\sigma_{\rm jet}}\xspace}
\newcommand{\sD} {\ensuremath{\sigma_{\rm dijet}}\xspace}
\newcommand{\sT} {\ensuremath{\sigma_{\rm trijet}}\xspace}
\newcommand{\sNC} {\ensuremath{\sigma_{\rm NC}}\xspace}
\newcommand{\sIN} {\ensuremath{\displaystyle\frac{\sI}{\sNC}}\xspace}
\newcommand{\sDN} {\ensuremath{\displaystyle\frac{\sD}{\sNC}}\xspace}
\newcommand{\sTN} {\ensuremath{\displaystyle\frac{\sT}{\sNC}}\xspace}

\def\Journal#1#2#3#4{{#1}~{\bf #2} (#3) #4}
\def\NPB{Nucl. Phys.~}
\def\PRL{Phys. Rev. Lett.~}
\def\EPJC{Eur. Phys. J.~}
\def\PLB{Phys. Lett.~}
\def\NIM{Nucl. Instrum. Meth.~}
\def\PRD{Phys. Rev.~}
\def\JHEP{JHEP~}
\def\PROC{Conf. Proc.~}
\def\CPC{Comp. Phys. Commun.~}

\begin{titlepage}

\noindent
\begin{flushleft}
{\tt DESY 16-200    \hfill    ISSN 0418-9833} \\
{\tt November 2016}                  \\
\end{flushleft}


\vspace{1.5cm}
\begin{center}
\begin{Large}

{\bf Measurement of Jet Production Cross Sections in Deep-inelastic \emph{ep} Scattering at HERA}

\vspace{1.5cm}

H1 Collaboration

\end{Large}
\end{center}

\vspace{1.5cm}

\begin{abstract}
A precision measurement of jet cross sections in neutral current deep-inelastic 
scattering for photon virtualities $5.5<\Qsq<80\,\GeVsq$ and 
inelasticities $0.2<y<0.6$ is presented, using data taken with the H1 
detector at HERA, corresponding to an integrated 
luminosity of $290\,\invpb$.
Double-differential inclusive jet, dijet and trijet cross sections are 
measured simultaneously and
are presented as a function of jet transverse momentum observables and as a function of \Qsq.
Jet cross sections normalised to the inclusive neutral current DIS cross section in the respective \Qsq-interval are also determined.
Previous results of inclusive jet cross sections in the range $150<\Qsq<15\,000\,\GeVsq$ are extended to low transverse jet momenta  $5<\ptjet<7\,\GeV$.
The data are compared to predictions from perturbative QCD in next-to-leading order in the strong coupling, in approximate next-to-next-to-leading order
and in full next-to-next-to-leading order. 
Using also the recently published H1 jet data at high values of \Qsq, the strong coupling constant \asmz\ is determined in next-to-leading order.
\end{abstract}

\vspace{1.0cm}

\begin{center} Published in EPJ C \end{center}

\vspace{1.0cm}

{
  {\bf Erratum.}
  The measured jet cross sections are compared to various predictions
  including the next-to-next-to-leading order (NNLO) QCD
  calculations.
  An implementation error in these NNLO predictions was
  found~\cite{Currie:2017tpe}, while the jet data and the other
  predictions remain unchanged. 
  In the present document, eight figures, one table and conclusions
  are adapted accordingly.
}
\vspace{1.0cm}

\begin{center} Erratum submitted to EPJ C \end{center}

\end{titlepage}

%
%
%
\begin{flushleft}

V.~Andreev$^{19}$,             
A.~Baghdasaryan$^{31}$,        
K.~Begzsuren$^{28}$,           
A.~Belousov$^{19}$,            
A.~Bolz$^{12}$,                
V.~Boudry$^{22}$,              
G.~Brandt$^{41}$,              
V.~Brisson$^{21}$,             
D.~Britzger$^{10}$,            
A.~Buniatyan$^{2}$
% ...



\bigskip{\it
% ...
$ ^{2}$ School of Physics and Astronomy, University of Birmingham,
          Birmingham, UK$^{ b}$ \\
% ...
$ ^{10}$ DESY, Hamburg, Germany \\
% ...
$ ^{12}$ Physikalisches Institut, Universit\"at Heidelberg,
          Heidelberg, Germany$^{ a}$ \\
% ...
$ ^{19}$ Lebedev Physical Institute, Moscow, Russia \\
% ...
$ ^{21}$ LAL, Universit\'e Paris-Sud, CNRS/IN2P3, Orsay, France \\
 $ ^{22}$ LLR, Ecole Polytechnique, CNRS/IN2P3, Palaiseau, France \\
% ...
$ ^{28}$ Institute of Physics and Technology of the Mongolian
          Academy of Sciences, Ulaanbaatar, Mongolia \\
% ...
$ ^{31}$ Yerevan Physics Institute, Yerevan, Armenia \\
% ...
$ ^{41}$ Now at II. Physikalisches Institut, Universit\"at G\"ottingen,
          G\"ottingen, Germany \\
% ...
$ ^a$ Supported by the Bundesministerium f\"ur Bildung und Forschung, FRG,
      under contract numbers 05H09GUF, 05H09VHC, 05H09VHF,  05H16PEA \\
 $ ^b$ Supported by the UK Science and Technology Facilities Council,
      and formerly by the UK Particle Physics and
      Astronomy Research Council \\
% ...
}
\end{flushleft}
%

\clearpage

\section{Introduction}
Jet production in neutral current (NC) deep-inelastic scattering (DIS)
at HERA is an important process to test 
perturbative calculations based on 
the theory of strong interactions,
which is described by Quantum Chromodynamics (QCD)~\cite{QCD0,QCD1,QCD2,QCD3,QCD4}. 
In contrast to inclusive DIS,
where QCD is probed by means of scaling violations,
jet production in the Breit frame~\cite{feynman,breit}
is a process which always involves at least one strong vertex even at
Born level and thus more directly probes QCD.

\begin{figure}[h]
\centering
 \includegraphics[height=3.5cm]{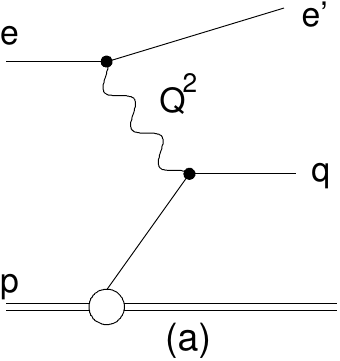}\hskip0.5cm
 \includegraphics[height=3.5cm]{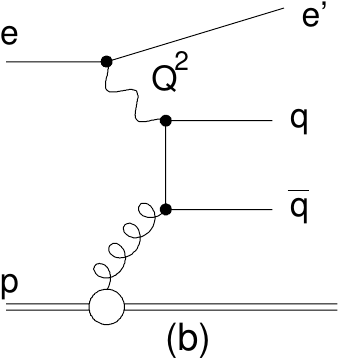}\hskip0.5cm
 \includegraphics[height=3.5cm]{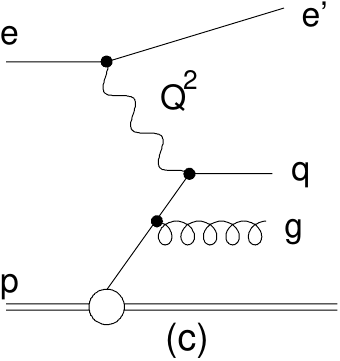}\hskip0.5cm
 \includegraphics[height=3.5cm]{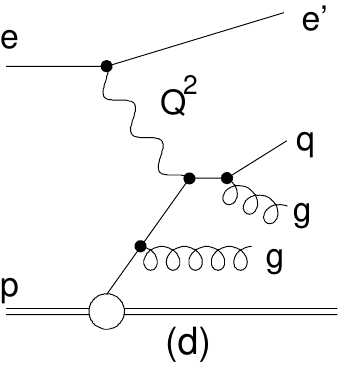}\hskip0.5cm 
\caption[Diagrams of different order in $\as$ in deep-inelastic lepton-proton scattering] %
{ Deep-inelastic $ep$ scattering at different orders in $\as$: 
  (a) Born contribution to inclusive NC DIS ($\mathcal{O}(\aem^2)$), 
  (b) photon-gluon fusion ($\mathcal{O}(\aem^2 \as)$),
  (c) QCD Compton scattering ($\mathcal{O}(\aem^2 \as)$) and 
  (d) a trijet process $\mathcal{O}(\aem^2 \as^2)$.
}
\label{fig:feynborn}
\end{figure}

In the Breit frame, where the virtual photon and the proton collide head on, 
the Born level contribution to DIS (figure~\ref{fig:feynborn}(a)) generates no transverse momentum. 
Significant transverse momentum of the outgoing partons, $P_{\mathrm{T}}$, can however be produced 
at leading order (LO) in the strong coupling $\as$ by the 
photon-gluon-fusion process (figure~\ref{fig:feynborn}(b)) and the QCD Compton process (figure~\ref{fig:feynborn}(c)). 
Photon-gluon fusion dominates jet production for the range of photon virtualities $Q^2$ accessible in this analysis
 and provides direct sensitivity to the gluon density function of the proton~\cite{Adloff2001}.
One of the diagrams of the next-to-leading order contribution is displayed in figure~\ref{fig:feynborn}(d),
which also illustrates one of the leading-order diagrams of the trijet perturbative QCD (pQCD) calculation.

About 25 years after next-to-leading order corrections to
 jet production cross sections in DIS have been studied for the first time~\cite{GraudenzThesis,Graudenz},
complete predictions at next-to-next-to-leading order in the strong coupling are now available for inclusive jet and dijet production in DIS~\cite{NNLO,NNLO17} and in hadron-hadron collisions~\cite{Currie:2016bfm}.
These new theoretical developments together with precise measurements and greater kinematic reach of the
 data allow the use of DIS jet cross sections for precise studies of QCD.

Measurements of jet production in NC DIS at HERA were performed by the H1 Collaboration~\cite{Adloff2001,H1_jets0,H1_jets1,H1_jets2,H1_jets3,H1_jets4,H1_jets9597,Adloff:2002ew,Aktas:2003ja,Aktas:2004px,H1_highq2,H1_lowq2,H1_highq2_H1andH2,H1Multijets} and
 the ZEUS Collaboration~\cite{ZEUS_jets0,ZEUS_jets1,Chekanov:2002be,ZEUS_jets2,ZEUS_jets3,Chekanov:2006xr,ZEUS_jets4,ZEUS_IJ_2010,ZEUS_2J_2010}.
In this paper new double-differential measurements of inclusive jet, 
dijet and trijet cross sections are presented,
extending the kinematic range of an earlier
 analysis \cite{H1Multijets} both to lower momentum transfer, $5.5<Q^2<80\,\GeVsq$, and to lower jet transverse momenta, as detailed in the following.

At low momentum transfer $5.5<Q^2<80\,\GeVsq$, the transverse momenta of jets in the Breit frame, \ptjet, are required to exceed 4\,\GeV. 
Inclusive jet cross sections are measured in the range $4.5<\ptjet<50\,\GeV$. 
Inclusive dijet cross sections are measured as a function of the average transverse momentum of the two jets with the highest \ptjet in an event, $\meanptdi= \tfrac{1}{2}(P_{\rm T}^{\rm jet1} + P_{\rm T}^{\rm jet2} )  $,
in the range $5<\meanptdi<50\,\GeV$, and trijet cross sections as a function 
of $\meanpttri= \tfrac{1}{3}(P_{\rm T}^{\rm jet1} + P_{\rm T}^{\rm jet2}+ P_{\rm T}^{\rm jet3} )$ in the range $5.5<\meanpttri<40\,\GeV$.
Compared to previous H1 jet cross section measurements in a similar kinematic domain~\cite{H1_lowq2} 
the overall uncertainty is reduced mainly due to the larger data set together with an 
improved calibration of the hadronic energy~\cite{H1Multijets,ThesisKogler}.
At large momentum transfer,  $150<\Qsq<15\,000\,\GeVsq$, an extension of previously published inclusive jet cross section measurements \cite{H1Multijets} to 
  lower transverse jet momenta, $5<\ptjet<7\,\GeV$, is presented.
In parallel to all cross section measurements, the corresponding normalised jet cross sections are presented, as the ratio of jet cross sections to 
inclusive NC DIS cross sections measured in the same \Qsq-ranges. 

The results are compared to pQCD predictions corrected 
for hadronisation effects.
The predictions include next-to-leading order (NLO) calculations \cite{Nagy2,Nagy3}, 
NLO calculations supplemented with two-loop threshold corrections (aNNLO)~\cite{aNNLO,aNNLOprivate} and next-to-next-to-leading order (NNLO)~\cite{NNLO,NNLOprivate} calculations. 
The experimental sensitivity to the strong coupling constant at the mass of the $Z$-boson, $\asmz$, is studied in a fit of NLO predictions to the data.
Together with the jet data at high $Q^2$~\cite{H1Multijets}, the data test the running of $\asmur$ 
in the range of the renormalisation scale $\mu_r$ between about 5 and 90\,\GeV.

This publication represents the first H1 analysis completely performed using the newly developed data preservation model~\cite{Akopov:2012bm}. Through a planned and documented programme, begun shortly after the end of HERA data taking, all aspects of H1 data analysis have been redeveloped into a framework suitable for continued use into the next decade and beyond~\cite{Ozerov:2013isa,Amerio:2015ipk}. In particular for this analysis, the continued ability to produce new Monte Carlo samples, including full detector simulation, access archived trigger information and documentation covering earlier analyses and create working event displays has proved to be crucial. This paper, therefore, also represents a proof-of-concept of the H1 data preservation model.

\section{Experimental method}
  In the following sections~\ref{sec:detector} to \ref{sec:jetreco}
  the analysis of jet cross sections 
  in the range $5.5<\Qsq<80\,\GeVsq$ is described.
  This kinematic range is denoted as `low-\Qsq' in contrast to the `high-\Qsq' regime,
  which is experimentally distinct by detecting the scattered electron in different
  detector components.
  In section~\ref{sec:pslow} the phase space of the new low-\Qsq\ cross
  sections is given. In section \ref{sec:pshigh} an extension of
  previous high-\Qsq\ inclusive jet cross section
  measurements~\cite{H1Multijets} to lower jet transverse momenta is
  described.

  For the low-\Qsq\ analysis, the data sample was collected with the H1 detector at HERA in the years $2005$ to $2007$, 
where electron or positron\footnote{The term `electron' 
is used in the following to refer to both electron and positron.} beams with an energy of $E_e = 27.6\,\GeV$
collided with protons of energy $E_p = 920\,\GeV$, resulting in a centre-of-mass energy of $\sqrt{s}=319\,\GeV$.
The integrated luminosity corresponds to 290\,\invpb.

\subsection{The H1 detector}
\label{sec:detector}
A full description of the H1 detector\footnote{The H1 detector uses a right-handed coordinate system, which is defined such that the positive $z$-axis points in the 
direction of the proton beam (also called `forward direction') and the nominal interaction point is located 
at $z = 0$. The polar angle $\theta$ is defined with respect to this axis. The pseudorapidity 
is defined as  $\eta=-\,\ln\tan(\theta/2)$.} can be found elsewhere~\cite{Abt1,Abt2,Appuhn,Andrieu}.  
The essential detector components used in the analysis are 
the Liquid Argon (LAr) calorimeter,
the lead-scintillating fibre calorimeter (SpaCal) 
and the inner tracking detectors.
The  central tracking system and the LAr calorimeter are surrounded by a superconducting solenoid providing a uniform 
field of 1.16\,T inside the tracking volume, thus enabling the measurement of transverse momenta of charged particles.

The central tracking system, covering $15^\circ < \theta < 165^\circ$, is located inside the LAr calorimeter. 
It consists of drift and proportional chambers and is complemented by a silicon vertex detector covering the 
range $30^\circ < \theta < 150^\circ$~\cite{Pitzl}. 
The trajectories of charged particles are measured 
with a transverse momentum resolution of $\sigma_{P_{\mathrm{T}} }/P_{\mathrm{T}} = 0.2\% {P_{\mathrm{T}}/\GeV}\oplus 1.5\%$.
The main tracking devices for this analysis 
are the Central Jet Chamber (CJC) and the Central (CST) and Backward (BST) Silicon Tracker.

The LAr calorimeter, covering the polar angular range $4^\circ<\theta<154^\circ$ over
the full azimuth~\cite{Andrieu}, is used in the analysis in the reconstruction of the 
hadronic final state. 
The LAr calorimeter consists of an electromagnetic section made of lead absorbers 
and a hadronic section with steel absorbers, with both sections equipped with highly 
segmented readout cells in the transverse and longitudinal directions. 
The energy resolution is $\sigma_E/E = 11\%/\sqrt{E/\GeV}\oplus 1\%$ for electrons and 
$\sigma_{E}/E \simeq 50\%/\sqrt{E/\GeV}\oplus 3\%$ for pions~\cite{LAr1,LAr2}. 

The lead-scintillating fibre calorimeter (SpaCal)~\cite{Appuhn} covers the region
$153^\circ<\theta<177.5^\circ$ with its electromagnetic and hadronic sections. The calorimeter is used to 
measure the scattered electron and hadronic energy in the backward region. The energy resolution for electrons 
in the electromagnetic section is $\sigma_{E}/E = 7.1\%/\sqrt{(E/\GeV)} \oplus 1\%$, as determined in test 
beam measurements~\cite{Andrieu,Nicholls}. 
The SpaCal also provides energy and time-of-flight information 
used for triggering purposes. 
The Backward Proportional Chamber (BPC) in front of the SpaCal assists the measurement of the electron scattering angle.

The luminosity is determined by measuring the event rate for the Bethe-Heitler process of QED 
bremsstrahlung \mbox{$ep\rightarrow ep\gamma$}, where the photon is detected in a calorimeter 
close to the beam pipe at $z=-103\,\mathrm{m}$. 
The overall normalisation is determined using a precision measurement of the QED Compton process 
with the electron and the photon detected in the SpaCal~\cite{AaronQED} ($e+p\rightarrow e+\gamma+p$).

\subsection{Event selection}
\label{sec:event_sel_rec}
The data sample of this analysis is obtained by reconstructing the scattered lepton, 
defined as the most energetic compact deposit (cluster) in the SpaCal, with an energy  $E_{e^\prime}>10.5\,\GeV$.
The cluster is required to be well contained within the acceptance of the SpaCal with a radial distance 
$R_{\rm clus}$ from the beam axis of $12.5<R_{\rm clus}<75\,\rm{cm}$.
The energy weighted cluster radius 
is required to be less than 4\,cm, and the energy deposit
associated to the cluster in the hadronic part must not exceed 0.5\,GeV,
both cuts following the expectations for an electromagnetic shower.
The event vertex position is obtained from tracks reconstructed in the tracking detectors~\cite{brokenlines}.
Its longitudinal position is restricted to the range $-35<z_{vtx}<35$\,cm. 

The polar angle of the scattered electron $\theta_{e^\prime}$ is determined from hits in the BPC, 
if these are associated with the SpaCal cluster~\cite{FL} and if
$R_{\rm clus}>20\,{\rm cm}$, otherwise from the aligned SpaCal position~\cite{AaronQED}.
The four-vector of the scattered lepton is calculated from the event vertex position, 
the hit position in the BPC or SpaCal and the measured energy in the SpaCal,  
assuming the charge to be equal to the beam charge~\cite{AaronQED}.
Four-vectors of hadronic final state (HFS) objects are formed from 
tracker and calorimeter measurements avoiding double-counting of energy~\cite{Peez,Portheault}.

The events are collected using time-dependent trigger conditions. 
The selection is based on the detection of a compact cluster in the SpaCal. 
This condition is fully efficient in the years 2006--2007 for both the selection of inclusive NC DIS events and for NC DIS events with jets.
For the year 2005, this condition is not available in the lower \Qsq\ range of about $\Qsq\lesssim 25\,\GeVsq$, and instead a mix of triggers is used.
At lower \Qsq, in addition to the SpaCal cluster a trigger signal originating
from the hadronic final state, either in the tracker or in the LAr is required~\cite{FTT1,FTT2,FTT3}.
Events triggered by the LAr calorimeter alone are also accepted.
This strategy is fully efficient for NC DIS events with jets, but has some
inefficiency for the inclusive NC DIS selection.
For this reason, the data samples for 2005 and 2006--2007 are used to measure the jet cross sections,
whereas only the data of 2006--2007 are used for the inclusive NC DIS measurement, in order 
to obtain normalised jet cross sections.
The data from the year 2005 correspond to an integrated luminosity of 106\,\invpb.

The total longitudinal energy balance, determined as the difference of the total 
energy $E_{\rm tot}$ and the longitudinal component of the total momentum $P_{z,{\rm tot}}$, 
calculated from all detected particles (HFS objects and the scattered electron) is restricted to  
$35 < E_{\rm tot}-P_{z,{\rm tot}}< 65\,\GeV$.
In a perfect detector without longitudinal energy loss the quantity $ E_{\rm tot}-P_{z,{\rm tot}}$ 
is equal to twice the electron beam energy, and this requirement thus reduces background from 
photoproduction events ($\Qsq\rightarrow 0\,\GeVsq$), where the scattered lepton remains undetected at small angles.
Events with significant initial state radiation are also removed.

Cosmic muon and beam induced backgrounds are reduced
to a negligible level after the application of dedicated background finder algorithms. 
A system of scintillators upstream and downstream of the interaction point and 
the SpaCal provide time-of-flight information to reject particles 
from non-$ep$ background at trigger level.
The resulting veto inefficiencies are about $0.8\,\%$ and are corrected for by applying time-dependent weights to the data.
Background from QED Compton processes is 
suppressed using a topological cut
against events with two azimuthally back-to-back electromagnetic clusters reconstructed
in the SpaCal~\cite{FL}.

The NC DIS kinematical variables are 
reconstructed from the four-momenta of the scattered electron $e^\prime$ and the hadronic final 
state particles using the I$\Sigma$ (`ISigma') method~\cite{sigma_method} as
\begin{equation}
  y = \frac{\sum_{i\in  \rm had}(E_i-P_{i,z})}{\sum_{ i\in \rm had}(E_i-P_{i,z}) + E_{e^\prime}(1-\cos\theta_{e^\prime}) }~, 
\label{eq:Sigma_y}
\end{equation}
\begin{equation}
  \Qsq = \frac{E^2_{e^\prime} \sin^2\theta_{e^\prime} }{1-y}~~~~{\rm and}~~~~
  x = \frac{E_{e^\prime}}{E_p} \frac{\cos^2(\theta_{e^\prime}/2)}{y}~,
\label{eq:ISigma_x}
\end{equation}
where the sum $\sum_{i\in  \rm had}$ runs over all reconstructed HFS objects.
This reconstruction is insensitive to initial state QED radiation off the electron beam, since the beam
energies do not enter the calculation of \Qsq\ and $y$. 
The energy of photons radiated collinearly to the scattered lepton is contained in the measured cluster energy.
The resulting  radiative correction factors are close to unity.

The kinematic region for the NC DIS event selection is defined by $3.0<\Qsq<120\,\GeVsq$ and $0.08<y<0.7$, 
which is larger than the final phase space of the cross sections in order to control migrations in the
variables \Qsq\ and $y$.
After this event selection, 24~million  events are kept for further analysis. 

\subsection{Reconstruction of jets}
\label{sec:jetreco}
In the selected NC DIS event sample, jets are constructed from the HFS objects
in the Breit frame. 
The Breit frame is defined by $2x\vec{P} + \vec{e} - \vec{e^\prime} = 0$, with $\vec{P}$ being 
the momentum of the beam proton, $\vec{e^\prime}$  the momentum of the scattered electron, 
$x$ as given in equation~\ref{eq:ISigma_x}, and $\vec{e}$ is the momentum of the incoming electron.
In order to account for collinear initial state radiation, the $z$-component of the incoming electron momentum in the
laboratory frame is calculated as $e_{z}= -E_{e} = -(E_{\rm tot}-P_{z,{\rm tot}})/2$.
The objects of the hadronic final state are clustered 
into jets using the inclusive $k_T$ 
algorithm with the massless $P_T$ recombination scheme and with the distance 
parameter $R = 1$ as implemented in the FastJet package~\cite{kt_algo,fastjet}. 
Monte Carlo studies indicate that this choice of the distance parameter ensures that 
the hadronisation corrections are close to one and that there is 
a good correspondence of jets reconstructed before and after the detector simulation. 
%
Jets are selected within an extended phase-space, in transverse momentum in the Breit frame $\ptjet > 3\,\GeV$, and in pseudorapidity 
of these jets boosted to the laboratory frame, $-1.5<\etalab<2.75$. 
The transverse momentum of these jets in the laboratory frame is required to exceed $P_{\rm T,lab}^{\rm jet}>2.5\,\GeV$, in order to remove jets which are not well measured.
The inefficiency of this requirement is small and is corrected for.
The jet with highest \ptjet\ is referred to as the leading jet.
Throughout this article $\pt$ denotes the observables \ptjet, \meanptdi\ or \meanpttri\ for 
the inclusive jet, dijet or trijet measurements, respectively.
Jet cross sections are denoted as `absolute', to make clear they are not normalised to NC DIS data.
Here, `differential' cross sections are bin-integrated  cross sections obtained in adjacent phase space regions (`bins').
All absolute jet cross section values are obtained in units of pb for given kinematic ranges.

\begin{boldmath}
\subsection{Phase space of cross section measurements at low $Q^2$}
\label{sec:pslow}
\end{boldmath}

The NC DIS and the jet phase space described for the event selection and data correction of the low-\Qsq\ analysis
refers to an `extended phase space' compared to the `measurement phase space' of the given cross sections. 
Extending the event selection to a larger phase space helps to describe 
migrations at the phase space boundaries, and thus stabilises the measurement and the 
determination of the uncertainties. The relevant quantities are summarised in table~\ref{tab:ps}.
\begin{table}[thbp]
  \footnotesize
\begin{center}
\begin{tabular}{l|cc|c}
\hline
\hline 
& \textbf{Low-\begin{boldmath}$Q^2$\end{boldmath} extended}
&\textbf{Low-\begin{boldmath}$Q^2$\end{boldmath} measurement}
&\textbf{High-\begin{boldmath}\Qsq\end{boldmath} measurement } \\
& \textbf{phase space}        &\textbf{phase space} &\textbf{phase
  space extension}\\
\hline 
Application & Used for event          & Phase space of     & Phase space of  \\
            & selection and unfolding & jet cross sections & jet cross sections \\
\hline 
  NC DIS phase space  &   $3 < \Qsq< 120\,\GeVsq$ & $5.5 < \Qsq < 80\,\GeVsq$   & $150 < \Qsq< 15\,000\,\GeVsq$\\
                      &     $0.08 < y < 0.7$      & $0.2 < y < 0.6$             & $0.2 < y < 0.7$               \\
\hline
 Phase space common  &     $-1.5 < \etalab < 2.75$  & $-1.0 < \etalab < 2.5$     & $-1.0 < \etalab < 2.5$           \\ 
 for all jets        &     $ \ptjet > 3\,\GeV$      & $\ptjet > 4\,\GeV$         &   \\ 
 \hline
Inclusive jet &     $ \ptjet > 3\,\GeV$    &  $4.5<\ptjet <50\,\GeV$       &   $ 5<\ptjet <7\,\GeV$                 \\ 
  & & &  \multicolumn{1}{r}{\scriptsize ($ 7<\ptjet <50\,\GeV$ published in \cite{H1Multijets})} \\ 
\hline  
 Dijet &  $N_{\text{jet}} \geq 2$   &  $N_{\text{jet}} \geq 2$     &     \\
       &  $\meanptdi > 3\,\GeV$     & $5<\meanptdi < 50\,\GeV$     & \\
\hline  
 Trijet &   $N_{\text{jet}} \geq 3$  &  $N_{\text{jet}} \geq 3$      &  \\
        &   $\meanpttri > 3\,\GeV$     & $5.5<\meanpttri < 40\,\GeV$   &  \\
\hline
\hline 
\end{tabular}
\caption{Summary of the measurement phase space of the jet cross sections and the extended analysis phase space of the low-\Qsq\ analysis, and the cross section phase space of the additional high-\Qsq\ data. 
}
\label{tab:ps}
\end{center}
\end{table}

The final cross sections for jet production are  measured in the NC DIS phase space
given by $5.5 < Q^2 < 80\,\GeVsq$ and $0.2 < y < 0.6$. 
To ensure that the jets are well contained in the LAr calorimeter, 
they are required to have $-1.0 < \etalab < 2.5$ and $\ptjet > 4.0\,\GeV$.

Cross sections for inclusive jet production are defined by counting all jets in a given 
event within the $\etalab$-range and $4.5 < \ptjet < 50\,\GeV$.
Cross sections for $\ptjet>50\,\GeV$ cannot be determined reliably due to 
small event counts.

Dijet events are defined by requiring at least two jets
passing the above criteria and are measured as a function of \meanptdi\ 
in the range $5.0<\meanptdi<50\,\GeV$.
Trijet events are defined by requiring at least three jets and
 are measured as a function of \meanpttri\ in the range $5.5<\meanpttri<40\,\GeV$.
The lower bounds on $\meanpt_{2,3}$ are set to the peak-region of the $\meanpt_{2,3}$ distributions in the extended 
phase space in order to reduce the  dependence on the requirement of $\ptjet>4\,\GeV$.
This asymmetry between the bound on \ptjet\ and the bound on $\meanpt_{2,3}$ furthermore 
 removes infrared sensitive parts of the phase space and thus secures 
stability of the pQCD calculations~\cite{NNLO,IRinsens,JetVip1}.

In order to reduce limitations of the theory at lower scales, e.g.\ from effects of the 
 heavy-quark masses or higher twist effects, the phase space may be 
restricted for phenomenological analyses.
For this purpose, the dijet and trijet bin boundaries are chosen such that 
only few percent of the events above $\meanptdi=7\,\GeV$ or $\meanpttri=8\,\GeV$ 
contain jets with $\ptjet<5\,\GeV$ which contribute to the measured observable.

\begin{boldmath}
\subsection{Phase space extension at high $Q^2$}
\label{sec:pshigh}
\end{boldmath}

Cross sections for jet production in NC DIS at high values of \Qsq\
have already been 
published earlier~\cite{H1Multijets} for a data taking period similar to the low \Qsq\ sample described above. For the high $Q^2$ sample,  
the scattered electron is identified in the LAr calorimeter.
The NC DIS kinematic region was defined as $150<\Qsq<15\,000\,\GeVsq$ and $0.2<y<0.7$, 
and jets were measured in the range of $7<\ptjet<50\,\GeV$.
These measurements were performed using a regularised unfolding procedure with  
data taken in an extended phase space. Migrations outside the measurement phase space 
were controlled by measuring additional columns in the migration matrix.
The migration matrix of the inclusive jet measurement contained side-bins 
$3<\ptjet<5\,\GeV$ and $5<\ptjet<7\,\GeV$ of the extended phase space.
With the improved understanding of the low-\ptjet\ region as already used in earlier works~\cite{DiffDijetsLRG,DiffDijetsVFPS}, 
the measurement phase space could be slightly extended, and cross section measurements for $5<\ptjet<7\,\GeV$ 
as a function of \Qsq\ are provided.

\section{Monte Carlo simulations}
\label{sect:mc_simulation}
The experimental data are corrected for effects of limited detector
acceptance and resolution in order to extract the cross sections at hadron level.
The hadron level of the Monte Carlo (MC) generator refers to all stable particles in an event
with a proper lifetime $c\tau>10\,{\rm mm}$.
The coefficients for this unfolding process are obtained from simulated NC DIS events. 
The generated NC DIS events are passed through a detailed simulation of the H1 detector and are subjected 
to the same reconstruction and analysis chain as the data.

Two Monte Carlo generators are used to model NC DIS events for $Q^2>2\,\text{GeV}^2$, both 
implementing Born-level matrix elements for the NC DIS, boson-gluon fusion and QCD Compton processes:
Djangoh~\cite{DJANGO}, which uses the Colour Dipole Model as 
implemented in Ariadne~\cite{ARIADNE} for higher order emissions, and Rapgap~\cite{RAPGAP},
which simulates parton showers in the leading-logarithmic approximation.
The hadronisation process is modelled in both programs with the Lund string fragmentation model~\cite{lundstring,lundstring2} using the ALEPH tune~\cite{alephtune}.
The QED effects on the leptonic tensor are simulated in both event generators
using the Heracles program~\cite{HERACLES}. 
With both Monte Carlo generators the CTEQ6L LO PDF set
\cite{Pumplin:2002vw} is used.

The effects of QED radiation, QED vertex corrections and self-energies of the
lepton lines, but not the running of $\aem(Q^2)$, are corrected for using bin-wise correction factors $c_i^{\rm rad}$ calculated
from generated events with these QED effects switched on or  off in Heracles.
In the generated events including QED radiation, which are also used for the unfolding,
the kinematics is calculated by merging the photon which is radiated off the final state electron with this electron.
The size of the resulting correction factors is 1.01 for NC DIS and ranges from 1.00 to 1.09 for inclusive jet,
0.96 to 1.12 for dijet and 1.01 to 1.08 for trijet cross sections.

To improve the reliability of the unfolding, the MC events are weighted to 
give a reasonable description of the data on detector level.
Weights are determined from observables on detector level for each MC model and 
applied to the respective hadron level quantities.
The weights are calculated from a linear interpolation of values obtained from two-dimensional histograms in order to have continuous functions.
The procedure is repeated for various jet and NC DIS observables, and weights are applied to all generated MC events used in the unfolding procedure.

The purity defined as the fraction of events reconstructed
in a bin that originate from that bin on hadron level, is found
to be around 70\,\% for the inclusive NC DIS measurement, while the acceptance,
defined as the fraction of events reconstructed in a bin to the number
of events generated in that bin, is around 80\,\%.
The lowest and highest \Qsq-ranges have somewhat reduced acceptances of about 60\,\% and 70\,\%, respectively, due to the SpaCal geometry.
The purity of the double-differential jet measurements is typically around
40 to 45\,\%. Lower values of 35\,\% are observed at the lowest \pt\ bin.

\begin{figure}[htb]
\begin{center}
   \includegraphics[width=0.48\textwidth]{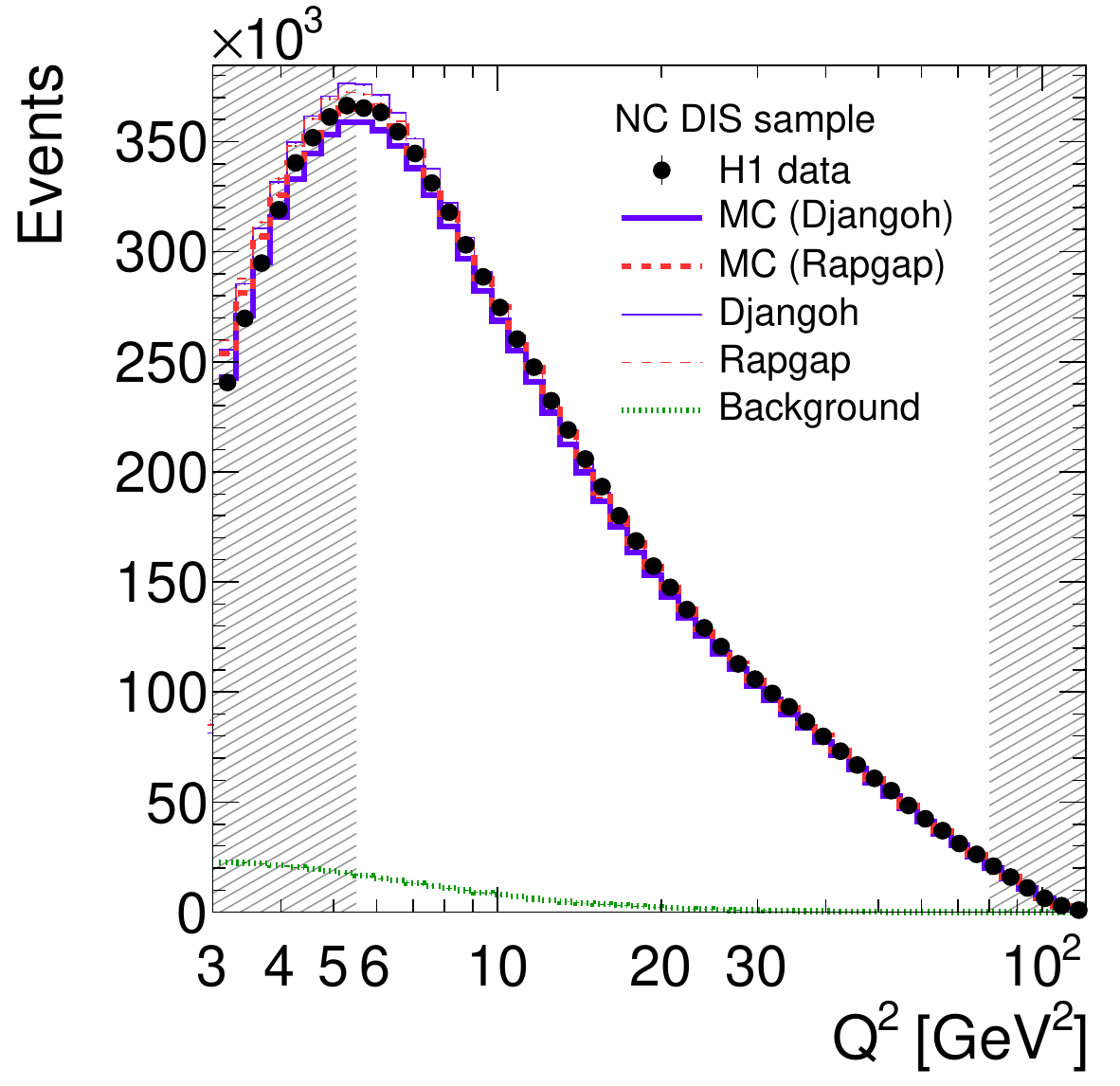}\hskip0.02\textwidth
   \includegraphics[width=0.48\textwidth]{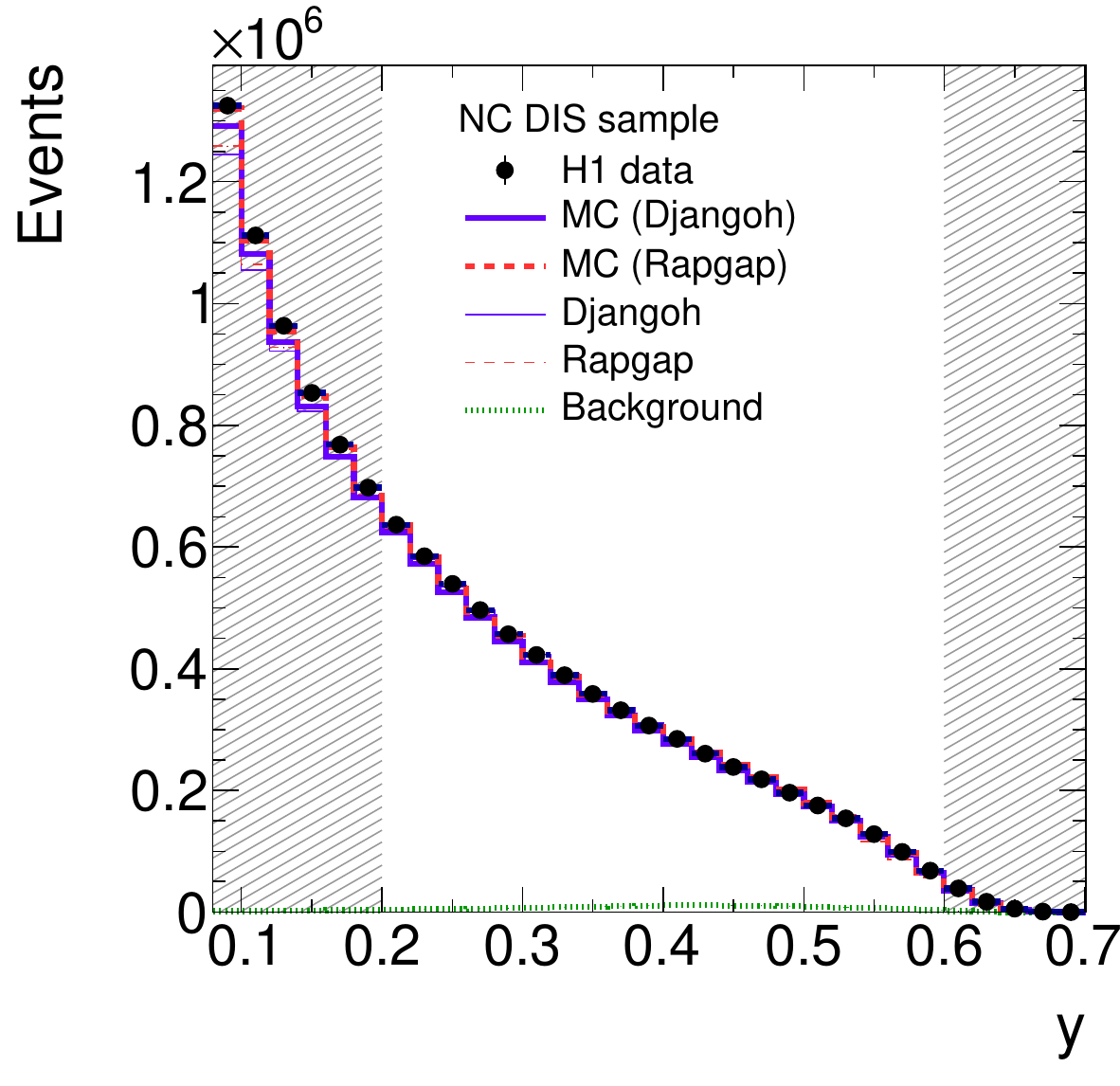}
\end{center}
\caption{
  Distributions of \Qsq\ and $y$ for the selected NC DIS data at detector level. 
The data are compared to predictions obtained from the Rapgap and Djangoh MC simulations, which are weighted to achieve a better description of the data (labelled as `MC').
The non-weighted predictions of the generators are shown as thin lines.
The background is obtained from simulated photoproduction events.
The shaded areas indicate kinematic regions which are considered in the extended phase space of the unfolding procedure.
}
\label{figControlDIS}
\end{figure}

\begin{figure}[htb]
\begin{center}
   \includegraphics[width=0.48\textwidth]{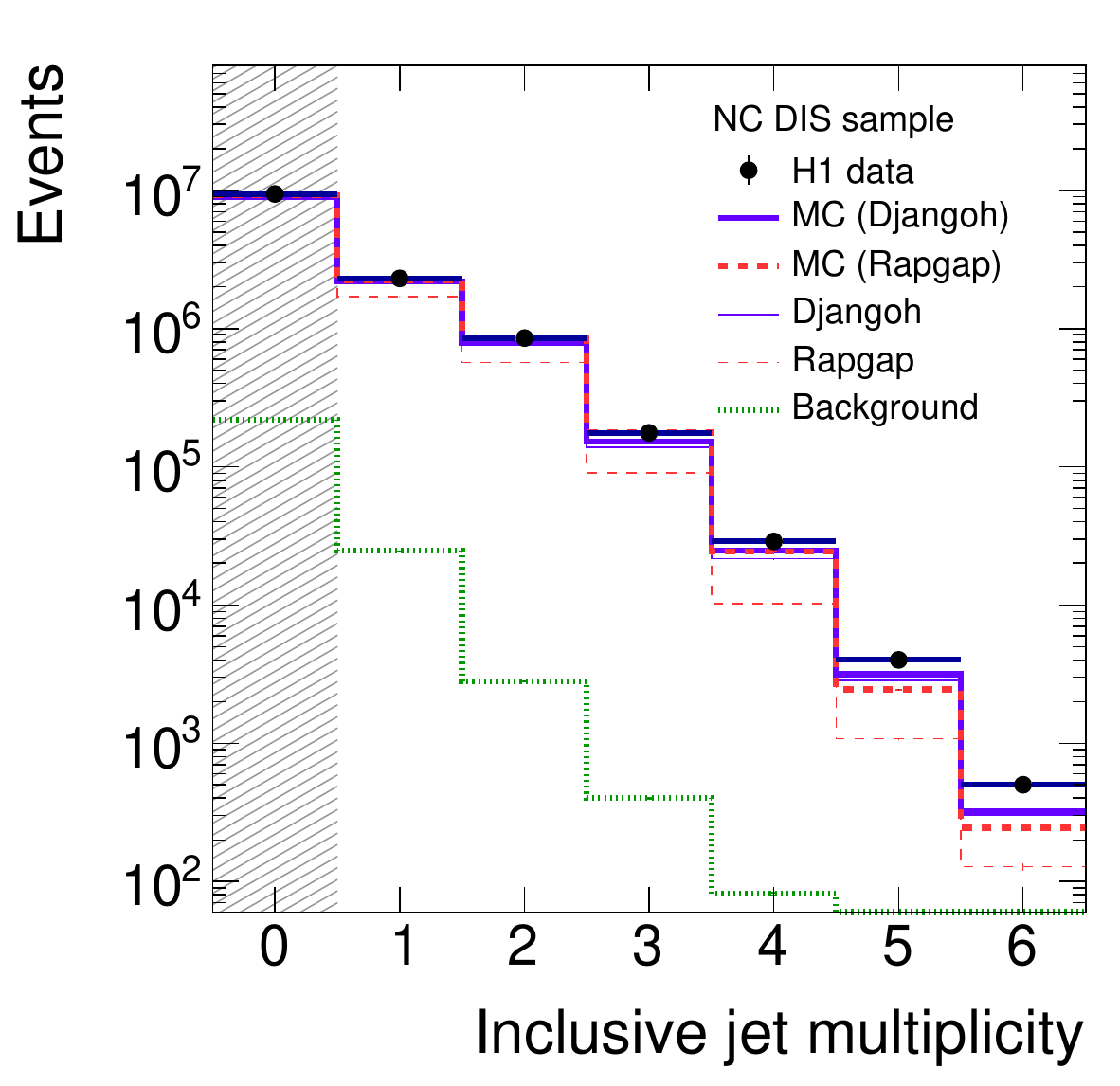}\hskip0.02\textwidth
   \includegraphics[width=0.48\textwidth]{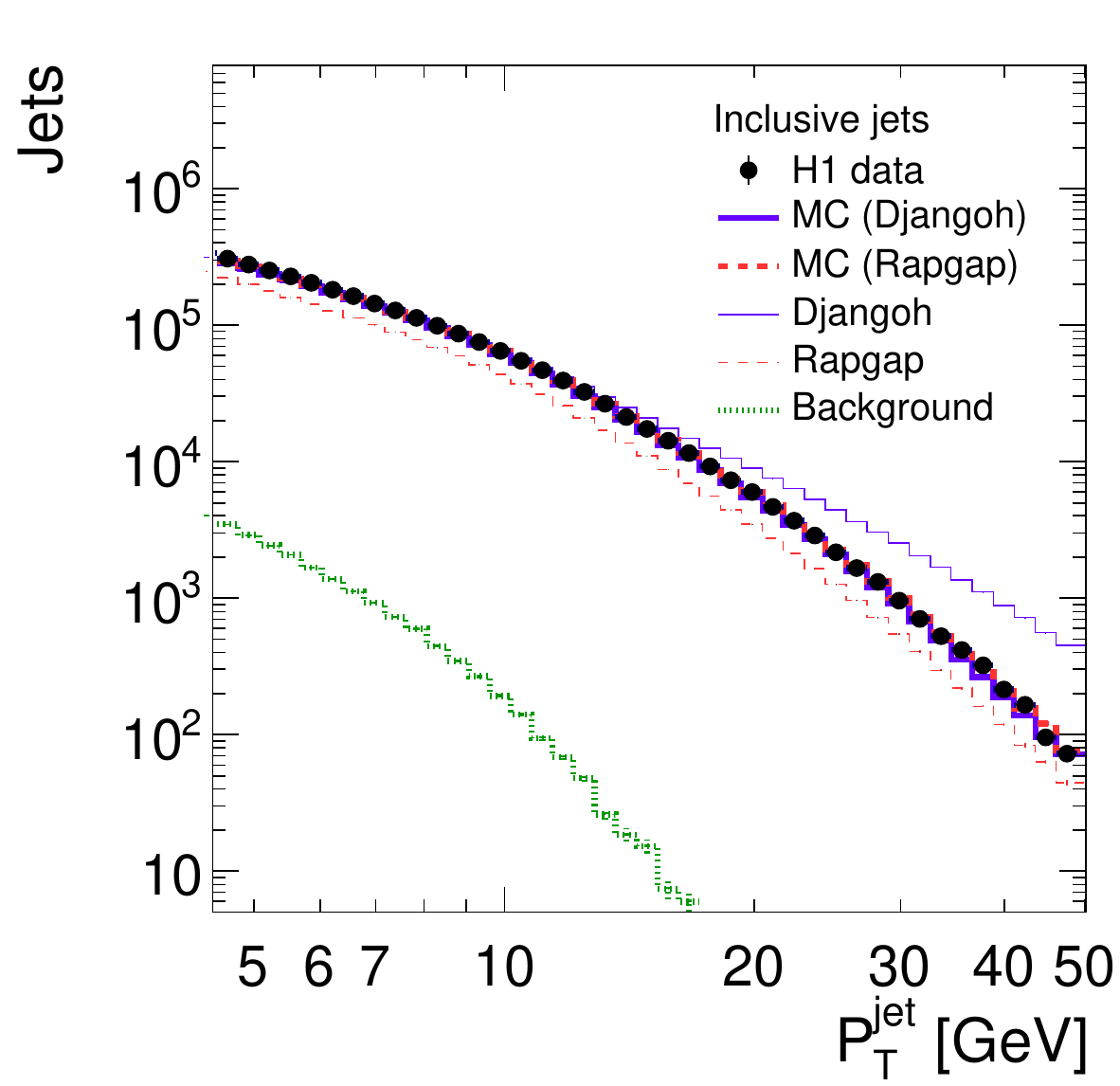}
\end{center}
\caption{
  Distributions of the inclusive jet multiplicity for the NC DIS data and the jet transverse momenta \ptjet\ of the inclusive jet measurement on detector level.
  Other details as in figure~\ref{figControlDIS}.
 } 
\label{figControlJets}
\end{figure}

\begin{figure}[htb]
\begin{center}
   \includegraphics[width=0.48\textwidth]{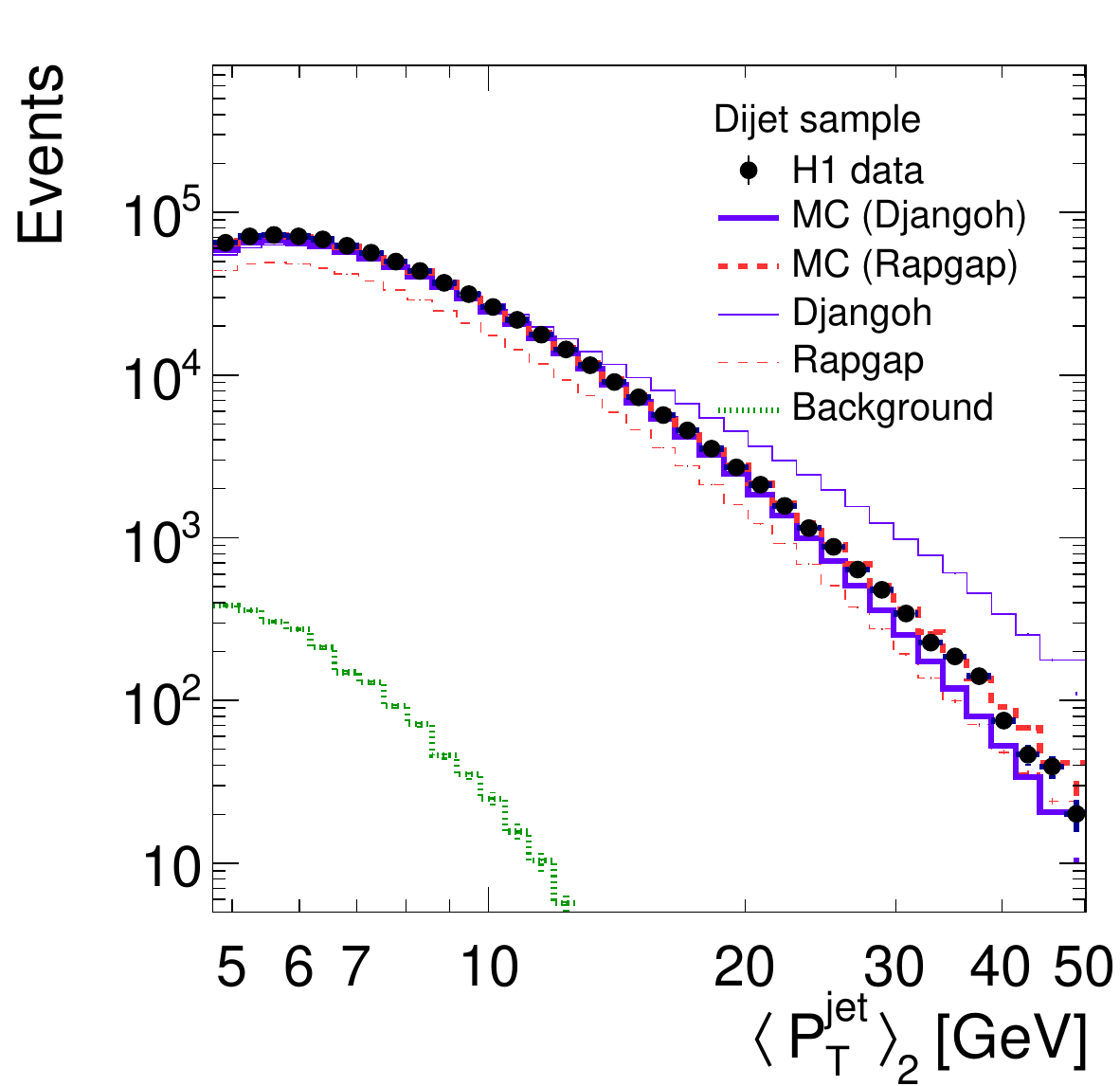}\hskip0.01\textwidth
   \includegraphics[width=0.48\textwidth]{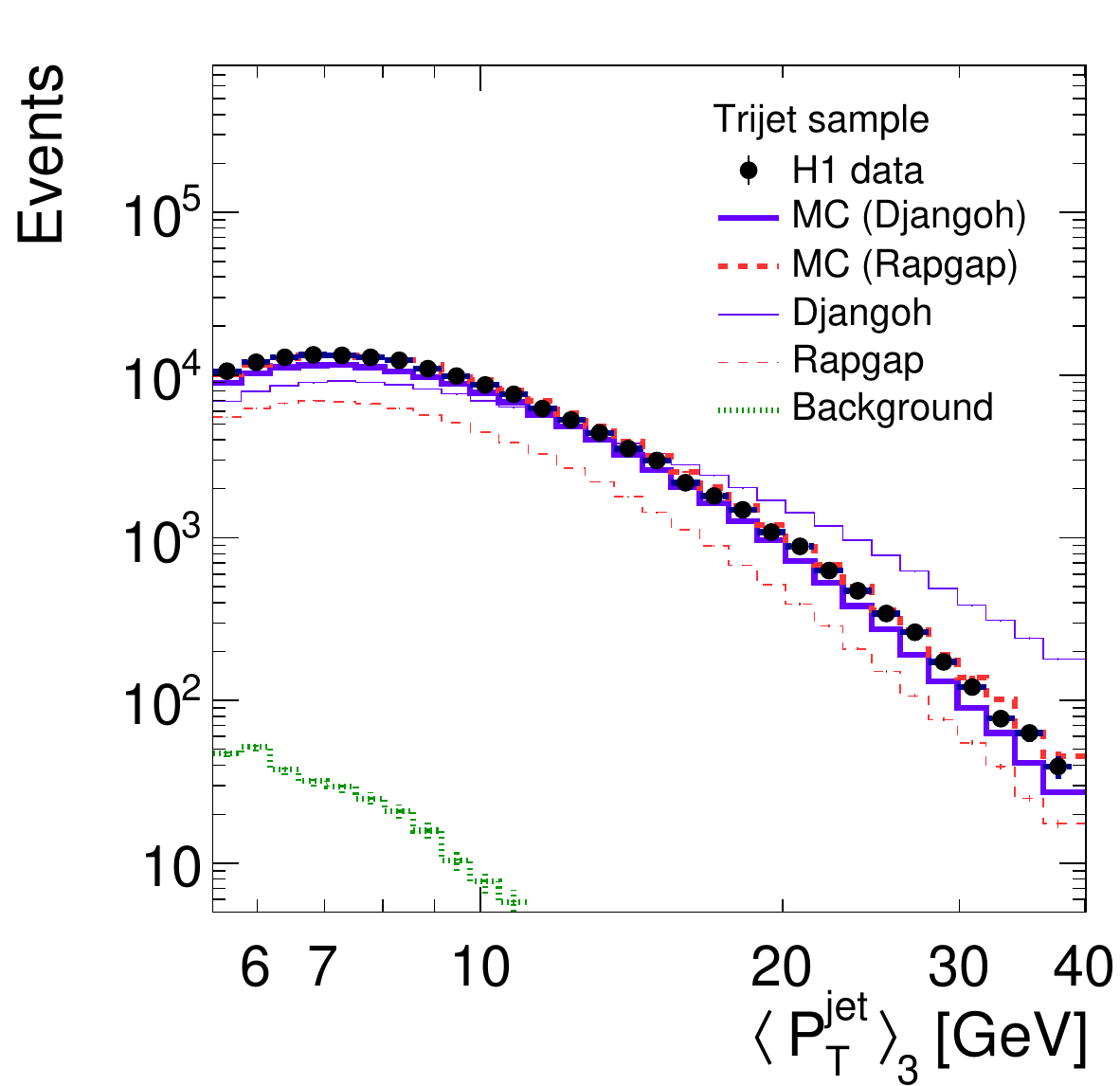}
\end{center}
\caption{
  Distributions of \meanptdi\ of the dijet and \meanpttri\ of the trijet data on detector level for the measurement phase space.
  Other details as in figure~\ref{figControlDIS}.
}
\label{figControlDijet}
\end{figure}

The distributions of the NC DIS kinematic of the weighted and non-weighted MC generators are compared to data in figure~\ref{figControlDIS}. 
The generators Rapgap and Djangoh provide a good description of the NC DIS 
quantities, but both generators have difficulties describing accurately all
jet observables prior to reweighting, in particular
at lower values of \Qsq\ or at higher values of \ptjet, as well as for events with several jets.
This is illustrated for instance in figure~\ref{figControlJets}, which shows the jet multiplicity, defined inclusively, and the distribution of the jet transverse momenta.
The non-weighted Rapgap event generator has in particular problems to describe the jet multiplicity and also predicts too few jets, whereas the overall shape of the \ptjet-distribution is reasonably well modelled.
In contrast, the non-weighted Djangoh prediction has too few jets at low values of \ptjet, and overshoots the data at high values of \ptjet\ significantly.
The distributions of \meanptdi\ and \meanpttri\ for a dijet and trijet event selection, corresponding to the measurement phase space on detector level, are displayed in figure~\ref{figControlDijet}.
Djangoh fails to describe these whereas Rapgap is off in normalisation, but gives a reasonable description of the shape.

Simulations from Pythia~\cite{Pythia} are used to estimate background contributions from
the photoproduction regime with $\Qsq<2\,\GeVsq$, where a hadron is misidentified as the scattered electron.
The normalisation of these events is determined from an event sample where the contribution from photoproduction is enriched.
This normalisation factor of 1.3 is validated with two alternative methods, one which reconstructs and makes use of the charge of the scattered electron candidate,
and one which uses data from the
Photon and Electron Tagger of the luminosity system~\cite{Abt1,etagger}.
The methods agree within 50\,\%.
This value is taken as the normalisation uncertainty on this background.


\section{Unfolding} 
\label{sect:unfolding}
The measured jet data are corrected for effects of detector acceptance, efficiency and resolution using a regularised unfolding as implemented in the TUnfold package~\cite{unfold}.
A detector response matrix $\A$, with elements $a_{ij}$ expressing the probability for an observable 
originating in the generated MC sample from an interval $i$ to be measured in an interval $j$, is 
determined using the average of the reweighted Djangoh and Rapgap MC simulations. 
It accounts for migration effects and efficiencies.
Migrations from the `extended analysis phase space' to the `measurement phase space',
 are included via additional rows and columns in the detector response matrix.
The unfolded hadron level distribution $x$ in the folding equation $y=\A x$, where $y$ is 
the detector level distribution after subtracting
background from photoproduction ($Q^2<2\,\text{GeV}^2$) and $\A$ is the detector response matrix, is given by minimising 
the $\chisq$ expression 
\begin{equation}
\chisq  := (y - \A x)^{\rm T} \V_y^{-1} (y-\A x) +  \tau^2 (x-x_0)^{\rm T}({\LL}^{\rm T}{\LL})(x-x_0)~,
\label{eq:UnfoldingChi2}
\end{equation}
where $\V_y$ is the covariance matrix on detector level, and the second term is a regularisation term to suppress fluctuations of the result.
The matrix ${\LL}$ contains the regularisation condition and is set to unity, and the bias vector $x_0$ represents the 
hadron level distribution of the MC model.
The regularisation parameter $\tau$ is a free parameter and is set to a small value, $\tau=10^{-6}$, where no significant dependence of the results on $\tau$ are observed.
The hadron level distribution $x$ is calculated as the stationary point of equation~\ref{eq:UnfoldingChi2} and thus given by
\begin{equation}
  x = (\A^T\V_y^{-1}\A+\tau^2{\LL}^T\LL)^{-1} \A^T\V_y^{-1}y~.
  \label{eq:UnfoldingX}
\end{equation}

The detector response matrix $\A$ is derived from another matrix ${\M}$~\cite{unfold},
called migration matrix throughout this article.
Here the matrix ${\M}$ is constructed from the observables of the inclusive NC DIS, the inclusive jet, the dijet and the trijet measurements simultaneously. 
The matrix $\V_y$ is constructed from data and takes into 
 account statistical correlations between the NC DIS, inclusive jet, 
 dijet and trijet measurements. 
For example, a trijet event may 
 create entries for each jet in inclusive jet bins and in 
 addition entries in three bins corresponding to the NC DIS, 
 dijet and trijet measurements, respectively. 
Such events populate  the corresponding off-diagonal elements of the matrix $\V_y$. 

The sub-matrix describing the inclusive NC DIS migrations is constructed from the observables \Qsq\ and $y$.
The sub-matrix of the dijet measurement is constructed from migrations in \meanptdi, \Qsq, $y$. 
Migrations of dijet events found in the analysis phase space extended in \etalab, $P_{\rm T}^{\rm jet2}$, \meanptdi, \Qsq\ and $y$ are considered as additional side-bins.
The sub-matrix of the trijet measurement  considers migrations in \meanpttri, \Qsq\ and $y$.
Trijet events found in the extended phase space, i.e.\ by successively enlarging the measurement phase space in the variables \etalab, $P_{\rm T}^{\rm jet2}$ and $P_{\rm T}^{\rm jet3}$ are considered as additional side-bins.
The sub-matrices for the inclusive NC DIS, dijet and trijet measurements are constructed from single entries for each MC event.
For the inclusive jet measurement a jet matching between the detector and hadron level is performed in the laboratory frame, applying a closest pair algorithm with a distance variable $R^2=(\Delta\etalab)^2+(\Delta\phi_{\rm lab}^{\rm jet})^2$, and the maximum distance is set to $R=0.9$.

The bins of the migration matrix are arranged similarly to those in reference~\cite{H1Multijets}.
The main differences are described in the following.
In contrast to the analysis in reference~\cite{H1Multijets}, the phase space of the 
dijets and trijets is not constrained by a cut on the dijet mass.
Therefore there is one variable less for which migrations have to be considered 
and this enables a finer binning in the remaining variables.
The migrations into the phase space from $y>0.6$ are considered as one column 
in the matrix and are constrained by data measured in the range $0.6<y<0.7$.
Large negative correlations between neighbouring bins of the unfolded cross sections are minimised by using 
 two bins on hadron level in \ptjet, \meanptdi\ or \meanpttri, 
which later are combined to obtain the final cross section bin.
In case of inclusive jets, two bins are also combined in \etalab.
The QED correction factors (see section~\ref{sect:mc_simulation}) are determined as bin-wise correction factors after the unfolding.

The entries representing the NC DIS measurement in the matrices $\A$ and $\V_y$, as well as the measurement vector $y$,  
do not include events from the year 2005.
In order to account for this excluded event sample, 
the events of the years 2006 and 2007 obtain an additional
 weight factor for the NC DIS entries, whereas 
the entries of the jet measurements remain unaffected (see section~\ref{sec:event_sel_rec}). 

The resulting migration matrix has a size of 3381 times 12\,300 elements, and the 
matrix $\V_y$ consists of 12\,300 times 12\,300 elements.
For the final cross sections, 320 unfolded values are used.
For the final inclusive jet cross sections, 168 unfolded values are used to calculate the 48 data points.
For the dijet and trijet cross sections, 88 and 56 unfolded values are used to obtain the 48 and 32 data points, respectively.

\section{Definition of the cross sections}
\label{sec:def_of_CS}

The jet cross sections presented are hadron level cross sections corrected for 
radiative QED effects. The cross section in bin $i$ is defined as
\begin{equation}
 \sigma_i = c_i^{\rm rad}\frac{n_i^{\rm unfolded}}{\mathcal{L}}\, ,
 \label{eq:CSdefinition}
\end{equation}
where $n_i^{\rm unfolded}$ is the sum of the unfolded number of events in bin $i$ calculated as the
sum of unfolded values $x_k$ as defined in section~\ref{sect:unfolding},
and $c_i^{\rm rad}$ denotes the correction for QED radiative effects.
The data correspond to an integrated luminosity of $\mathcal{L} = 290\,\invpb$,
where about half of the data was taken with electron beams, otherwise with positrons.

The simultaneous unfolding of the inclusive NC DIS measurement together with the jet measurements, 
respecting all statistical correlations, allows for the determination 
of jet cross sections normalised to the inclusive NC DIS cross section. 
Normalised jet cross sections are defined as the ratio of the 
double-differential absolute jet cross section to the NC DIS cross 
section in the respective \Qsq-bin $i_{q}$:
\begin{equation}
 \sigma_i^{\rm norm} = \frac{\sigma_i}{\sigma_{i_{q}}^{\rm NC}}
 \label{eq:CSNormDefinition}
\end{equation}
The covariance matrix of the statistical uncertainties is determined taking the 
statistical correlations between NC DIS and the jet measurements into account.
The systematic experimental uncertainties are correlated 
between the NC DIS and the jet measurements.
Consequently, the systematic uncertainties cancel to some extent.

\section{Experimental uncertainties}
\label{sec:exp_unc}
Statistical uncertainties, \DStat{}, are determined on detector level and are propagated through the unfolding equations.
The statistical covariance matrix $\V_y$ 
includes all correlations of the inclusive jet data and between the different observables.

The following systematic uncertainties are estimated by determining alternative migration matrices, 
i.e.\ by varying the detector response in the simulations by one standard deviation (reported as `up' and `down' variations where appropriate) of each uncertainty described below:
\begin{itemize}

\item The energy of the scattered lepton is measured with a precision of 0.5\,\% in the SpaCal~\cite{DstarF2cc}, which defines the electron energy uncertainty, \DEe{}.

\item The azimuthal angle of the scattered lepton is measured with a precision of 0.5\,mrad~\cite{DstarF2cc}, which defines the electron angle uncertainty, \DThe{}.

\item The calibration procedure for HFS objects~\cite{H1Multijets,ThesisKogler} results in two independent uncertainty contributions:
the `jet energy scale uncertainty' (JES), \DJES{}, for HFS objects contained in a laboratory jet with high transverse momentum, 
and the `remaining cluster energy scale' (RCES) for the remaining clusters, \DHFS{}. 
Both uncertainties are determined by varying the energy of the HFS object by 1\,\%~\cite{H1Multijets}.
The JES uncertainty is a dominant uncertainty at high values of \pt, and the RCES uncertainty is sizable at low values of \pt.

\item The model uncertainty, \DMod{}, accounts for uncertainties in
  the MC simulation stemming from differences in the modelling of the
  hadronic final states in Djangoh and Rapgap. 
  It is obtained as the difference between the nominal result and the result obtained using the migration matrix from the reweighted Djangoh or the reweighted Rapgap predictions alone.

\item The reweighting uncertainty, \DModRW{}, accounts for changes of the result due to the weighting of the simulations to data. 
Three different sets of weighting constants are determined: 
Two sets with only few weights, where one is focused on more exclusive jet quantities, like e.g.\ $\ptjet$ or $\etalab$ of the leading, sub-leading or third jet in an event, and the other focused on more inclusive jet quantities like the invariant mass \Mjj\ of the two leading jets or \meanptdi\ and \meanpttri.
The nominal set of weights contains a mixture of both and there are altogether sixteen 2D reweighting functions.
The reweighting uncertainty is then defined as the difference of the nominal set to one of the alternative sets.
Using the other alternative set leads to very similar uncertainties.

\end{itemize}
The uncertainty on the cross sections is obtained by propagating the difference to the nominal response matrix $\A$ 
to the hadron level in equation~\ref{eq:UnfoldingX}.
This calculation is performed with simulated events in order
to avoid fluctuations caused by the limited data statistics.
The quoted relative uncertainties are obtained by dividing the estimated absolute uncertainties by the data cross sections.

The following additional systematic uncertainties are further assigned to the jet cross section without having to alter the response matrix:
\begin{itemize}
\item The uncertainty of the luminosity measurement and the overall normalisation is known with a precision of 2.5\,\%~\cite{AaronQED} (denoted as \DNorm{}), 
  which includes a 1.5\,\% normalisation uncertainty on each data taking period~\cite{AaronQED,H1NCDISHighQ2}. 
  The latter contribution does not fully cancel for normalised jet cross sections, and thus the normalised jet cross sections obtain a luminosity uncertainty of 0.8\,\%.

\item A correlated uncertainty due to the algorithm to suppress electronic noise in the LAr is found to change by 0.5\,\% the inclusive jet, by 0.6\,\% the dijet and by 0.9\,\% the trijet cross sections~\cite{H1Multijets} (denoted as \DLAr{\csdsub}).

\item An uncorrelated uncertainty of 1\,\% is assigned to each data point to account for various smaller sources of uncertainties, such as the momentum resolution of the electron, uncertainties introduced due to combining of cluster and track information, inefficiencies of SpaCal clusters or e.g.\ uncertainties on the track or vertex reconstruction.

\item The uncertainty on the QED radiative corrections, \DRad{}, based on Heracles, is estimated as half of the difference between the correction factors obtained with Djangoh or Rapgap. 
The statistical uncertainty on that correction factor reaches up to 5\,\% for high \pt\ data points and is added quadratically to obtain the quoted uncorrelated uncertainties.
\end{itemize}

The analysis also takes into account:
\begin{itemize}
\item The normalisation uncertainty on the background events, as well as their statistical uncertainty,
are added to the covariance matrix prior to the unfolding.
The resulting statistical and normalisation uncertainty of the background is found to be smaller than 0.1\,\% and is not included in the results.

\item The estimated statistical uncertainty of the MC simulations (\DMCSt{}) on the elements $a_{ij}$ of the detector response matrix are propagated to the
cross sections~\cite{unfold} and are typically around one third of the data statistical uncertainty. 
The effect of statistical correlations of an element 
$a_{ij}$ with another element $a_{kl}$, similar to the effect 
giving rise to the off-diagonal elements in the matrix $\V_y$, is 
neglected in this analysis. 
Given that the MC statistical uncertainty, when neglecting these effects, is only one third of the 
data statistical uncertainty, the effect is expected to be small.
\end{itemize}

The calculation of the statistical uncertainty of the normalised jet cross sections takes the
 statistical correlations between the jet observable and the NC DIS measurement into account.
Systematic uncertainties of normalised jet cross sections are calculated separately for each source,
 where numerator and denominator are varied simultaneously.

The quadratic sum of all experimental uncertainties is denoted as systematic uncertainty {\DSys{\csdsub}}, 
where uncertainties of the same sign are summed up in order to calculate the `plus' and `minus' variations.

The jet cross sections measured at higher values of \Qsq\ are statistically independent 
of the low-\Qsq\ data points, discussed above.
For a common analysis of the low- and high-\Qsq\ jet measurements, the uncertainties on the reconstruction of the scattered lepton
 are uncorrelated between the two kinematic regimes, since different detectors were used for the lepton reconstruction.
The uncertainties on the reconstruction of jets are taken to be correlated.

For the calculation of the covariance matrix of the experimental uncertainties,
the normalisation, the LAr noise, and uncertainties on the electron reconstruction 
are taken as correlated between the data points.
Uncertainties on the cross sections due to the MC models as well as the jet energy 
scale uncertainties are divided into two equally sized components\footnote{Whenever an 
uncertainty is split up into a correlated and uncorrelated part, the quadratic sum of 
those two yield the initial value of the uncertainty.}, where one part is treated
as correlated and the other as uncorrelated.
The statistical uncertainty of the MC models, the uncorrelated uncertainty of 1\,\% and the
uncertainty on the radiative corrections are treated as uncorrelated for the calculation of the covariance matrix.

%
\section{QCD calculations}
\label{sec:nlo}
The absolute and normalised jet cross sections are compared to theoretical
predictions in next-to-leading order (NLO), approximate next-to-next-to-leading order (aNNLO)
and full NNLO in pQCD.
The pQCD calculations are corrected for hadronisation effects by applying multiplicative bin-wise hadronisation correction factors.
Normalised jet cross section predictions are calculated by dividing predictions
for jet cross sections in the numerator by inclusive NC DIS cross sections in the denominator.
QED radiation is not included in the theoretical predictions, but the running of the 
electromagnetic coupling~\cite{EPRC,AemParam} with \Qsq\ is taken into account.

The different theoretical calculations are summarised in table~\ref{tab:theo} and are explained in the following:
\begin{table}[t!hbp]
  \footnotesize
  \begin{center}
    \begin{tabular}{l|ccc}
      \hline
      \hline
      Predictions     &\textbf{NLO} & \textbf{aNNLO}  & \textbf{NNLO}  \\
      \hline
      Program for jet cross sections        &  nlojet++    &   JetViP  &  NNLOJET \\
      pQCD order           & NLO      & approximate NNLO  & NNLO \\
      Calculation detail   & Dipole subtraction & Phase space slicing  & Antenna subtraction\\
      & & NNLO contributions \\
      & & from unified threshold &   \\
      & & resummation formalism \\
      \hline 
      Program for NC DIS   &   QCDNUM   &   APFEL  &   APFEL \\
      Heavy quark scheme  & ZM-VFNS &  FONLL-C & FONLL-C \\
      Order       & NLO      & NNLO & NNLO \\
      \hline
      PDF set   &  NNPDF3.0\_NLO   &   NNPDF3.0\_NNLO  &  NNPDF3.0\_NNLO \\
      \hline
      \asmz    &  0.118   &  0.118   &  0.118 \\
      \hline
      Hadronisation corrections   &     \multicolumn{3}{c}{Djangoh and Rapgap}  \\
      \hline
      Available for \\
      (Normalised) Inclusive jet    &  \checkmark   &   \checkmark  & \checkmark \\
      (Normalised) Dijet            &  \checkmark   &   \checkmark  & \checkmark \\
      (Normalised) Trijet           &  \checkmark   &     &  \\
      \hline
      \hline
    \end{tabular}
    \caption{Summary of the theory predictions for the normalised jet cross sections.
}
    \label{tab:theo}
  \end{center}
\end{table}
\begin{itemize}
  \item Predictions for jet cross sections in NLO are obtained using the program nlojet++~\cite{Nagy1,Nagy2,Nagy3}.
    The matrix elements are calculated in the $\overline{\mbox{\rm MS}}$ scheme for
    five massless quark flavours.
    The calculations are interfaced to FastNLO~\cite{fastNLO1,fastNLO2} and a value of the strong coupling constant
    of $\asmz=0.118$ and the PDF set NNPDF3.0\footnote{The LHAPDF~\cite{lhapdf} PDF set `NNPDF30\_nlo\_as\_0118' is used for NLO calculations and the set `NNPDF30\_nnlo\_as\_0118' for aNNLO and NNLO calculations.}~\cite{NNPDF30} is used. 
    The PDF set NNPDF3.0 was chosen, because for its determination, besides inclusive DIS data, jet production data from the LHC~\cite{ATLASjets7,CMSjets7,CMSjets2p76} and the Tevatron~\cite{CDFjetsRunI,CDFjetsRunII,D0jets} were used.
  \item  Predictions for inclusive jet and dijet cross sections 
    in approximate next-to-next-to-leading order, applying the unified threshold resummation formalism~\cite{aNNLO},
    are obtained \cite{aNNLOprivate} using the program JetViP~\cite{JetVip1,JetVip2,JetVip3,JetVip4,JetVip5}.
The approximation is expected to agree with exact calculations at very large jet transverse momenta, possibly beyond the reach of this analysis.
  \item  Predictions in full next-to-next-to-leading order pQCD (NNLO) are obtained \cite{NNLOprivate} using the program NNLOJET~\cite{NNLO,NNLO17,NNLOJET}
    for inclusive jet and dijet production, where the infrared and collinear singularities are cancelled using the antenna subtraction formalism~\cite{antenna1,antenna2,antenna3,antenna4}.
    These predictions are available for this analysis only at a fixed
choice of PDF (NNPDF3.0 NNLO) and at fixed $\asmz$. The scale is given
by $\mu_r^2=\mu_f^2=(Q^2+P_T^2)/2$, only allowing for variations of the scale up and down by a factor of two.
    \item In order to account for small numerical differences of different fixed order predictions for jet cross sections, 
      originating from effects such as limited statistical precision
      or small numerical 
      differences of input constants~\cite{IRinsens}, the aNNLO and NNLO cross sections presented here are calculated 
      as multiplicative corrections to the NLO predictions obtained from nlojet++.
This procedure is only of
      relevance for the JetVip predictions at high values of \ptjet.
  \item   For inclusive NC DIS cross sections in NLO the program
    QCDNUM~\cite{qcdnum} with the zero-mass variable-flavour-number
    scheme (ZM-VFNS)~\cite{zmvfns} is used, with 
    the PDF set NNPDF3.0 and $\asmz=0.118$.
  \item   For inclusive NC DIS cross sections in NNLO the program APFEL~\cite{apfel1,apfel2} with the FONLL-C heavy quark scheme~\cite{fonllc} is used.
    These inclusive NC DIS cross sections are calculated as in the NNPDF3.0 PDF extraction~\cite{apfel2}.
    The predictions employing the FONLL-C scheme are about 2.5\,\% higher at $\Qsq\approx 6\,\GeVsq$ and 1\,\% higher at $\Qsq\approx  70\,\GeVsq$ than predictions employing the ZM-VFNS in NNLO~\cite{zmnnlo1,zmnnlo2,zmnnlo3,zmnnlo4}.
\end{itemize}

    The dependence of the jet cross section predictions on the choice of the PDF set is shown in figure~\ref{figPDFs}.
    Predictions obtained with the NNPDF3.0 are compared to predictions
    using the PDF sets MMHT~\cite{MMHT}, CT14~\cite{CT14},
    HERAPDF2.0~\cite{HERA20} and ABMP16\cite{Alekhin:2016uxn}.
    All PDF sets shown are obtained in NNLO precision\footnote{Since the NNLO predictions for the jet cross sections could not be obtained with varying NNLO PDFs, the PDFs are convoluted with NLO matrix elements for this study.} with a value for the strong coupling constant of $\asmz=0.118$.
    For comparison, also the NNPDF3.0 PDF set extracted at NLO is displayed in figure~\ref{figPDFs}.
    Most predictions are consistent within $2\%$, but the ABMP16
    prediction differs by up to $10\%$ at high $\ptjet$. 

\begin{figure}[htb]
\begin{center}
  \includegraphics[width=1.\textwidth]{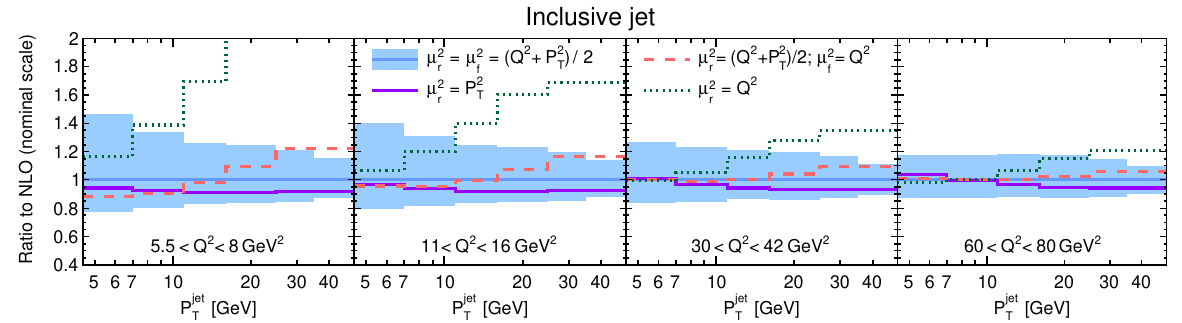}
  \includegraphics[width=1.\textwidth]{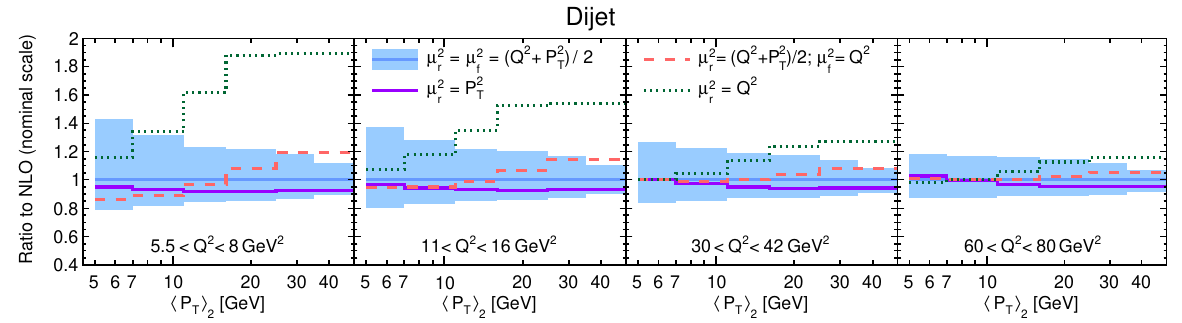}
   \includegraphics[width=1.\textwidth]{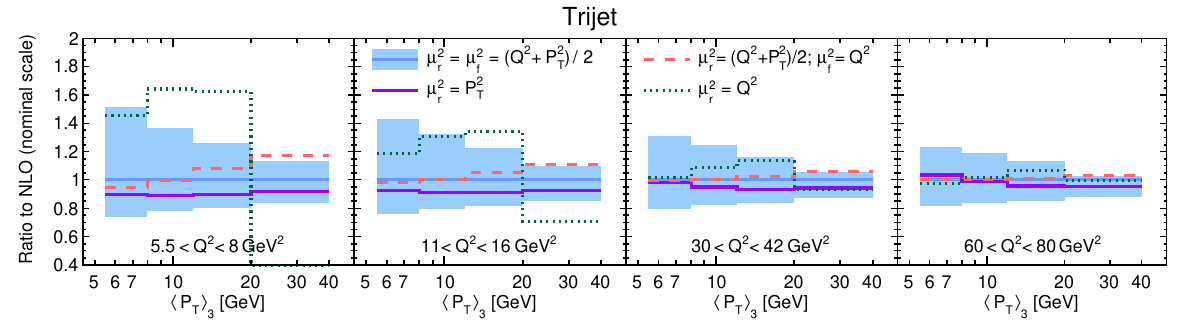}
\end{center}
\caption{
  Comparison of NLO predictions obtained with scale choices of $\mur^2=\muf^2=\tfrac{1}{2}(\Qsq+\pt^2)$, $\mur^2=\muf^2=\pt^2$, $\mur^2=\muf^2=\Qsq$, 
and $\mur^2=\tfrac{1}{2}(\Qsq+\pt^2)$ with $\muf^2=\Qsq$ for selected \Qsq\ bins of the inclusive jet, dijet
  and trijet cross sections,
  using the NNPDF3.0 PDF set.
  The shaded area around the theory predictions indicates the scale uncertainty on the nominal scale choice of $\mur^2 = \muf^2 = \tfrac{1}{2}(Q^{2}+ P_{T}^2)$ as described in the text.
}
\label{figScales}
\end{figure}

    For the calculation of the jet cross sections, the squares of the factorisation and
    the renormalisation scales, $\muf^2$ and $\mur^2$, are taken to be $\tfrac{1}{2}(Q^2 + \pt^{2})$. 
    This choice ensures that the squared  scales are always greater than 12\,\GeVsq, where the
    PDFs are well in the perturbative regime~\cite{HERA20,PertStab}.
    Predictions in NLO using other choices for $\mur^2$ and $\muf^2$ are displayed in figure~\ref{figScales}.
    The predictions with scales which involve $\pt$ are  covered within the scale uncertainty (see below).
    A scale choice of $\mu^2=\Qsq$ results in very high cross sections at high values of \pt\ as 
    compared to other scale choices, because \Qsq\ may be small compared to $\pt^2$. 
    Such a choice for the two scales is  disfavoured by the data (see section~\ref{sec:res}).

    The uncertainty on the pQCD predictions is estimated by varying the renormalisation and factorisation
    scales by factors of 0.5 and 2, using the 6-point scale variation prescription~\cite{sixpoint1,sixpoint2}
    where opposite variations of the two scales are excluded.
    The highest and the lowest cross sections out of the resulting six variations are displayed as scale uncertainty.
    For the calculations of the normalised jet cross sections scale uncertainties on the NC DIS cross sections are neglected, 
    because they are expected to be small compared to the scale uncertainties of the jet cross section predictions.
    The uncertainty originating from the PDFs is small compared 
    to the size of the scale uncertainty.

Corrections to the fixed order pQCD predictions for hadronisation 
effects are calculated for each data point
using the Monte Carlo event generators Djangoh and Rapgap with 
QED radiative effects switched off.
These corrections are defined as the ratio of the 
cross section at hadron level to the cross 
section at the parton level, i.e.\ after parton showers and with coloured partons as input to the jet algorithm. 
The correction factors are determined
as the average values obtained from Djangoh and Rapgap simulations.
The hadronisation uncertainty is taken as 
half of the difference between Djangoh and Rapgap.
For inclusive and dijet cross sections the hadronisation 
corrections are typically around 0.86 to 0.97 and agree to better 
than 5\,\% between the two MC programs. 
For trijet cross sections the hadronisation correction factors are in the range 0.73 to 0.86
with up to $8\%$ differences between the two  MC models.

\section{Cross section measurements}
\label{sec:res}
In the following, the differential cross sections,
corresponding to the measurement phase space given in
Table~\ref{tab:ps}, are presented for absolute and normalised inclusive jet,
dijet and trijet production at hadron level and are compared to the predictions.
The measurements are shown in tables~\ref{tab:IncJet} to \ref{tab:NormIncJetHQ}
and figures~\ref{figCorrelations} to \ref{figNormTrijetRatio}.
The agreement of the various predictions with the data is judged by calculating values of $\chi^2/\ndf$~\cite{H1Multijets}.
Here \ndf\ is the number of data points in the calculation.
The covariance matrix is calculated from the experimental and the hadronisation uncertainty,
while half of the hadronisation uncertainty is taken as uncorrelated and the other half is taken to be correlated.
PDF uncertainties are not included, because they were not available for the aNNLO and NNLO jet cross section predictions.
The values of $\chisq/\ndf$ for the absolute and the normalised jet cross sections are listed in table~\ref{tab:chisq}.
\begin{table}[t!hbp]
  \footnotesize
  \begin{center}
    \begin{tabular}{lccccccc}
      \hline
      \hline
              &  ~~\ndf~~ &\multicolumn{6}{c}{Value of $\chi^2/\ndf$} \\
              &   & NLO & aNNLO  & NNLO & NLO & aNNLO  & NNLO \\
       \hline
                     &&\multicolumn{3}{c} {Absolute jet cross sections} & \multicolumn{3}{c} {Normalised jet cross sections} \\
      Inclusive jet at low-\Qsq           & 48 &1.7&2.1&0.7&1.9&1.6&1.0\\ 
      Inclusive jet at low- and high-\Qsq & 78 &1.7&2.0&1.1&1.9&2.2&1.5\\ 
      Dijet  at low-\Qsq                  & 48 &1.4&1.9&0.4&1.6&1.7&0.6\\ 
      Trijet at low-\Qsq                  & 32 &0.6&   &   &0.6&   & \\
      \hline
      \hline
    \end{tabular}
    \caption{Summary of values of $\chi^2/n_{\rm dof}$ for absolute and normalised jet cross sections for the NLO, aNNLO and NNLO predictions, whenever those are available. }
    \label{tab:chisq}
  \end{center}
\end{table}

All calculations provide a reasonable value of $\chisq/\ndf$, taking into account 
the fact that uncertainties on the theory predictions, such as scale variations 
or the PDF uncertainties, are not included.
Having neglected these uncertainties may explain the larger value of $\chisq/\ndf$ observed
for the experimentally more precise normalised jet cross sections as compared to the absolute ones.
In the following, the individual measurements are discussed in
 greater detail.

\subsection{Statistical correlations of cross sections}
\begin{figure}
  \centering
  \includegraphics[width=0.85\textwidth]{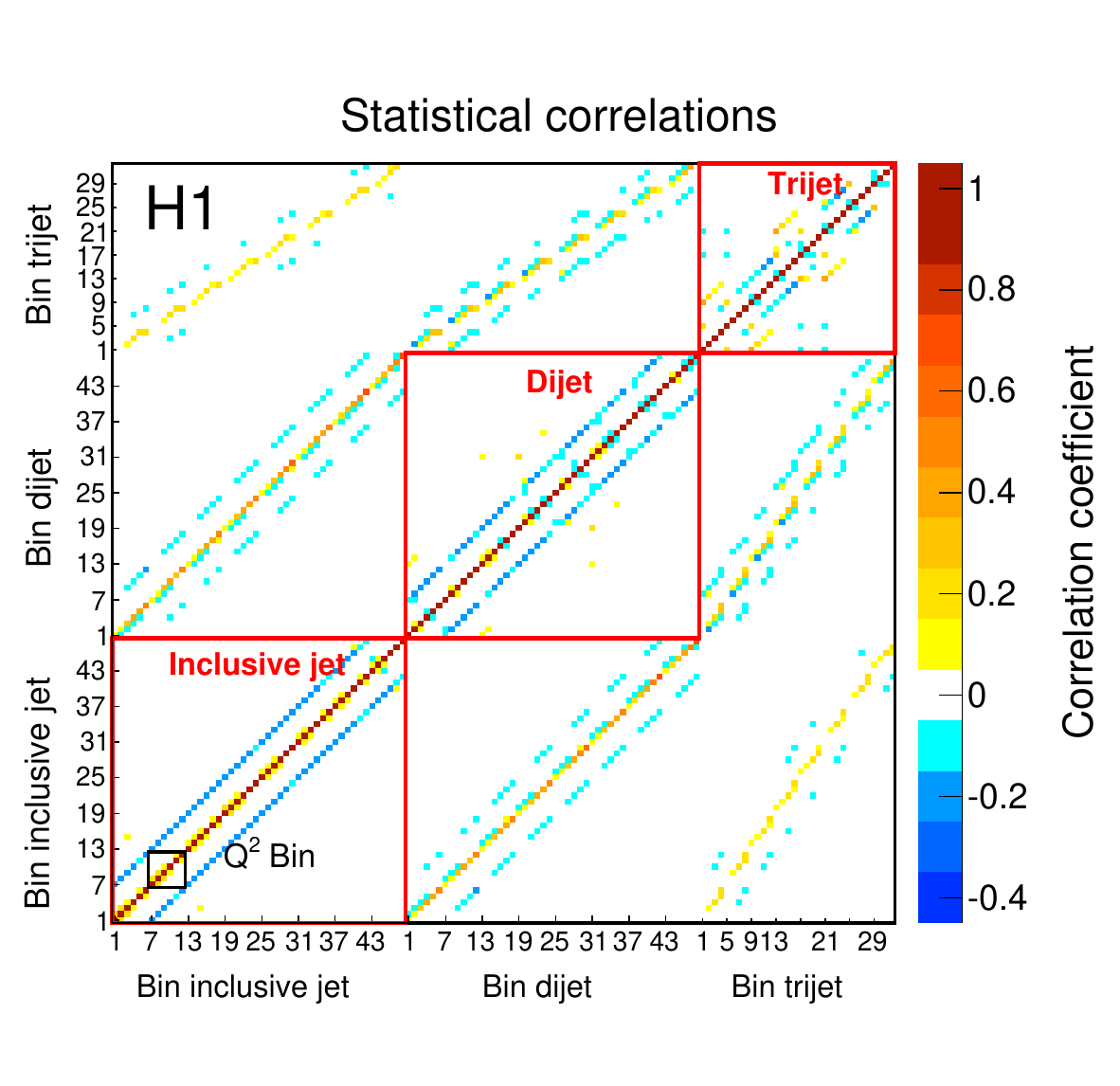}
  \caption{
    Matrix of statistical correlation coefficients of the unfolded cross sections.
    The bin labels are specified in table~\ref{tab:binlabels}.
  }
  \label{figCorrelations}
\end{figure}

Statistical correlations of the data points are displayed in figure~\ref{figCorrelations}.
Large positive correlations are present in particular in the highest \pt\ bins between the inclusive jet,
dijet and trijet cross sections.
Negative correlations of typical size $-0.2$ are present between adjacent bins in \Qsq.
The correlations between adjacent \pt-bins are small,
because the final data points are formed by combining unfolded smaller bins, 
which have sizable negative correlations.
Positive correlations due to the counting of multiple jets in events are found to be small as compared 
to the predominant negative correlations arising in the unfolding procedure
caused by the finite detector resolution.
The individual entries of the correlation matrix are provided elsewhere~\cite{results}. 
The correlations of the normalised jet cross section measurements are 
very similar to those displayed in figure~\ref{figCorrelations}.

\subsection{Inclusive jet cross section}
\label{sec:Ijets}
\begin{figure}[htbp]
  \centering
  \includegraphics[width=0.95\textwidth]{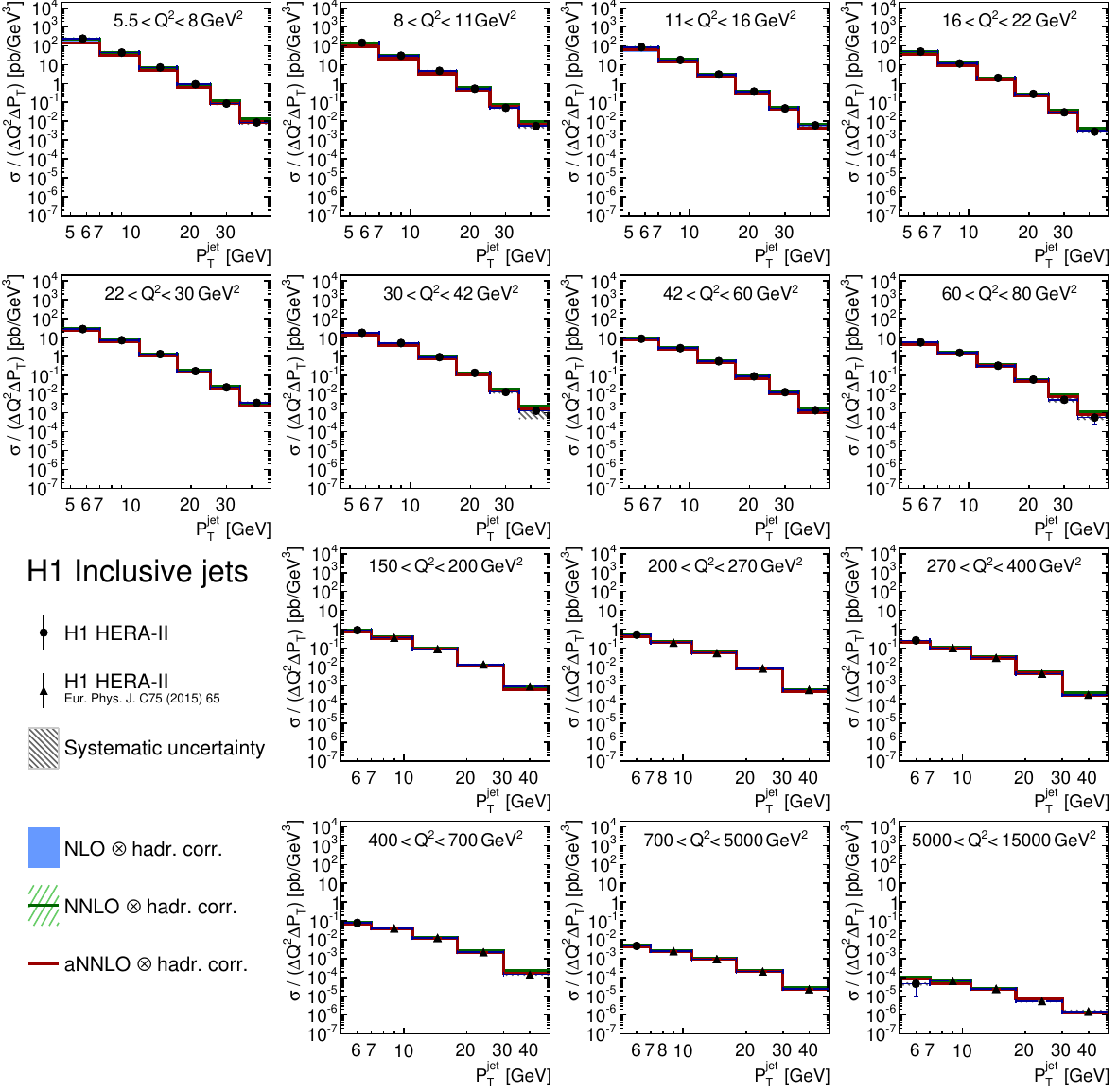}
  \caption{
    Cross sections for inclusive jet production in NC DIS as a function of \ptjet\ for different \Qsq\ ranges.
    The new data are shown as full circles whereas full triangles indicate previously published data.
    The error bars indicate statistical uncertainties.
    The hatched area indicates all other experimental uncertainties added in quadrature.
    The NLO and NNLO QCD predictions corrected for hadronisation effects together with their uncertainties from scale variations are shown by the shaded and hatched band, respectively.
    The aNNLO calculations are shown as full red line.
    The cross sections in each bin are divided by the bin-size in \ptjet\ and \Qsq.
  }
  \label{figInclJet}
\end{figure}

\begin{figure}[htbp]
  \centering
  \includegraphics[width=0.95\textwidth]{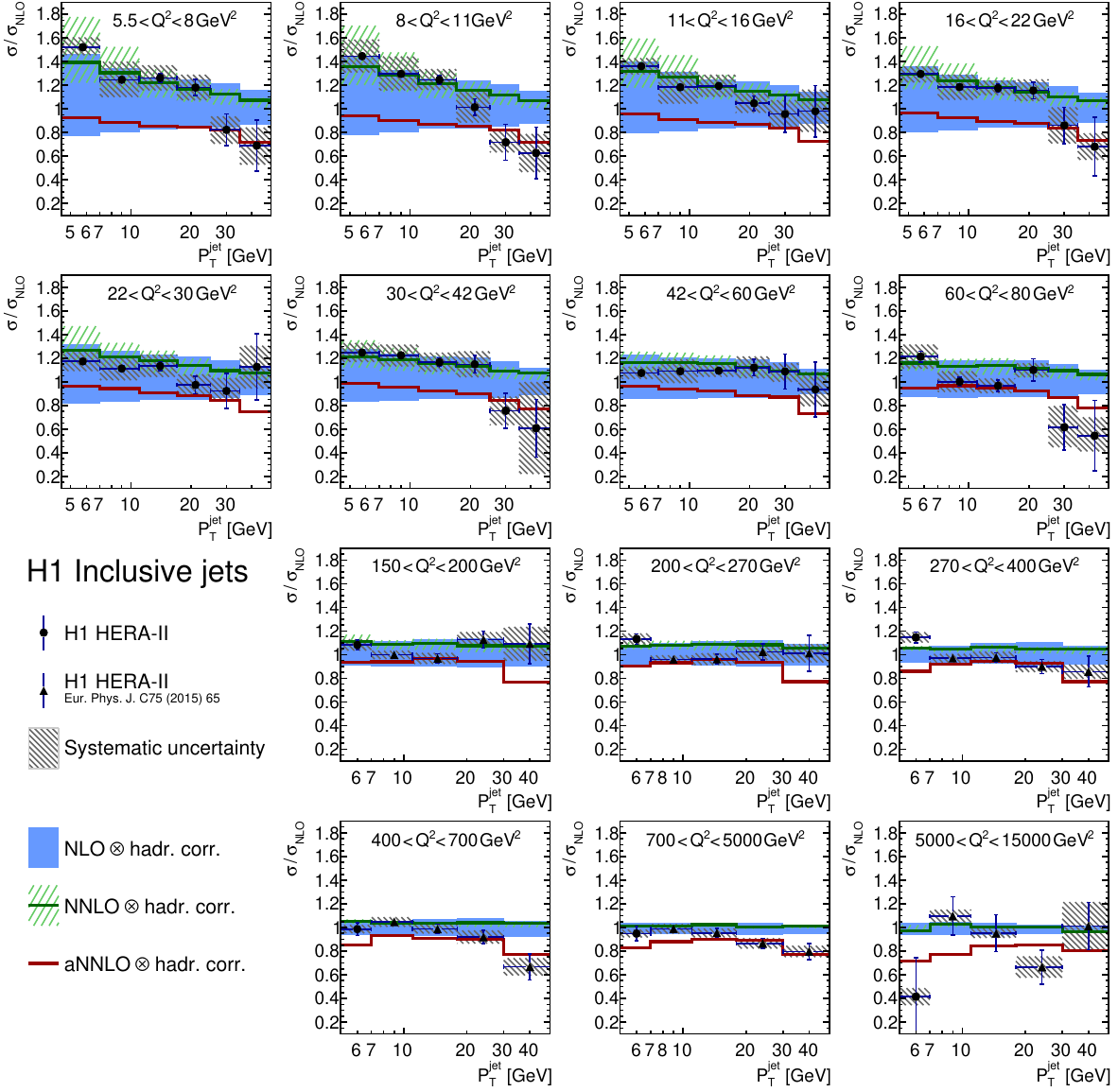}
  \caption{
    Ratio of inclusive jet cross sections to the NLO predictions and ratio of aNNLO 
    and NNLO to NLO predictions as function of \Qsq\ and \ptjet.
    More details are given in the caption of figure~\ref{figInclJet}.
  }
  \label{figInclJetRatio}
\end{figure}

The measured double-differential inclusive jet cross sections as function of \ptjet\  and \Qsq\ for low and high values of \Qsq\
are compared to different theoretical predictions in figure~\ref{figInclJet}. 
Ratios of the data and of the predictions in aNNLO and full NNLO to the NLO predictions are provided in figure~\ref{figInclJetRatio}.

\begin{boldmath}
\subsubsection{Inclusive jet cross sections at low-$\Qsq$ ($\Qsq<80\,\GeVsq$)}
\end{boldmath}

The data points at lower values of \ptjet\ have a significantly lower statistical than systematic uncertainty. 
At higher values of \ptjet\ these two uncertainties are of about equal size.
The dominant experimental uncertainties are \DJES{}, \DHFS{}, \DMod{} and \DModRW{}.

In general, the data are well described by the predictions within experimental and theoretical uncertainties.
The central values of the  NLO and the aNNLO predictions are lower than the data in most bins, while the NNLO predictions 
have a tendency to lie above the data.
At lower values of \Qsq, NLO predicts harder \ptjet-spectra than observed.
The NNLO predictions give a good description of the \ptjet-distributions
following the excellent value of $\chisq/\ndf$ (table~\ref{tab:chisq}).
The aNNLO predictions provide a reasonable description of the shape of the \ptjet-distributions.

\begin{figure}[htb]
\begin{center}
  \includegraphics[width=1.\textwidth]{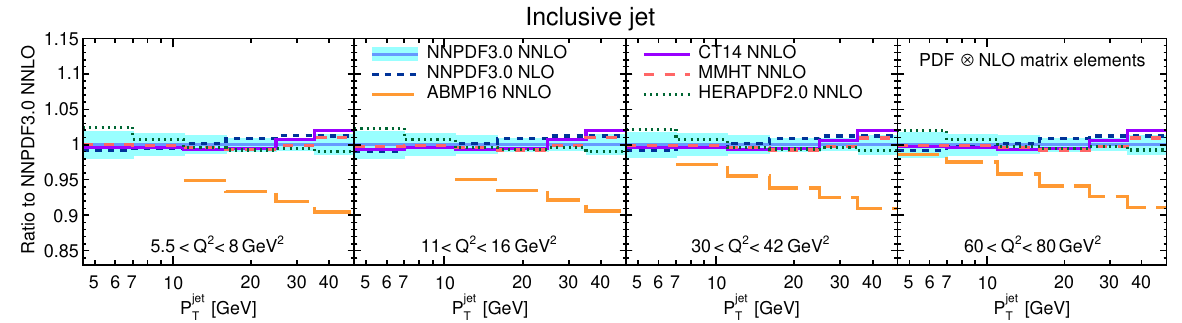}
  \includegraphics[width=1.\textwidth]{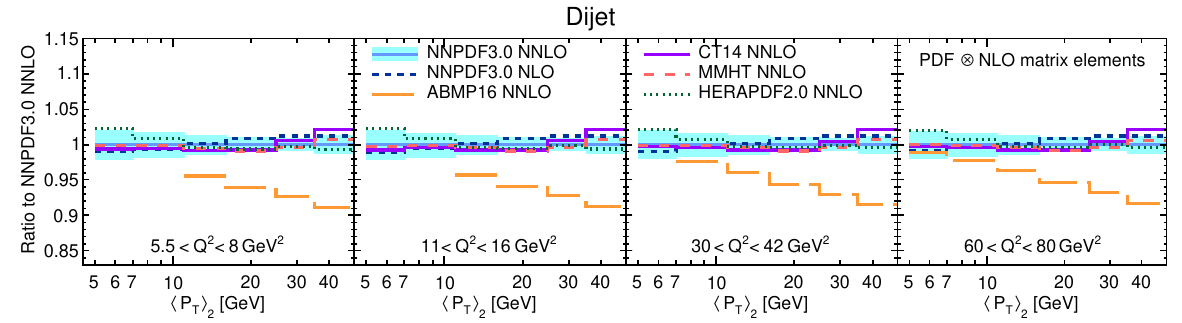}
   \includegraphics[width=1.\textwidth]{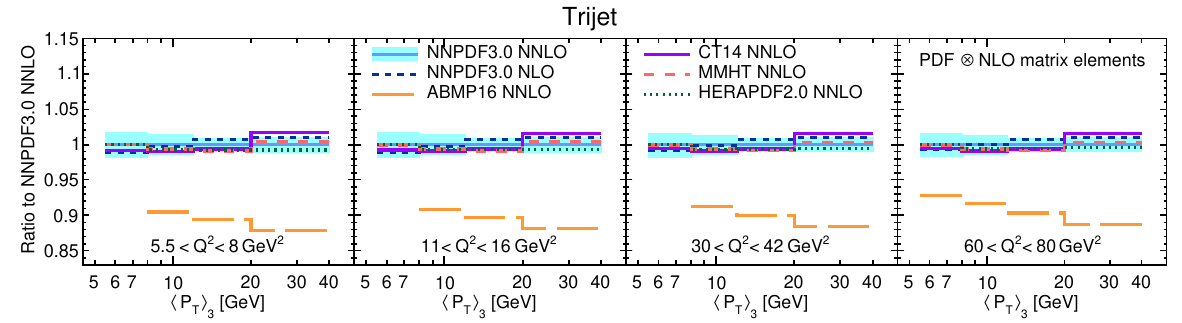}
\end{center}
\caption{
  Comparison of predictions obtained with different PDF sets for selected \Qsq\ bins of the inclusive jet, dijet
  and trijet cross sections.
  The ratios of
  predictions obtained using 
  the CT14, MMHT, HERAPDF2.0 and ABMP16 to the NNPDF3.0 NNLO PDF set are displayed, where NLO matrix elements have been used in all cases.
  For comparison, also the ratio of NNPDF3.0 PDF set extracted at NLO precision  to NNPDF3.0 NNLO is shown.
  The shaded area indicates the PDF uncertainty determined using NNPDF3.0 NNLO.
  The predictions based on ABMP16 shown here are valid only for five active flavours, and hence are shown only for $(Q^2 + \pt^{2})/2>25\,\text{GeV}^2$. 
}
\label{figPDFs}
\end{figure}

Some of the differences between data and predictions may be attributed to the PDFs which are evaluated in the range $x>0.08$ ($x>0.04$) to obtain the predictions for $\ptjet>35\,\GeV$ ($\ptjet>25\,\GeV$). 
In this high-$x$ domain the gluon PDF is not well known and sizable differences
between PDF sets are present as shown in figure~\ref{figPDFs}.

The NNLO corrections to the cross section predictions, which are defined as  
ratios of NNLO to NLO predictions and are displayed in figure~\ref{figInclJetRatio}, 
are particularly large at 
low values of \ptjet\ or at low values of \Qsq, equivalent to low values of
the renormalisation and factorisation scales \mur\ and \muf.
The NNLO predictions themselves have significantly smaller scale uncertainties than the NLO predictions.
At low values of \ptjet\ and small \Qsq, where the data are most precise, the uncertainties from scale variations 
of all predictions, however, are significantly larger than the experimental uncertainties. 
At higher values of \ptjet\ the relative theoretical uncertainties are becoming smaller, 
but the data uncertainties, both statistical and systematic, increase and overshoot the uncertainties from scale variations.

\begin{boldmath}
\subsubsection{Measurement of inclusive jets at high-$\Qsq$ ($\Qsq>150\,\GeVsq$)}
\end{boldmath}

In this section, the extension of the measurement of inclusive jet production at high-\Qsq~\cite{H1Multijets} to lower transverse momenta is described. 

The phase space of additional inclusive jet cross sections at high values of \Qsq\ is extended to the region $\ptjet<7\,\GeV$ by adding an extra bin at low \ptjet\ as outlined in section~\ref{sec:pshigh}.
These additional cross section points as a function of \Qsq\ for
inclusive jet production in the range
$5<\ptjet<7\,\GeV$
are shown in figures~\ref{figInclJet} and ~\ref{figInclJetRatio}.
The statistical correlations of these data points to other data points~\cite{H1Multijets} 
are listed elsewhere~\cite{results}.

The new low-\ptjet\ inclusive jet cross sections at high-\Qsq are underestimated by the 
NLO and aNNLO predictions, while the NNLO predictions give a good description of these new data points.
For each of the \Qsq-bins there is a sizable negative correlations of around -0.55 between the new measurement at $5<\ptjet<7\,\GeV$ and the previously published measurement at $7<\ptjet<11\,\GeV$.
In the high-\Qsq domain 
the NNLO predictions have significantly smaller scale uncertainties 
than the NLO calculations, and the NNLO scale uncertainties typically are smaller than the experimental uncertainties.
Figure ~\ref{figInclJetRatio} and
the values of $\chisq/\ndf$ in table~\ref{tab:chisq} indicate that the aNNLO predictions
have difficulties describing the previously published
high-\Qsq\ inclusive jet data \cite{H1Multijets} accurately \cite{NNLO}.
The NNLO predictions provide a good description of both, the low- and high-\Qsq\ inclusive jet data.
As shown in figure~\ref{figPDFs}, the jet cross sections at high
\ptjet\ depend significantly on the PDF set, which is a possible
explanation of part of the discrepancies observed.
The H1 jet
data thus can be used to improve future PDF determinations.

\subsection{Normalised inclusive jet cross section}
\label{sec:NormIjets}
\begin{figure}
  \centering
  \includegraphics[width=0.60\textwidth]{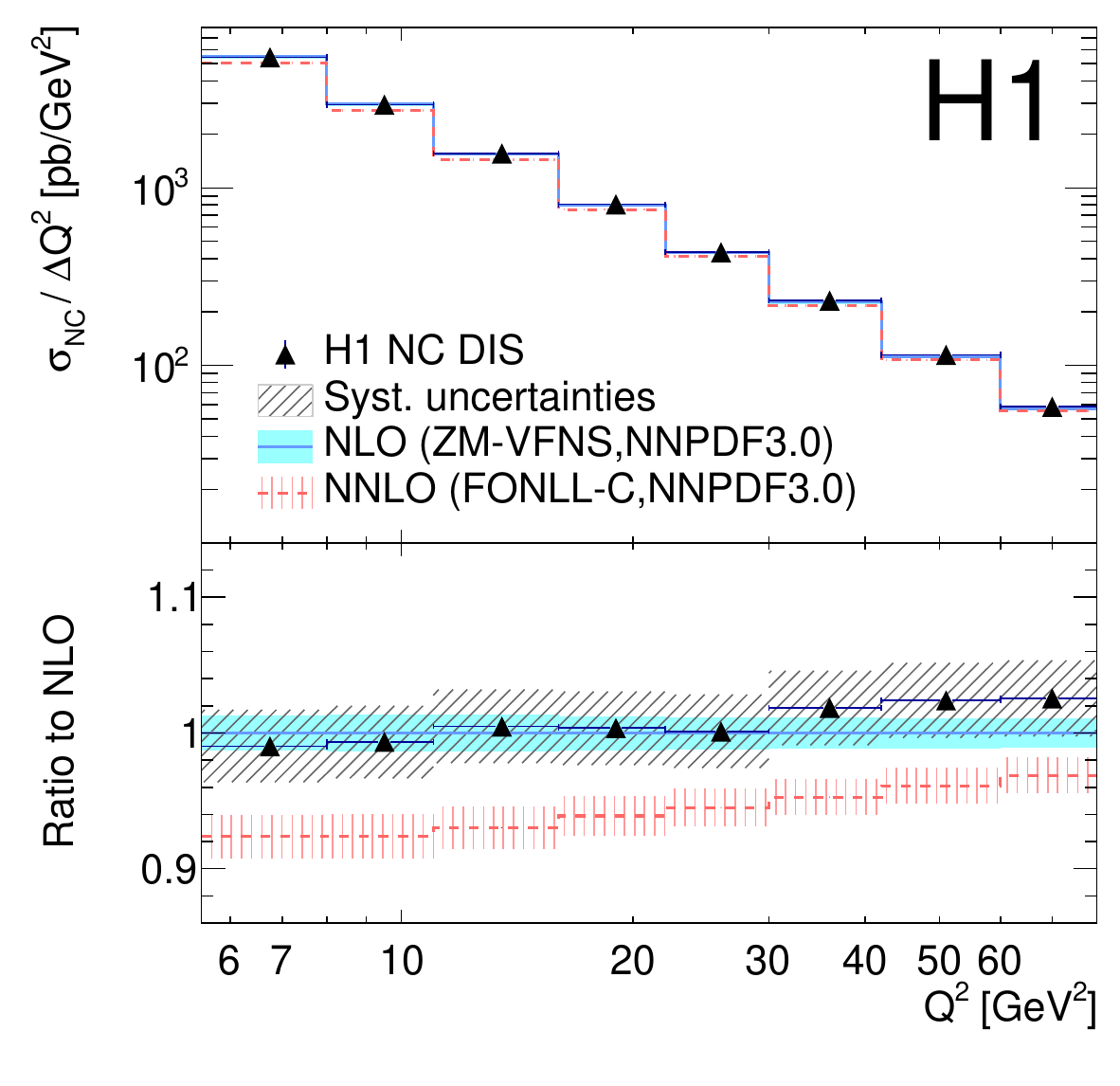}
  \caption{
    Cross sections for inclusive NC DIS in the \Qsq\ bins as for the jet cross sections 
    for $0.2<y<0.6$ compared to NLO and NNLO predictions as used for the calculation of the normalised jet cross sections.
    The cross sections are divided in each bin by the bin size in $\Qsq$.
    The experimental uncertainties are indicated by a dark hatched area
    and are dominated by the luminosity uncertainty of 2.5\,\%. The bands
    around the predictions, visible in the ratio, indicate the PDF
    uncertainties.
  }
  \label{figNCDIS}
\end{figure}

In order to obtain the normalised jet cross sections, cross sections for inclusive NC DIS 
are measured for $0.2<y<0.6$ in the \Qsq\ bins in the range $5.5<\Qsq<80\,\GeVsq$.
The single-differential inclusive NC DIS cross sections are displayed in figure~\ref{figNCDIS} 
and are compared to predictions in NLO and NNLO, which are used for the predictions of the normalised jet cross sections.
The statistical uncertainties on these data are almost negligible compared to the dominant 
luminosity uncertainty of 2.5\,\% and the other experimental uncertainties of 1 to 2\,\%.
The inclusive NC DIS cross sections are well described by the NLO predictions within the experimental uncertainties. 
The NNLO predictions for the inclusive NC DIS cross section undershoot the data by about 6 to 8\,\%. 

\begin{figure}[htbp]
  \centering
  \includegraphics[width=0.95\textwidth]{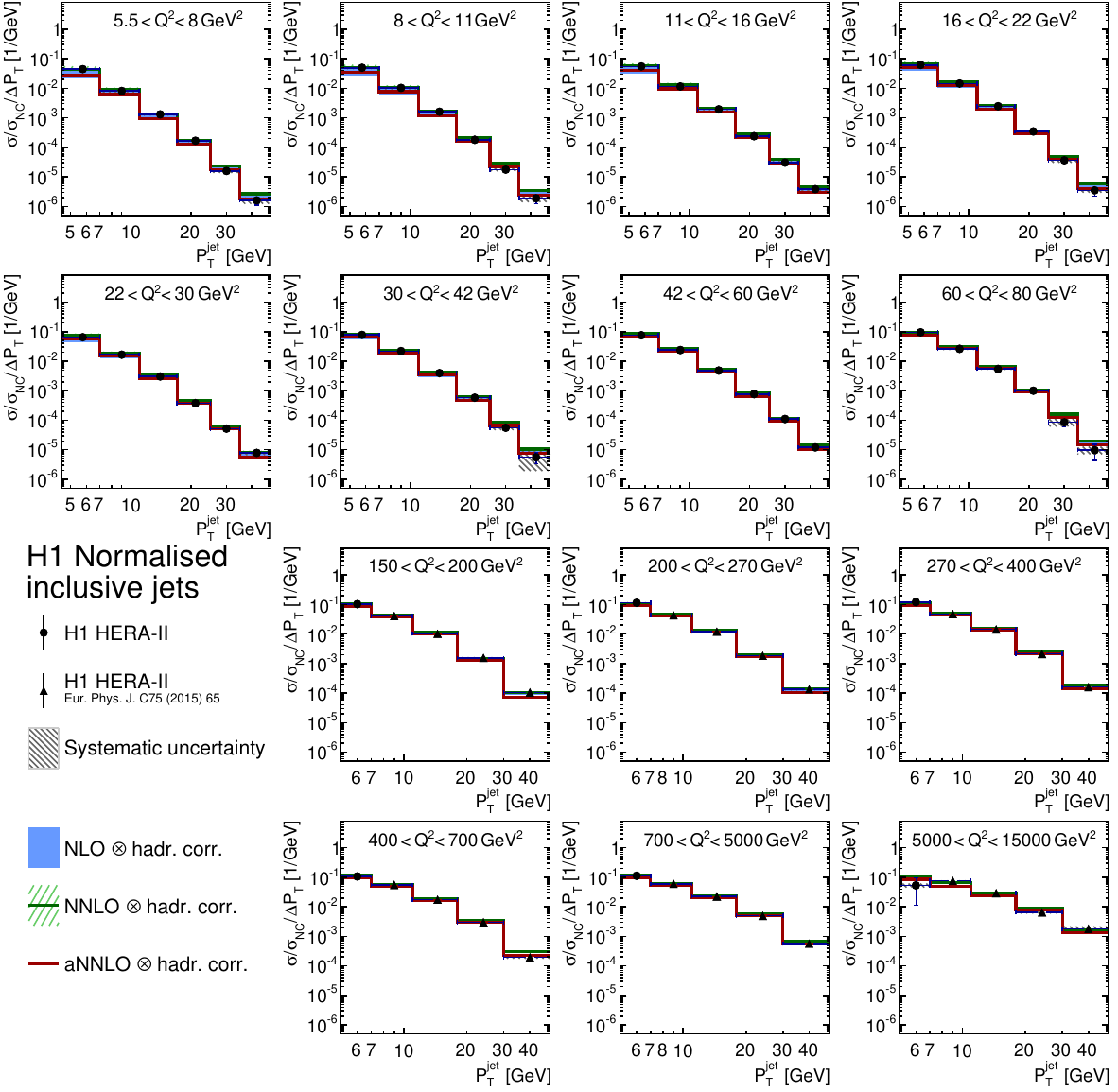}
  \caption{
    Normalised inclusive jet cross sections compared to NLO, aNNLO and NNLO predictions as a function of \Qsq\ and \ptjet.
    The cross sections are divided in each bin by the bin size in $\ptjet$.
    Further details can be found in the caption of figure~\ref{figInclJet}.
  }
  \label{figNormInclJet}
\end{figure}

The normalised inclusive jet cross sections, derived using the inclusive NC DIS and the 
absolute inclusive jet cross sections, are displayed together with theoretical predictions
in figure~\ref{figNormInclJet}.
The normalised jet cross sections increase as a function of \Qsq\ for a given interval in \ptjet.
This effect is most pronounced at high values of \ptjet.
The ratio of normalised inclusive jet cross sections to NLO prediction and the predictions in aNNLO and full NNLO to the NLO predictions is shown in figure~\ref{figNormInclJetRatio}.
The dominating systematic uncertainties do not cancel in the normalisation, and the systematic uncertainty is significantly reduced only in bins where the overall systematic error is small, typically at low \ptjet.
The normalised jet cross sections hence do not lead to stronger conclusions
when confronted with theoretical predictions, as compared to the absolute cross sections.

\subsection{Dijet cross sections}
\label{sec:2jets}
\begin{figure}
  \centering
  \includegraphics[width=0.85\textwidth]{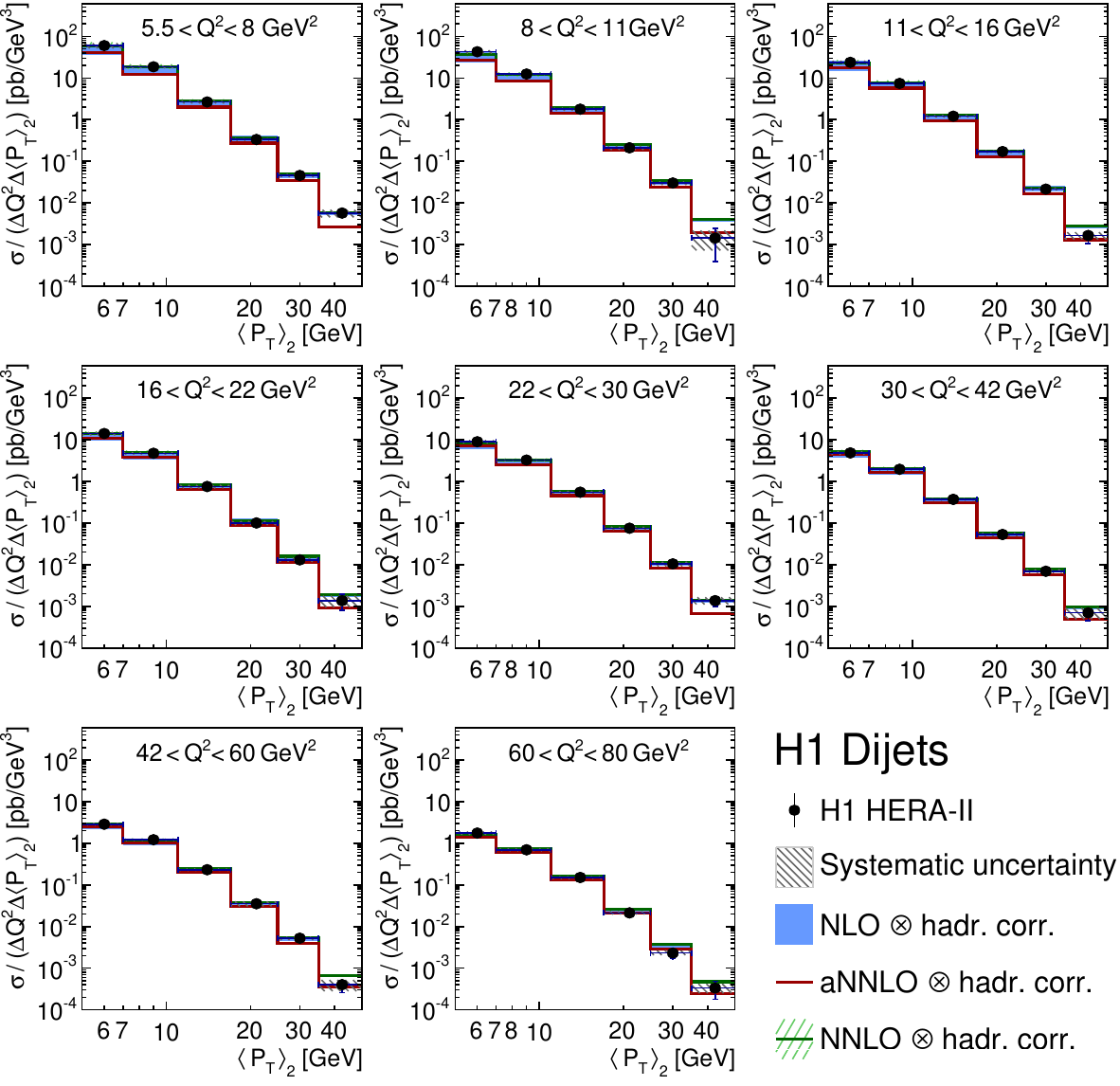}
  \caption{
    Dijet cross sections compared to NLO, aNNLO and  NNLO predictions as a function of \Qsq\ and \meanptdi.
    The cross sections in each bin are divided by the bin-size in
    \meanptdi{} and \Qsq{}.
    Further details can be found in the caption of figure~\ref{figInclJet}.
  }
  \label{figDijet}
\end{figure}

The double-differential dijet cross sections as function of \meanptdi\ and \Qsq\ 
are displayed in figure~\ref{figDijet} and compared to theoretical
predictions in NLO, aNNLO and NNLO. 
A comparison of the ratio of data to NLO predictions is provided in figure~\ref{figDijetRatio}
together with the predictions in NNLO.

Within the scale uncertainties, the data are described well by the NLO predictions.
The aNNLO and NNLO predictions provide a better description of the shapes,
while the NNLO predictions provide an overall accurate description of the normalisation of the dijet data.
The uncertainty from scale variations of the NLO predictions is larger than 
the experimental uncertainty for  $\meanptdi<35\,\GeV$,
while the scale uncertainty of the NNLO calculations is reduced compared to the 
NLO predictions and is larger than the experimental uncertainties only
at lower \Qsq\ values and for $\meanptdi<11\,\GeV$.

\begin{figure}
  \centering
  \includegraphics[width=0.85\textwidth]{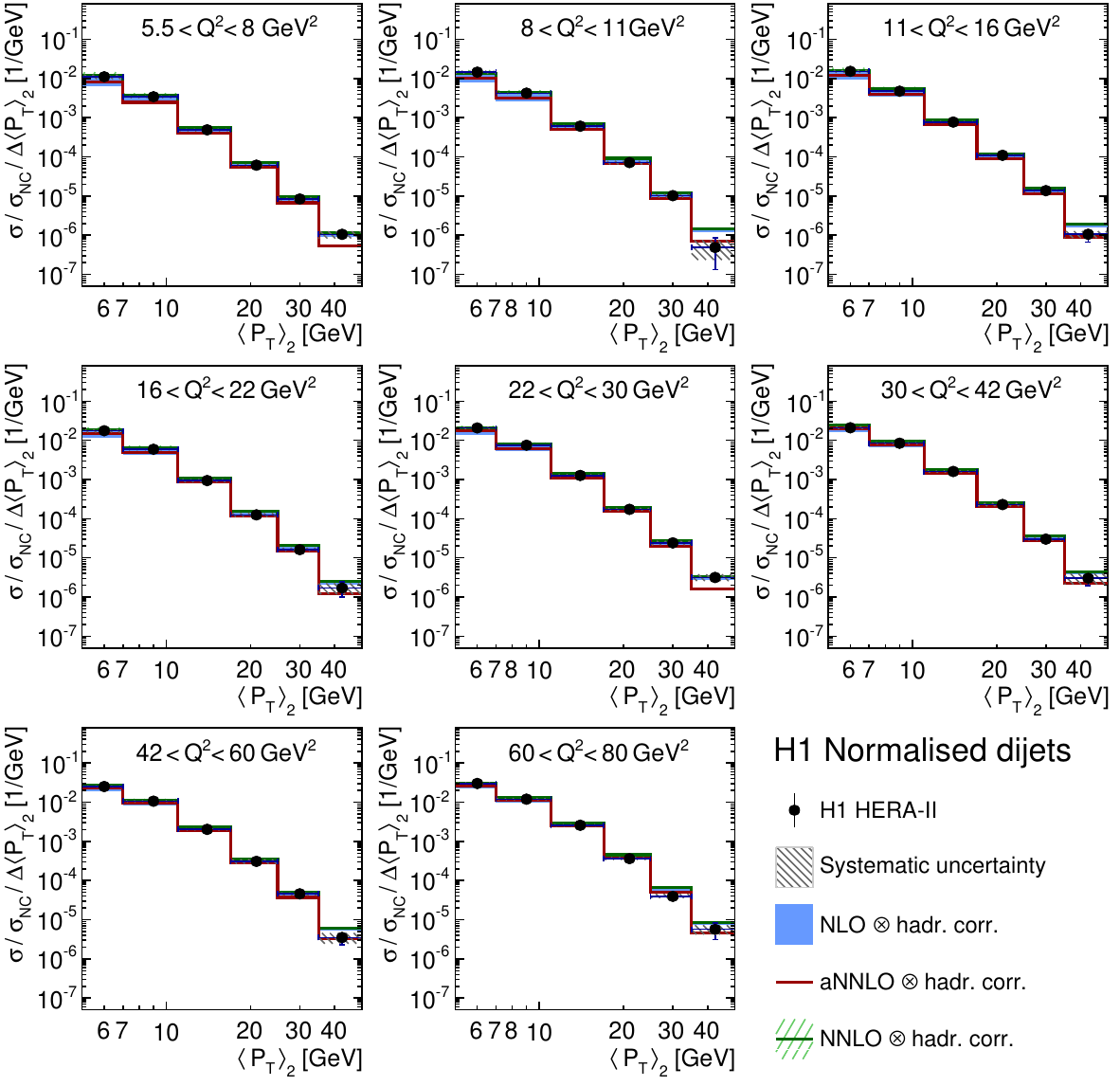}
  \caption{
    Normalised dijet cross sections compared to NLO, aNNLO and NNLO predictions as a function of \Qsq\ and \meanptdi.
    The cross sections are divided in each bin by the bin size in \meanptdi.
    Further details can be found in the caption of figure~\ref{figInclJet}.
  }
  \label{figNormDijet}
\end{figure}

The normalised dijet cross sections are displayed together with theoretical predictions
in figure~\ref{figNormDijet}, and the ratio to NLO predictions is shown in figure~\ref{figNormDijetRatio}.
The relative experimental uncertainties of the normalised jet cross sections are of similar size
as the ones of the absolute jet cross sections.
When comparing normalised dijet cross sections to theory predictions, the features 
observed with the absolute dijet cross sections are confirmed. 

\subsection{Trijet cross sections}
\label{sec:3jets}
\begin{figure}
  \centering
  \includegraphics[width=0.85\textwidth]{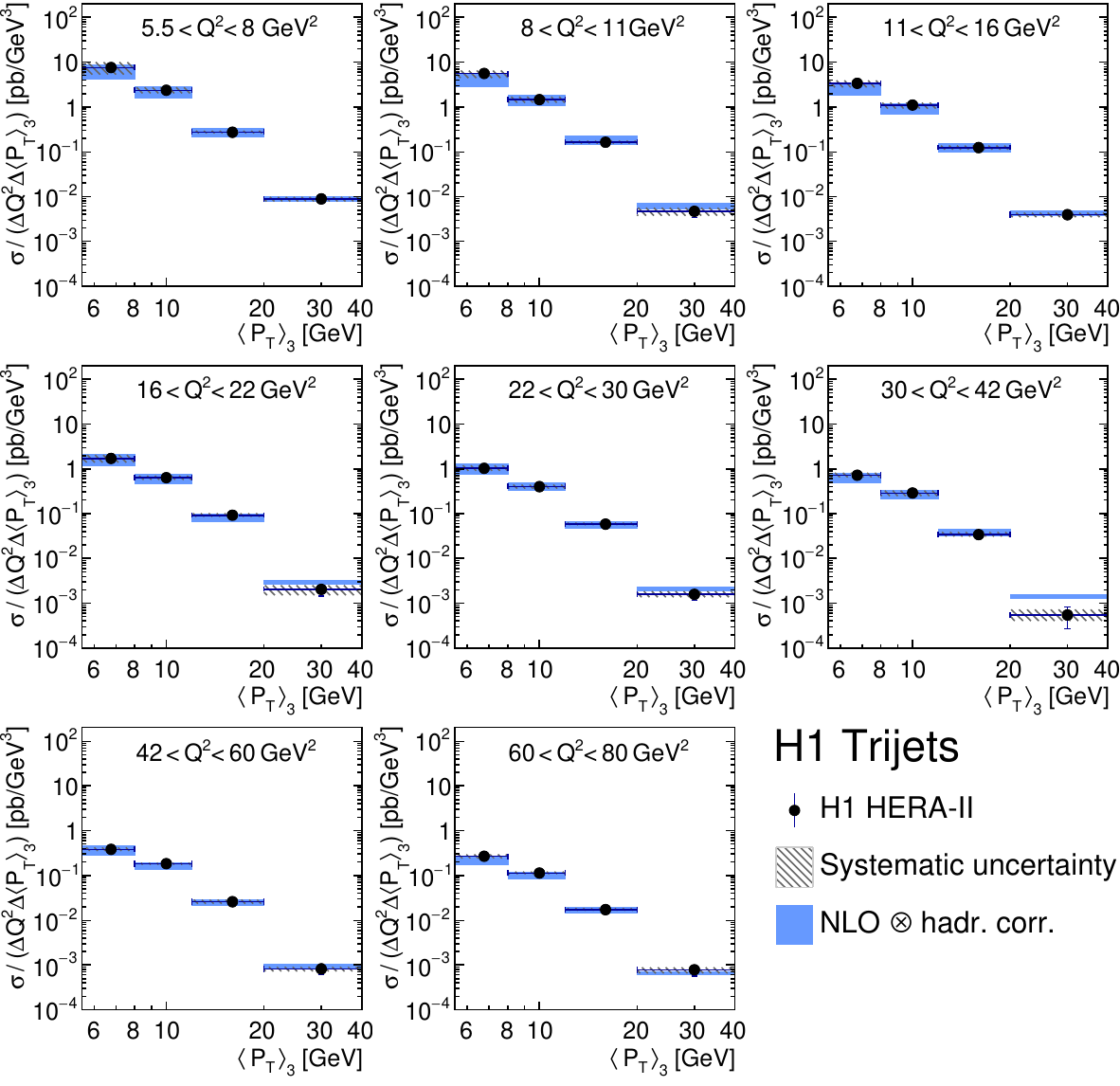}
  \caption{
    Trijet cross sections compared to NLO predictions as a function of \Qsq\ and \meanpttri.
    The cross sections in each bin are divided by the bin-size in \meanpttri\ and \Qsq{}.
    Further details can be found in the caption of figure~\ref{figInclJet}.
  }
  \label{figTrijet}
\end{figure}

\begin{figure}[tbhp]
  \centering
  \includegraphics[width=0.85\textwidth]{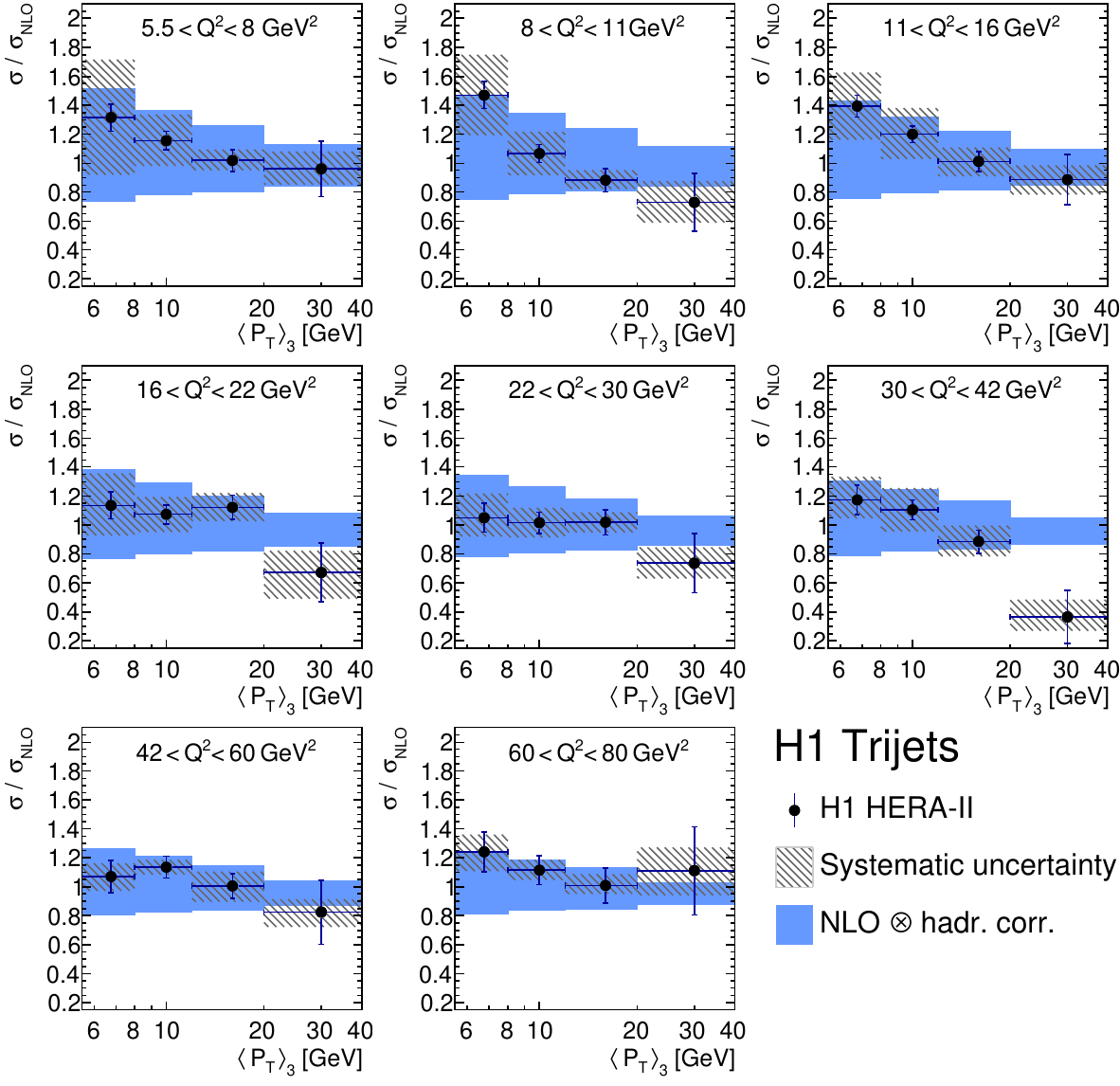}
  \caption{
    Ratio of trijet cross sections to NLO predictions  as a function of \Qsq\ and \meanpttri.
    Further details can be found in the caption of figure~\ref{figInclJet}.
  }
  \label{figTrijetRatio}
\end{figure}

The double-differential trijet cross sections as function of \meanpttri\ and \Qsq\ 
are displayed in figure~\ref{figTrijet} and compared to NLO predictions.
A comparison of the ratio of data to NLO predictions is provided in figure~\ref{figTrijetRatio}.

The experimental uncertainty is smaller than the uncertainty on the NLO predictions from scale variations for most of the data points.
The NLO calculations give an overall good agreement with the data over the full phase space, and the overall good description is confirmed by the value of $\chisq/\ndf=0.8$.
However, a trend to undershoot the data at lower values of \meanpttri\ is observed, which is
more pronounced at lower values of \Qsq, while the NLO predictions tend to overshoot the data at higher values of \meanpttri.

\begin{figure}[tbhp]
  \centering
  \includegraphics[width=0.85\textwidth]{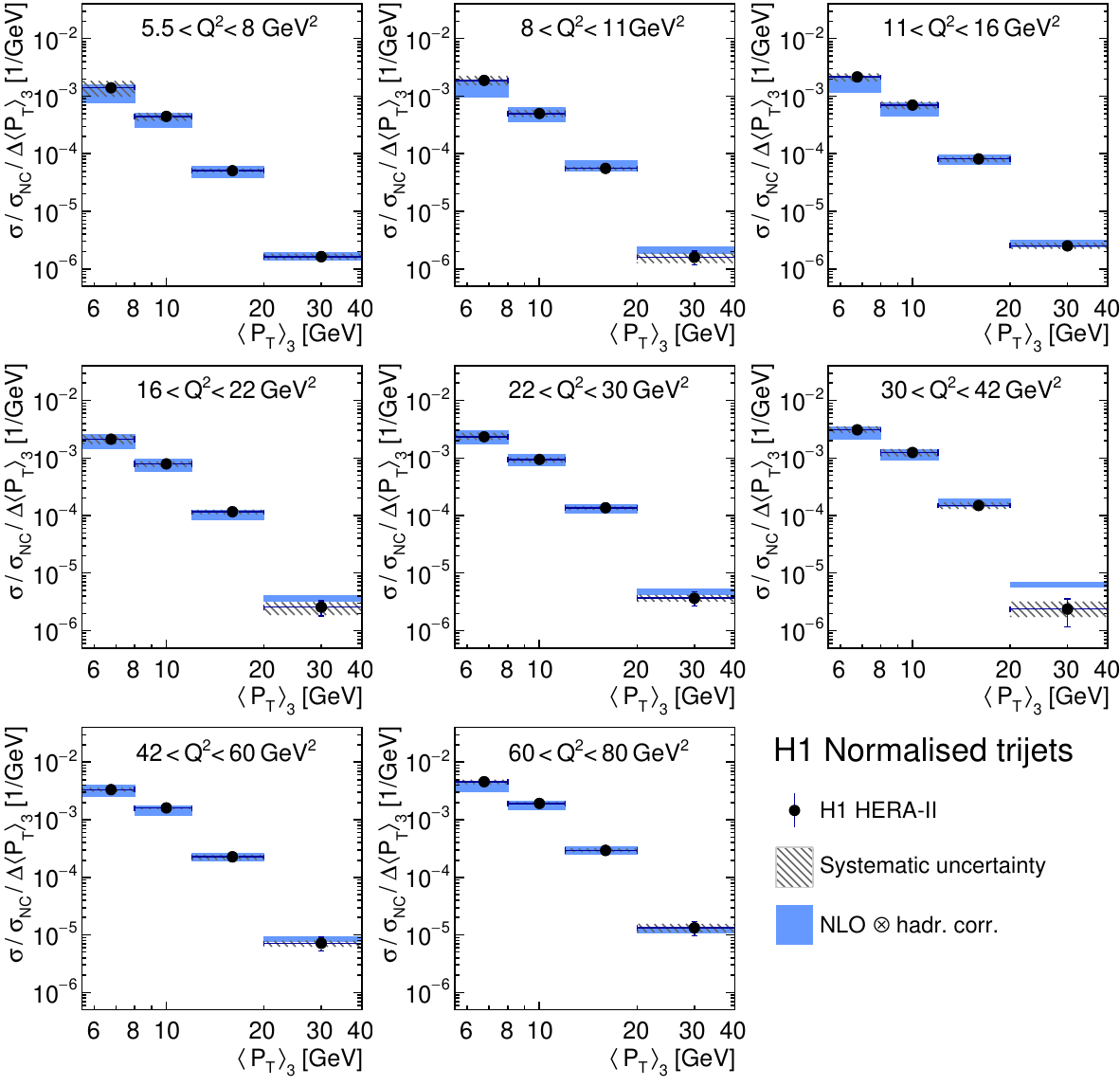}
  \caption{
    Normalised trijet cross sections compared to NLO predictions as a function of \Qsq\ and \meanpttri.
    The cross sections are divided in each bin by the bin size in \meanpttri.
    Further details can be found in the caption of figure~\ref{figInclJet}.
  }
  \label{figNormTrijet}
\end{figure}

\begin{figure}
  \centering
  \includegraphics[width=0.85\textwidth]{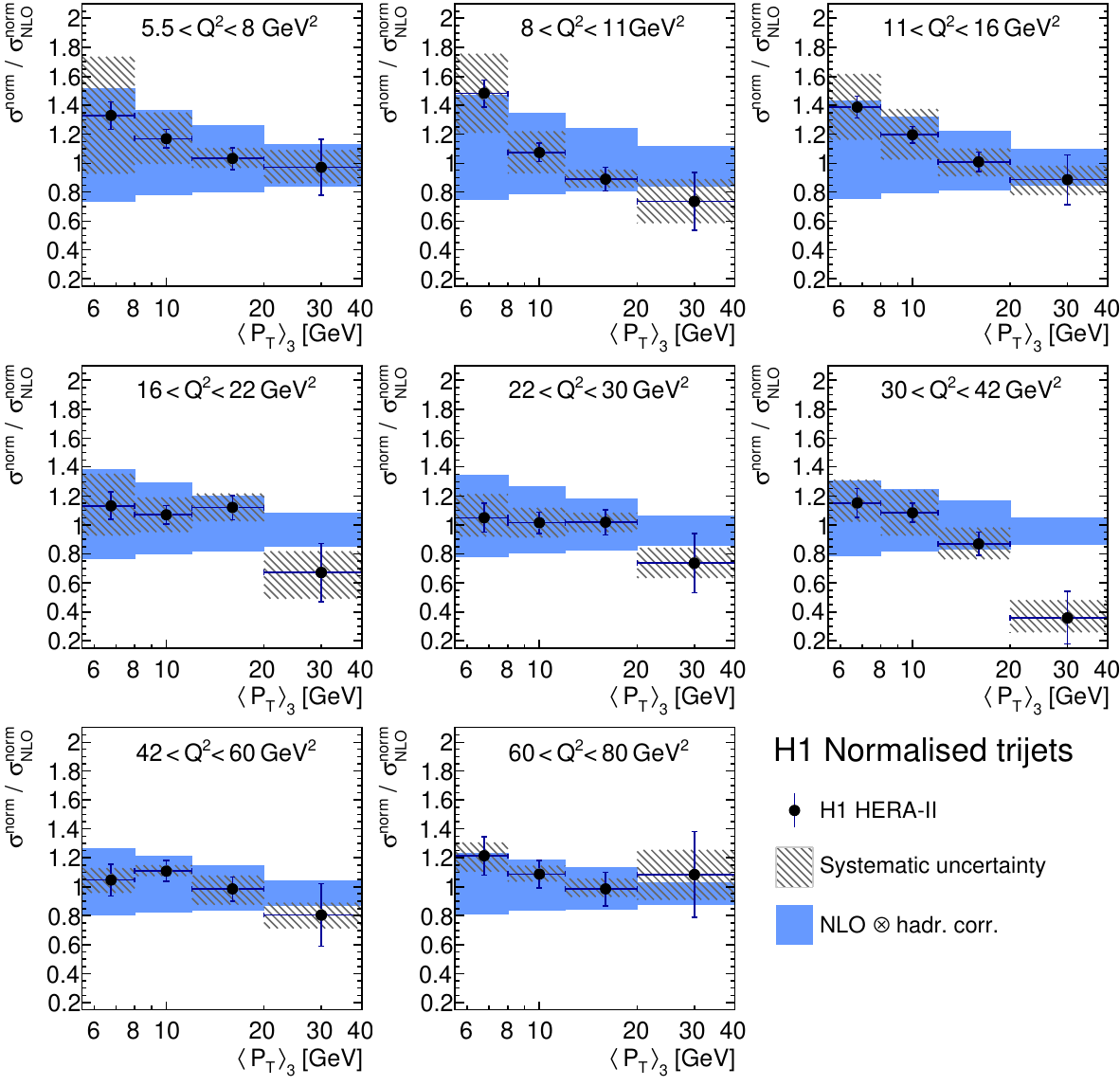}
  \caption{
    Ratio of normalised trijet cross sections to NLO predictions  as a function of \Qsq\ and \meanpttri.
    Further details can be found in the caption of figure~\ref{figInclJet}.
  }
  \label{figNormTrijetRatio}
\end{figure}

The normalised trijet cross sections are displayed together with theoretical predictions
in figure~\ref{figNormTrijet}, and the ratio to NLO predictions is shown in figure~\ref{figNormTrijetRatio}. 
A similar level of agreement of predictions and data is observed as for the absolute trijet cross sections.

As for the case of normalised inclusive jet and dijet cross sections, 
the normalised trijet cross sections increase  as a function of \Qsq\ for a given interval in \meanpttri.
This effect is sizable even at the lowest values of \meanpttri.

\section{Strong coupling determination}
\label{sec:alphas}
The presence of a strong vertex in leading order for jet production in the Breit frame allows a precision extraction of the strong coupling constant.
Since the full NNLO calculations are not yet available for an extraction of the strong coupling,
the sensitivity of the data to \asmz\ is studied in fits of NLO predictions to the data.
The sensitivity is quantified in terms of the experimental uncertainty on the fit result.
Beyond experimental uncertainties, uncertainties in the theory predictions from PDFs, the strong coupling used in the PDF extraction, the uncertainty on the hadronisation correction and uncertainties from the scale choice and missing higher orders also have to be considered.

The strong coupling constant is extracted in a $\chisq$ minimisation procedure of NLO predictions with respect to data, where
the NLO predictions are obtained as described in section~\ref{sec:nlo}.
The covariance matrix in the \chisq\ minimisation is calculated using the experimental uncertainties,
the PDF uncertainties, as determined from NNPDF3.0 replicas, and the uncertainty on the hadronisation corrections.

To improve the sensitivity to the strong coupling, data points from the high-\Qsq\ domain are also considered in the fit~\cite{H1Multijets}.
The uncertainties of these data are treated as described in \cite{H1Multijets}.
The correlated components of the JES, RCES and model uncertainty as well as the luminosity uncertainty are considered to be correlated between
the low- and high-\Qsq\ data set in the $\chisq$-calculation.

\begin{figure}[htbp]
  \centering
  \includegraphics[width=0.95\textwidth]{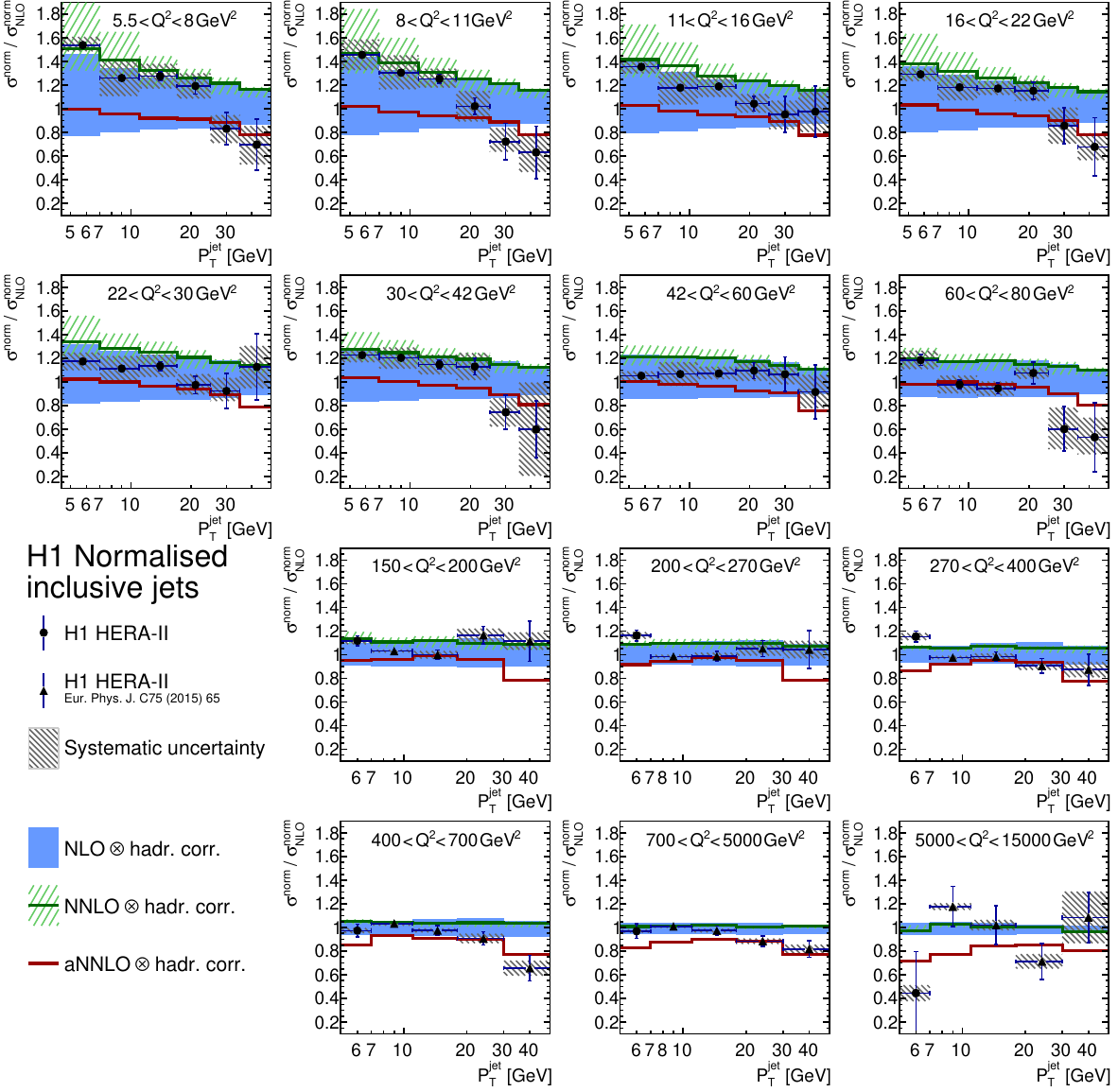}
  \caption{
    Ratio of normalised inclusive jet cross sections to NLO predictions and ratio of the NNLO and aNNLO to the NLO predictions as a function of \Qsq\ and \ptjet.
    Further details can be found in the caption of figure~\ref{figInclJet}.
  }
  \label{figNormInclJetRatio}
\end{figure}

\begin{figure}
  \centering
  \includegraphics[width=0.85\textwidth]{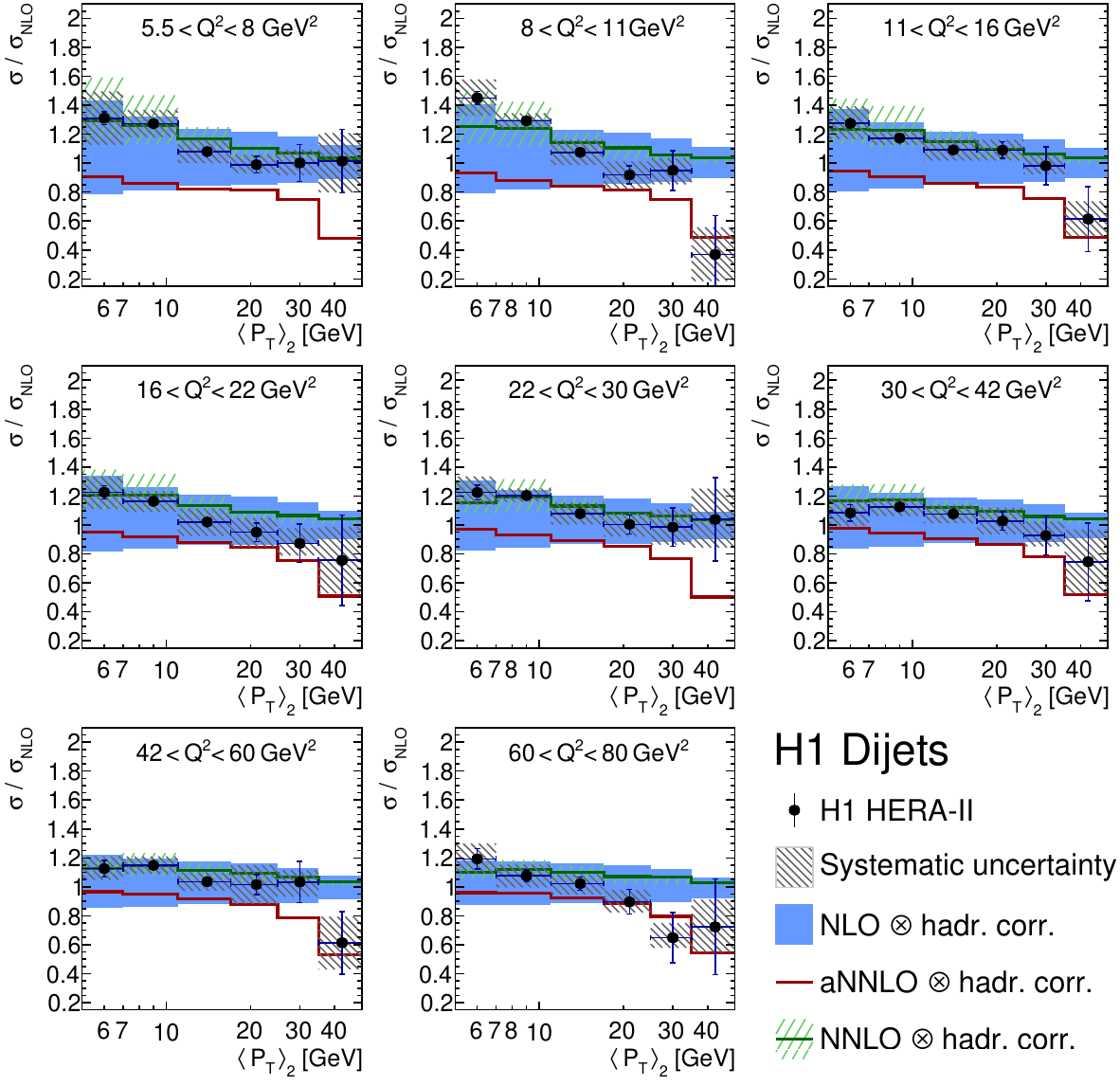}
  \caption{
    Ratio of dijet cross sections to NLO predictions and ratio of the aNNLO and NNLO to the NLO predictions as a function of \Qsq\ and \meanptdi.
    Further details can be found in the caption of figure~\ref{figInclJet}.
  }
  \label{figDijetRatio}
\end{figure}

\begin{figure}
  \centering
  \includegraphics[width=0.85\textwidth]{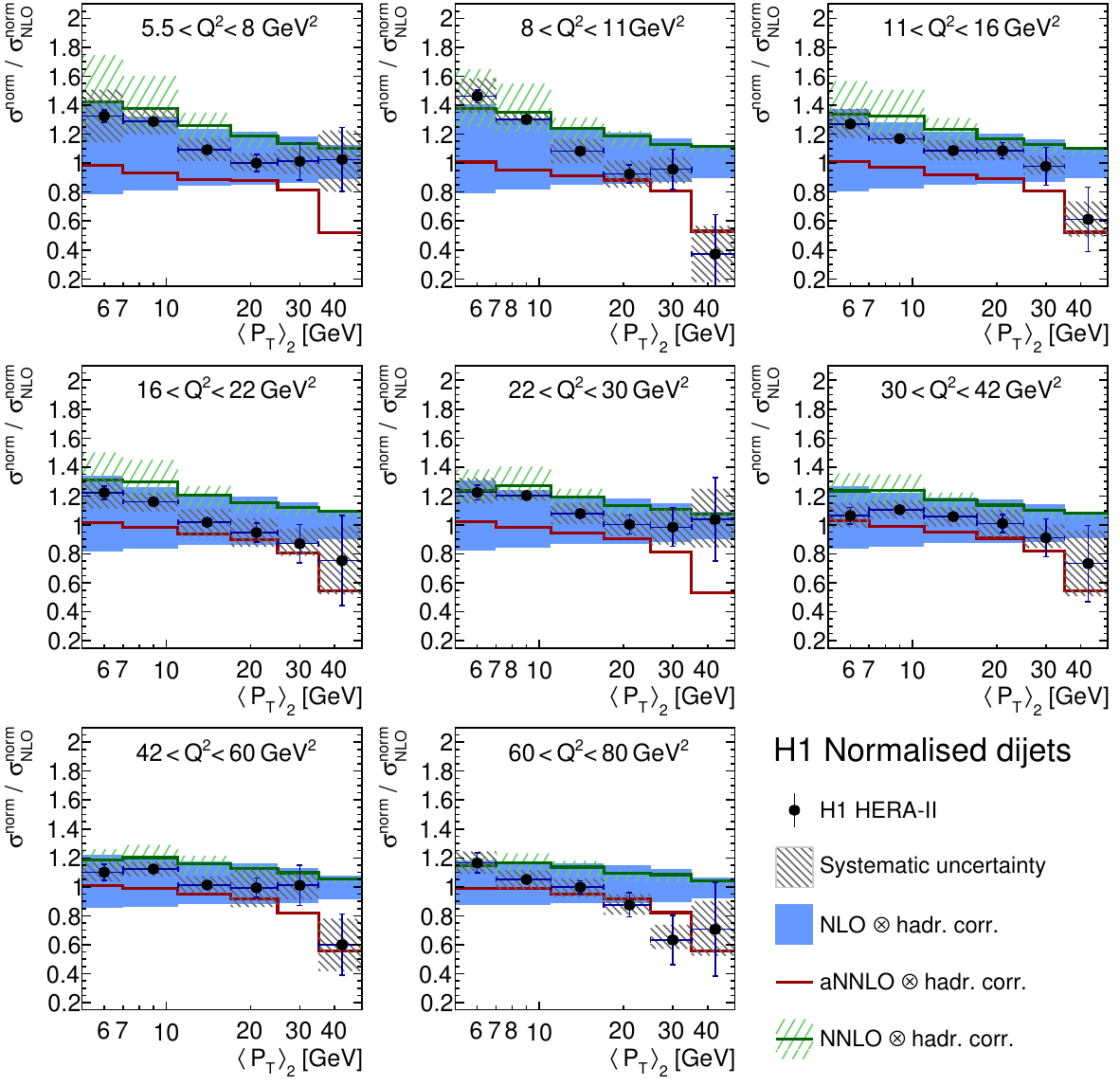}
  \caption{
    Ratio of normalised dijet cross sections to NLO predictions and ratio of the aNNLO 
    and NNLO to the NLO predictions as a function of \Qsq\ and \meanptdi.
    Further details can be found in the caption of figure~\ref{figInclJet}.
  }
  \label{figNormDijetRatio}
\end{figure}

The experimental uncertainties on \asmz\ from the fit of NLO predictions to different sets of data points are displayed in table~\ref{tab:Alphas}.
The fits are repeated using only the data points with $\Qsq\geq
11\,\GeVsq$ and  $\pt\geq 7\,\GeV$, since the pQCD predictions do not include
quark masses in the matrix elements. This may become
 important at low $\ptjet$, close to the beauty quark production threshold.
\begin{table}[thbp]
  \footnotesize
  \center
  \begin{tabular}{l | c c | c c }
    \hline
    \hline
    Dataset    & \multicolumn{2}{c|}{Low-\Qsq jet data} & \multicolumn{2}{c}{Low-\Qsq and high-\Qsq jet data}  \\
    \hline
    Kinematic range     & All data points & $\Qsq\geq 11\,\GeVsq$ and & All data points & $\Qsq\geq 11\,\GeVsq$ and \\
                        &                 & $\pt\geq 7\,\GeV$         &                 & $\pt\geq 7\,\GeV$ \\
    \hline
    Inclusive jet        & $\pm(0.0013)_{\rm exp}$  &   $\pm(0.0022)_{\rm exp}$ &   $\pm(0.0012)_{\rm exp}$  &  $\pm(0.0018)_{\rm exp}$ \\
    Dijet                & $\pm(0.0014)_{\rm exp}$  &   $\pm(0.0016)_{\rm exp}$ &   $\pm(0.0014)_{\rm exp}$  &  $\pm(0.0015)_{\rm exp}$ \\
    Trijet               & $\pm(0.0015)_{\rm exp}$  &   $\pm(0.0018)_{\rm exp}$ &   $\pm(0.0012)_{\rm exp}$  &  $\pm(0.0013)_{\rm exp}$ \\
    \hline
    Norm.\ inclusive jet  & $\pm(0.0008)_{\rm exp}$  &   $\pm(0.0019)_{\rm exp}$ &   $\pm(0.0005)_{\rm exp}$  &  $\pm(0.0007)_{\rm exp}$ \\
    Norm.\ dijet          & $\pm(0.0009)_{\rm exp}$  &   $\pm(0.0011)_{\rm exp}$ &   $\pm(0.0007)_{\rm exp}$  &  $\pm(0.0007)_{\rm exp}$ \\
    Norm.\ trijet         & $\pm(0.0012)_{\rm exp}$  &   $\pm(0.0015)_{\rm exp}$ &   $\pm(0.0008)_{\rm exp}$  &  $\pm(0.0009)_{\rm exp}$ \\
    \hline
    Incl.\ jet, dijet and trijet             & $\pm(0.0011)_{\rm exp}$  &   $\pm(0.0013)_{\rm exp}$ &   $\pm(0.0009)_{\rm exp}$  &  $\pm(0.0011)_{\rm exp}$ \\
    Norm.\ incl.\ jet, dijet and trijet  & $\pm(0.0007)_{\rm exp}$  &   $\pm(0.0009)_{\rm exp}$ &   $\pm(0.0004)_{\rm exp}$  &  $\pm(0.0005)_{\rm exp}$ \\
    \hline
    \hline
  \end{tabular}
  \caption{
    Experimental uncertainties on \asmz\ in a fit of NLO predictions to low-\Qsq\ and high-\Qsq\ jet data~\cite{H1Multijets}.
  }
  \label{tab:Alphas}
\end{table}
The values of $\chisq/\ndf$ range from 0.7 to 2.0, where $\ndf$ is defined as the number of data points minus one.
The value of $\chisq/\ndf$ is observed to become sizable whenever data at lower values of \Qsq\ and \pt\ and thus
at low values of the renormalisation scale, have high experimental precision.
In this kinematic region the contributions beyond NLO are sizable, as can be observed in the 
ratio NNLO to NLO in figures~\ref{figInclJetRatio},~\ref{figNormInclJetRatio}, ~\ref{figDijetRatio} and~\ref{figNormDijetRatio}.

The inclusive jet, dijet and trijet cross sections separately show a similar sensitivity to \asmz, if all data points are considered.
The restriction of the data points to $\Qsq\geq 11\,\GeV$ and $\pt\geq 7\,\GeV$, significantly degrades the experimental precision of \asmz\
in particular for the inclusive jets.
The inclusion of the high-\Qsq\ data improves the experimental precision in all cases.
As expected, the highest precision is obtained from a combined fit to all the normalised 
jet cross sections at low- and high-\Qsq, with an `experimental'
 precision on \asmz\ of better than 0.5\,\%.

The value of the strong coupling,  extracted in NLO from the normalised inclusive jet, dijet and trijet cross sections at low- and high-\Qsq, is
\begin{equation}
  \asmz = 0.1172\,(4)_{\rm exp}~ (3)_{\rm PDF}~ (7)_{\rm PDF(\as)} ~(11)_{\rm PDFset} ~ (6)_{\rm had} ~ (^{+51}_{-43})_{\rm scale}\,.
 \label{eq:alphasresult}
\end{equation}

The following uncertainties on the NLO predictions are considered:
\begin{itemize}
  \item The PDF uncertainty  (denoted as `PDF') is calculated as the square root of the difference of the quadratic uncertainties of the nominal result and of a fit where the PDF uncertainties are not included in the covariance matrix.
  \item The uncertainty on the choice of the PDF set (denoted as `PDFset') is estimated by calculating half of the maximum difference of fits employing the ABM11~\cite{ABM11}, CT14, HERAPDF2.0, NNPDF3.0 or MMHT PDF sets.
  \item The uncertainty due to the value of \asmz\ as input to the PDF extraction (denoted as `PDF(\as)') is estimated
    by repeating the fit with PDF sets for two available \asmz\ values of 0.117 and 0.121,
corresponding to a variation by $\pm0.002$ \cite{Lai:2010nw,Botje:2011sn}.
    The uncertainty is calculated as half the difference of the corresponding fit results.
  \item The hadronisation uncertainty (denoted as `had') is obtained as the square root of the difference of the quadratic uncertainty of the nominal result and of a fit where the hadronisation uncertainty is not included in the covariance matrix.
  \item The scale uncertainty is obtained from refits with scale-factors for the renormalisation and factorisation scale according to the 6-point prescription. 
\end{itemize}

A value of $\chi^2/\ndf = 1.36$ is obtained in the fit of $198$ data points using NNPDF3.0. 
Fits using ABM11, CT14, HERAPDF2.0 or MMHT result in an improved description of the data with values of $\chi^2/\ndf$ of about 1.1, where in each case the respective PDF uncertainties are used in the $\chisq$ calculation.

The extracted value of \asmz\ is compatible with the world average value~\cite{Bethke2016} of $0.1181\,(11)$.
The uncertainty on \asmz\ from scale variations is much larger than
the experimental uncertainty.
Given the reduction of scale uncertainties observed for the NNLO
as compared to the NLO predictions, future fits of the
inclusive jet and dijet data at NNLO may reduce the theoretical
uncertainties on the extracted \asmz\ significantly.

The sensitivity of the normalised jet cross section data, at low and high values of \Qsq, to the 
running of $\alpha_{\rm s}(\mu_r)$ is studied in a fit of NLO predictions to the data.
The data points for normalised inclusive jet, dijet and trijet production are grouped into ten groups with
comparable values of $\mu_r$, and the value of \asmz\ is obtained from minimising $\chi^2$ separately for each group.
The value of $\alpha_{\rm s}(\mu_r)$ is calculated from \asmz\ by applying the solution for
the evolution equation of $\alpha_{\rm s}(\mu_r)$ using a representative value of \mur\ for each group.
The scale uncertainty is obtained by repeating the fits 
using the 6-point scale variation prescription (see section~\ref{sec:nlo}).

\begin{figure}
  \centering
  \includegraphics[width=0.85\textwidth]{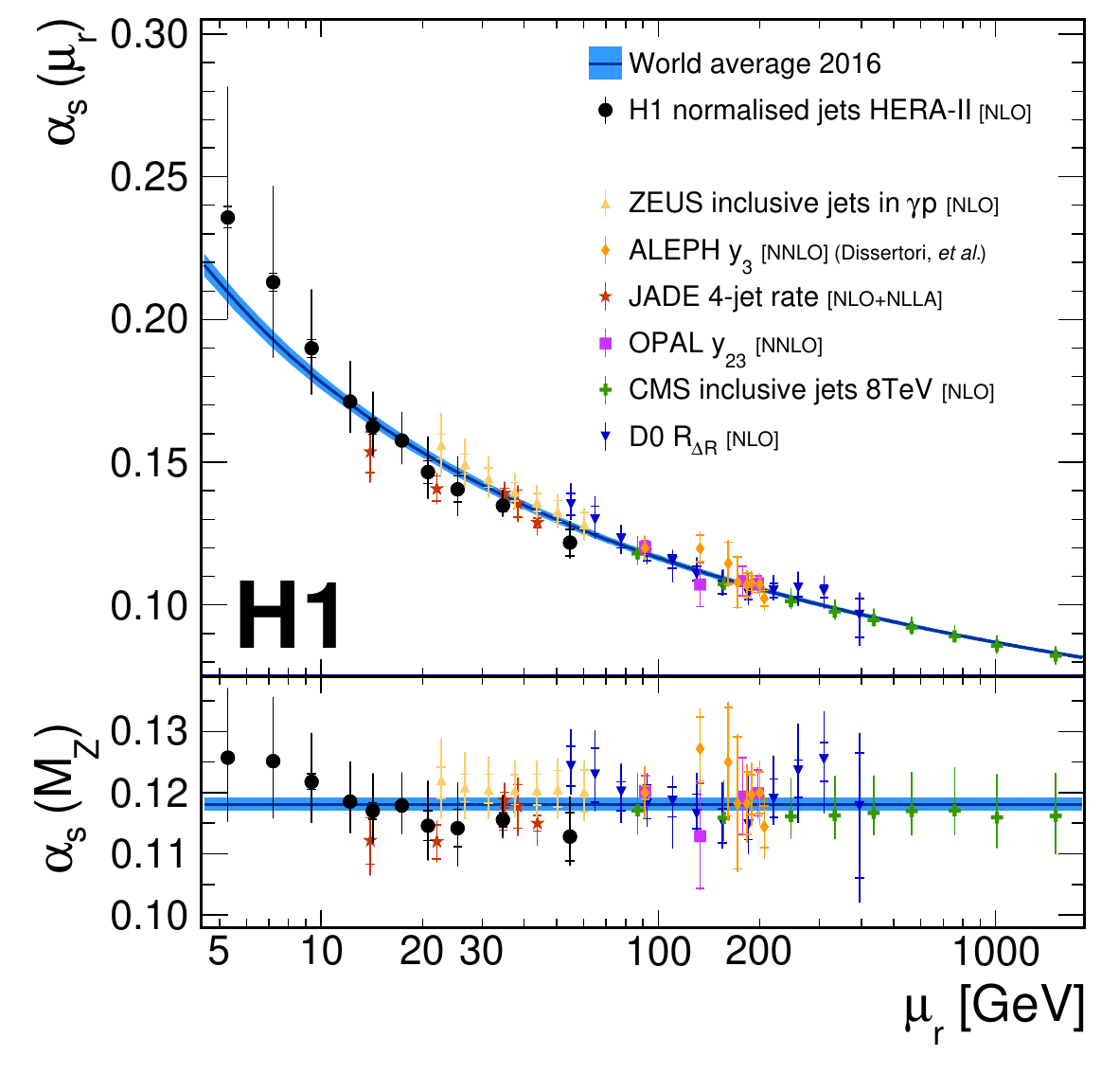}
  \caption{
  Values of \asmz\ extracted from the normalised inclusive jet, dijet and trijet cross sections
  using NLO predictions compared to values extracted from other jet data.
  The upper panel shows the values of the strong coupling $\alpha_{\rm s}(\mu_r)$ and the
  lower panel the equivalent values of \asmz\ for all measurements.
  The full circles show the extracted values from the low- and high-\Qsq\ normalised inclusive jet, dijet and trijet
  data as outlined in the text.
  The inner error bars indicate the experimental uncertainty, while the full error bars indicate the total uncertainty,
  including the experimental and theoretical contributions.
  The solid line shows the world average value of $\asmz = 0.1181\pm0.0011$ and its value evolved to $\mu_r$ using the solution of the QCD renormalisation
  group equation.
  Also shown are the values of \as\
  from inclusive jet measurements in photoproduction by the ZEUS experiment (upper triangles),
  from the 3-jet rate $y_3$ in a fit to ALEPH data taken at LEP (diamonds),
  from the 4-jet rate measured by the JADE experiment at PETRA (stars),
  from the jet transition value $y_{23}$ measured by OPAL at LEP (squares),
  from inclusive jet cross sections as measured by the CMS experiment at the LHC (crosses)
  and
  from jet angular correlations $R_{\Delta R}$ by the D0 experiment at the Tevatron (lower triangles).
  }
  \label{figAlphasRunning}
\end{figure}

The results are shown in table~\ref{tab:AlphasRunning} and are compared to extractions from other jet data~\cite{zeusgp,aleph,jade,opal,cmsIncJet8TeV,d0} in figure~\ref{figAlphasRunning}.
The H1 jet data probe the running of the strong coupling in the range $5 < \mu_r < 90\,\GeV$.


\section{Summary}
\label{sect:Conclusion}
Measurements of the inclusive jet, dijet and trijet cross sections
in neutral current deep-inelastic electron-proton scattering in the
range $5.5<\Qsq<80\,\GeVsq$ 
as well as these jet cross sections normalised to the NC DIS cross sections are reported.
At values of $150<\Qsq<15\,000\,\GeVsq$ new cross section measurements for inclusive jet cross sections for jet transverse momenta of $5<\ptjet<7\,\GeV$ are presented, extending the kinematic reach of previously published results.

The jets are reconstructed using the inclusive $k_T$ algorithm 
in the Breit frame and are required to have a minimum
transverse momentum of 4\,\GeV.
The precision of the measurements is in the range of $6\%$ to $20\%$,
with the exception of the highest $\ptjet$ bins.
Calculations at NLO QCD, corrected for hadronisation effects, provide a
reasonable description of the double differential cross sections as
functions of the jet transverse momentum \pt\ and the boson virtuality
$Q^2$.

New theoretical calculations in the threshold resummation formalism with 2-loop contributions
give a good description of the data overall.

New predictions in  next-to-next-to-leading order in perturbative QCD  improve the descriptions of the inclusive jet
and dijet cross sections compared to NLO predictions, and give an
overall good description of the new data at low and high $Q^2$.
However, at high $\ptjet$ the predictions also depend significantly on
the PDF set used. The H1 jet data thus may be useful to further
constrain PDF parametrisations.

The strong coupling \asmz\ is determined from a fit of NLO predictions 
to the measured normalised jet cross sections as
$ \alpha_s(M_Z) = 0.1172\,(4)_{\rm exp}\,(^{+53}_{-45})_{\rm th}$, which is
consistent with other extractions and demonstrates the high
experimental precision of the data. The running of $ \alpha_s$ with
the scale $\mur$ is probed by the H1 data in the range $5 <
\mur < 90\,\GeV$ and is found to be consistent with other data and with
expectations.



\section*{Acknowledgements}
We are grateful to the HERA machine group whose outstanding
efforts have made this experiment possible.
We thank the engineers and technicians for their work in constructing and
maintaining the H1 detector, our funding agencies for
financial support, the
DESY technical staff for continual assistance
and the DESY directorate for support and for the
hospitality which they extend to the non--DESY
members of the collaboration.

We would like to give credit to all partners contributing to the EGI computing infrastructure for their support for the H1 Collaboration. 

We are thankful to T.~Gehrmann, J.~Niehues and G.~Kramer for providing their predictions and for helpful discussions.

We express our thanks to all those involved in securing not only the H1 data but also the software and working environment for long term use allowing the unique H1 dataset to continue to be explored in the coming years. The transfer from experiment specific to central resources with long term support, including both storage and batch systems has also been crucial to this enterprise. We therefore also acknowledge the role played by DESY-IT and all people involved during this transition and their future role in the years to come.



\clearpage

\begin{flushleft}

\end{flushleft}


\clearpage
\begin{table}[t!hbp]
  \footnotesize
  \begin{center}
    \begin{tabular}{l|cccccccc}
      \multicolumn{9}{c} {\textbf{Bin boundaries and labels}}      \\
      \hline
      \hline
      \Qsq-range $[\GeVsq ]$ & $[5.5\,;8]$& $[8\,;11]$& $[11\,;16]$& $[16\,;22]$& $[22\,;30]$& $[30\,;42]$& $[42\,;60]$& $[60\,;80]$\\
      \hline
      \ptjet-range [\GeV ] & \multicolumn{8}{c} {Inclusive jet}      \\
      $[4.5\,;7]$ & 1&  7& 13& 19& 25& 31& 37& 43\\
      $[7\,;11]$  & 2&  8& 14& 20& 26& 32& 38& 44\\
      $[11\,;17]$ & 3&  9& 15& 21& 27& 33& 39& 45\\
      $[17\,;25]$ & 4& 10& 16& 22& 28& 34& 40& 46\\
      $[25\,;35]$ & 5& 11& 17& 23& 29& 35& 41& 47\\
      $[35\,;50]$ & 6& 12& 18& 24& 30& 36& 42& 48\\
      \hline
      \meanptdi-range [\GeV ]& \multicolumn{8}{c} {Dijet}      \\
      $[5\,;7]$   & 1&  7& 13& 19& 25& 31& 37& 43\\
      $[7\,;11]$  & 2&  8& 14& 20& 26& 32& 38& 44\\
      $[11\,;17]$ & 3&  9& 15& 21& 27& 33& 39& 45\\
      $[17\,;25]$ & 4& 10& 16& 22& 28& 34& 40& 46\\
      $[25\,;35]$ & 5& 11& 17& 23& 29& 35& 41& 47\\
      $[35\,;50]$ & 6& 12& 18& 24& 30& 36& 42& 48\\
      \hline
      \meanpttri-range [\GeV ]& \multicolumn{8}{c} {Trijet}      \\
      $[5.5\,;8]$  & 1&  5&  9& 13& 17& 21& 25& 29\\
      $[8\,;12]$   & 2&  6& 10& 14& 18& 22& 26& 30\\
      $[12\,;20]$  & 3&  7& 11& 15& 19& 23& 27& 31\\
      $[20\,;40]$  & 4&  8& 12& 16& 20& 24& 28& 32\\
      \hline
      \hline
    \end{tabular}
    \caption{Overview of bin labels and bin boundaries for cross section and correlation tables. }
    \label{tab:binlabels}
  \end{center}
\end{table}

\clearpage
\begin{table}
  \tiny
  \setlength\tabcolsep{5pt} 
  \center
\begin{tabular}{ccrr@{\hskip0pt}rrc@{\hskip0pt}r@{\hskip0pt}c@{\hskip0pt}r@{\hskip0pt}rr@{\hskip0pt}rr@{\hskip0pt}rr@{\hskip0pt}rr@{\hskip0pt}r@{\hskip0pt}r@{\hskip0pt}c|rr|r}
\multicolumn{24}{c}{ \textbf{Inclusive jet cross sections in bins of \begin{boldmath}$\Qsq$ and $\ptjet$\end{boldmath}} } \\
\hline
\hline
Bin & \multicolumn{1}{c}{\CS} & \multicolumn{1}{c}{\DStat{\csdsub}} & \multicolumn{2}{c}{\DSys{\csdsub} [\%]} & \multicolumn{1}{c}{\DMod{\csdsub}} & \multicolumn{3}{c}{\DModRW{\csdsub}} & \multicolumn{2}{c}{\DJES{\csdsub} [\%]} & \multicolumn{2}{c}{\DHFS{\csdsub} [\%]} & \multicolumn{2}{c}{\DEe{\csdsub} [\%]} & \multicolumn{2}{c}{\DThe{\csdsub} [\%]} & \multicolumn{3}{c}{\DMCSt{\csdsub}} & \DRad & \cHad & \DHad & \cRad  \\
label & \multicolumn{1}{c}{[pb]} & \multicolumn{1}{c}{[\%]} & \multicolumn{1}{c}{plus} & \multicolumn{1}{c}{minus} & [\%] & &[\%] & & up & down & up & down & up & down & up & down  & & [\%] & &[\%] &  & [\%] &  \\
\hline
1 &$1.50 \cdot 10^{3}$ &$ 2.0$  & ${+5.5~}$&${~-4.4}$ & $ 0.7$  & &$ 1.1$&&${+1.0~}$&${~-1.1}$ & ${-2.7~}$&${~+4.2}$ & ${+0.7~}$&${~-0.0}$ & ${+0.6~}$&${~+0.1}$ & &$ 1.3$&&$ 0.4$  & $0.86$  & $ 6.1$  & $1.01$   \\ 
2 &$4.38 \cdot 10^{2}$ &$ 2.2$  & ${+12.3~}$&${~-11.7}$ & $ 3.0$  & &$ 2.5$&&${-8.1~}$&${~+9.2}$ & ${+6.4~}$&${~-6.9}$ & ${+0.3~}$&${~+0.2}$ & ${+0.7~}$&${~-0.3}$ & &$ 1.3$&&$ 0.8$  & $0.90$  & $ 4.5$  & $1.02$   \\ 
3 &$1.06 \cdot 10^{2}$ &$ 2.8$  & ${+8.5~}$&${~-8.4}$ & $ 4.9$  & &$ 3.4$&&${-4.9~}$&${~+5.1}$ & ${-0.3~}$&${~-0.9}$ & ${+0.1~}$&${~+0.2}$ & ${+0.4~}$&${~-0.4}$ & &$ 1.3$&&$ 0.9$  & $0.93$  & $ 3.0$  & $1.02$   \\ 
4 &$1.81 \cdot 10$ &$ 5.7$  & ${+9.2~}$&${~-9.1}$ & $ 4.2$  & &$ 4.9$&&${-5.1~}$&${~+5.3}$ & ${+0.8~}$&${~-0.9}$ & ${+0.5~}$&${~-0.3}$ & ${-0.2~}$&${~+0.3}$ & &$ 2.4$&&$ 0.9$  & $0.95$  & $ 2.2$  & $1.02$   \\ 
5 &$2.16$ &$16.4$  & ${+13.7~}$&${~-14.4}$ & $ 6.5$  & &$ 6.1$&&${-8.4~}$&${~+7.6}$ & ${+0.6~}$&${~-3.1}$ & ${+0.4~}$&${~+0.8}$ & ${+0.8~}$&${~-0.3}$ & &$ 6.3$&&$ 1.5$  & $0.95$  & $ 0.9$  & $1.03$   \\ 
6 &$3.26 \cdot 10^{-1}$ &$31.1$  & ${+22.1~}$&${~-23.5}$ & $12.0$  & &$-0.7$&&${-13.4~}$&${~+10.1}$ & ${+2.5~}$&${~-2.8}$ & ${+1.4~}$&${~-0.0}$ & ${-1.1~}$&${~+3.8}$ & &$14.2$&&$ 2.8$  & $0.95$  & $ 0.2$  & $1.05$   \\ 
7 &$1.08 \cdot 10^{3}$ &$ 2.0$  & ${+9.7~}$&${~-9.5}$ & $ 6.6$  & &$ 4.6$&&${+0.3~}$&${~-0.9}$ & ${-3.9~}$&${~+4.4}$ & ${+0.5~}$&${~-0.6}$ & ${+0.2~}$&${~-0.2}$ & &$ 1.2$&&$ 0.4$  & $0.87$  & $ 5.4$  & $1.02$   \\ 
8 &$3.61 \cdot 10^{2}$ &$ 2.1$  & ${+11.4~}$&${~-10.8}$ & $ 6.0$  & &$ 2.5$&&${-6.9~}$&${~+8.2}$ & ${+3.3~}$&${~-4.1}$ & ${+0.7~}$&${~-0.8}$ & ${+0.3~}$&${~-0.1}$ & &$ 1.1$&&$ 0.3$  & $0.90$  & $ 4.2$  & $1.02$   \\ 
9 &$8.48 \cdot 10$ &$ 3.0$  & ${+7.4~}$&${~-6.8}$ & $ 4.9$  & &$ 1.2$&&${-3.1~}$&${~+4.4}$ & ${-1.1~}$&${~+0.2}$ & ${+0.5~}$&${~-1.0}$ & ${-0.4~}$&${~-0.0}$ & &$ 1.2$&&$ 0.7$  & $0.93$  & $ 2.8$  & $1.02$   \\ 
10 &$1.27 \cdot 10$ &$ 6.9$  & ${+13.0~}$&${~-12.9}$ & $ 9.4$  & &$ 6.1$&&${-5.1~}$&${~+5.6}$ & ${-0.0~}$&${~-1.1}$ & ${-0.4~}$&${~-0.1}$ & ${+0.2~}$&${~-0.6}$ & &$ 2.2$&&$ 0.6$  & $0.95$  & $ 1.9$  & $1.03$   \\ 
11 &$1.56$ &$21.2$  & ${+16.0~}$&${~-11.9}$ & $-1.9$  & &$ 2.9$&&${-7.5~}$&${~+12.2}$ & ${-0.9~}$&${~+1.8}$ & ${+3.1~}$&${~-3.4}$ & ${-0.3~}$&${~+4.6}$ & &$ 7.2$&&$ 1.3$  & $0.95$  & $ 0.9$  & $1.04$   \\ 
12 &$2.51 \cdot 10^{-1}$ &$34.8$  & ${+25.5~}$&${~-25.7}$ & $17.8$  & &$ 4.3$&&${-8.4~}$&${~+11.0}$ & ${+1.3~}$&${~-5.8}$ & ${-3.6~}$&${~+1.0}$ & ${+1.4~}$&${~-4.4}$ & &$13.2$&&$ 2.8$  & $0.96$  & $ 2.0$  & $1.03$   \\ 
13 &$1.06 \cdot 10^{3}$ &$ 1.9$  & ${+5.1~}$&${~-5.0}$ & $ 1.2$  & &$ 2.0$&&${+0.9~}$&${~-1.1}$ & ${-3.1~}$&${~+3.3}$ & ${+0.6~}$&${~-0.6}$ & ${-0.3~}$&${~+0.3}$ & &$ 1.0$&&$ 0.2$  & $0.88$  & $ 4.7$  & $1.02$   \\ 
14 &$3.58 \cdot 10^{2}$ &$ 1.9$  & ${+11.4~}$&${~-11.7}$ & $ 4.5$  & &$ 5.9$&&${-6.9~}$&${~+7.1}$ & ${+3.9~}$&${~-5.1}$ & ${+0.3~}$&${~-0.6}$ & ${-0.2~}$&${~-0.1}$ & &$ 0.9$&&$ 0.6$  & $0.91$  & $ 3.8$  & $1.02$   \\ 
15 &$9.04 \cdot 10$ &$ 2.7$  & ${+7.9~}$&${~-7.7}$ & $ 4.9$  & &$ 3.1$&&${-4.0~}$&${~+4.5}$ & ${-0.0~}$&${~-0.9}$ & ${+0.5~}$&${~-0.4}$ & ${-0.1~}$&${~+0.1}$ & &$ 1.0$&&$ 0.8$  & $0.93$  & $ 2.8$  & $1.03$   \\ 
16 &$1.49 \cdot 10$ &$ 5.9$  & ${+8.3~}$&${~-7.7}$ & $ 5.2$  & &$ 2.1$&&${-4.0~}$&${~+5.2}$ & ${+0.2~}$&${~-0.6}$ & ${+0.7~}$&${~-0.8}$ & ${-0.2~}$&${~-0.0}$ & &$ 1.7$&&$ 0.8$  & $0.95$  & $ 2.0$  & $1.04$   \\ 
17 &$2.38$ &$16.0$  & ${+12.4~}$&${~-13.0}$ & $ 8.8$  & &$ 5.6$&&${-4.6~}$&${~+2.8}$ & ${+0.5~}$&${~-1.5}$ & ${-0.4~}$&${~+0.6}$ & ${+0.5~}$&${~-0.9}$ & &$ 5.2$&&$ 1.3$  & $0.96$  & $ 1.6$  & $1.05$   \\ 
18 &$4.48 \cdot 10^{-1}$ &$22.1$  & ${+18.6~}$&${~-17.7}$ & $12.5$  & &$ 6.8$&&${-4.8~}$&${~+6.8}$ & ${+1.3~}$&${~+2.0}$ & ${+2.1~}$&${~-1.1}$ & ${-0.3~}$&${~+1.8}$ & &$ 7.2$&&$ 5.2$  & $0.96$  & $ 0.6$  & $1.00$   \\ 
19 &$7.41 \cdot 10^{2}$ &$ 2.2$  & ${+5.0~}$&${~-5.5}$ & $ 2.1$  & &$-0.9$&&${+0.5~}$&${~-0.7}$ & ${-3.9~}$&${~+3.1}$ & ${+0.0~}$&${~-0.2}$ & ${+0.8~}$&${~-0.6}$ & &$ 1.1$&&$ 0.2$  & $0.89$  & $ 4.0$  & $1.02$   \\ 
20 &$2.78 \cdot 10^{2}$ &$ 2.2$  & ${+8.9~}$&${~-9.1}$ & $ 3.6$  & &$ 2.0$&&${-6.6~}$&${~+6.7}$ & ${+2.8~}$&${~-3.8}$ & ${+0.6~}$&${~-0.3}$ & ${+0.7~}$&${~-0.4}$ & &$ 0.9$&&$ 0.5$  & $0.92$  & $ 3.4$  & $1.02$   \\ 
21 &$7.13 \cdot 10$ &$ 3.0$  & ${+5.6~}$&${~-5.1}$ & $ 1.8$  & &$-0.1$&&${-3.7~}$&${~+4.4}$ & ${-0.2~}$&${~-0.6}$ & ${+0.2~}$&${~-0.4}$ & ${+0.5~}$&${~-0.5}$ & &$ 1.0$&&$ 0.5$  & $0.94$  & $ 2.7$  & $1.03$   \\ 
22 &$1.33 \cdot 10$ &$ 6.1$  & ${+8.1~}$&${~-7.7}$ & $ 4.9$  & &$ 3.2$&&${-3.2~}$&${~+3.9}$ & ${-0.2~}$&${~-0.2}$ & ${+0.3~}$&${~-0.1}$ & ${+0.3~}$&${~+0.1}$ & &$ 1.7$&&$ 2.3$  & $0.95$  & $ 1.7$  & $1.03$   \\ 
23 &$1.77$ &$17.9$  & ${+17.8~}$&${~-17.7}$ & $12.7$  & &$ 7.8$&&${-7.2~}$&${~+7.2}$ & ${-0.5~}$&${~+1.5}$ & ${+1.0~}$&${~-0.0}$ & ${+1.8~}$&${~-1.2}$ & &$ 4.9$&&$ 2.5$  & $0.96$  & $ 1.5$  & $1.05$   \\ 
24 &$2.55 \cdot 10^{-1}$ &$36.3$  & ${+16.6~}$&${~-16.3}$ & $-2.2$  & &$-5.1$&&${-8.5~}$&${~+8.9}$ & ${+1.1~}$&${~-0.2}$ & ${-0.1~}$&${~-2.3}$ & ${+3.8~}$&${~-2.6}$ & &$11.5$&&$ 3.3$  & $0.94$  & $ 0.5$  & $1.06$   \\ 
25 &$5.59 \cdot 10^{2}$ &$ 2.6$  & ${+7.0~}$&${~-7.0}$ & $ 4.3$  & &$ 2.2$&&${+0.8~}$&${~-1.1}$ & ${-4.0~}$&${~+4.0}$ & ${+0.6~}$&${~-0.5}$ & ${+0.2~}$&${~+0.2}$ & &$ 1.1$&&$ 0.3$  & $0.89$  & $ 3.3$  & $1.02$   \\ 
26 &$2.30 \cdot 10^{2}$ &$ 2.4$  & ${+9.4~}$&${~-8.9}$ & $ 2.7$  & &$ 4.0$&&${-5.8~}$&${~+6.7}$ & ${+3.4~}$&${~-3.7}$ & ${+0.1~}$&${~-0.4}$ & ${-0.4~}$&${~+0.2}$ & &$ 1.0$&&$ 0.3$  & $0.93$  & $ 2.9$  & $1.03$   \\ 
27 &$6.31 \cdot 10$ &$ 3.3$  & ${+8.5~}$&${~-8.1}$ & $ 5.0$  & &$ 4.5$&&${-3.3~}$&${~+4.2}$ & ${+0.7~}$&${~-1.1}$ & ${+0.4~}$&${~-0.3}$ & ${-0.1~}$&${~+0.2}$ & &$ 1.0$&&$ 0.7$  & $0.94$  & $ 2.8$  & $1.02$   \\ 
28 &$1.05 \cdot 10$ &$ 7.7$  & ${+9.2~}$&${~-8.1}$ & $ 4.0$  & &$ 4.7$&&${-3.8~}$&${~+5.6}$ & ${+1.7~}$&${~-1.1}$ & ${+0.6~}$&${~-0.3}$ & ${+0.3~}$&${~+0.4}$ & &$ 2.0$&&$ 0.7$  & $0.95$  & $ 2.1$  & $1.03$   \\ 
29 &$1.81$ &$16.0$  & ${+11.9~}$&${~-9.2}$ & $ 4.5$  & &$ 4.0$&&${-3.4~}$&${~+8.1}$ & ${+1.5~}$&${~-1.5}$ & ${+0.2~}$&${~-0.8}$ & ${-0.9~}$&${~+1.9}$ & &$ 4.8$&&$ 1.8$  & $0.96$  & $ 0.9$  & $1.01$   \\ 
30 &$4.06 \cdot 10^{-1}$ &$24.9$  & ${+16.0~}$&${~-16.2}$ & $-8.9$  & &$-6.8$&&${-3.4~}$&${~+3.6}$ & ${-0.7~}$&${~+0.1}$ & ${+1.1~}$&${~-1.3}$ & ${-2.5~}$&${~-0.1}$ & &$ 8.4$&&$ 6.3$  & $0.95$  & $ 1.4$  & $1.00$   \\ 
31 &$5.36 \cdot 10^{2}$ &$ 2.5$  & ${+6.5~}$&${~-6.4}$ & $-0.1$  & &$-4.7$&&${+0.9~}$&${~-0.8}$ & ${-2.9~}$&${~+3.2}$ & ${-0.0~}$&${~-0.1}$ & ${+0.4~}$&${~-0.5}$ & &$ 1.0$&&$ 0.3$  & $0.89$  & $ 2.6$  & $1.03$   \\ 
32 &$2.43 \cdot 10^{2}$ &$ 2.3$  & ${+7.5~}$&${~-7.7}$ & $ 1.8$  & &$-0.2$&&${-6.1~}$&${~+6.2}$ & ${+2.4~}$&${~-3.2}$ & ${+0.1~}$&${~-0.3}$ & ${+0.3~}$&${~-0.4}$ & &$ 0.8$&&$ 0.1$  & $0.94$  & $ 2.3$  & $1.02$   \\ 
33 &$6.56 \cdot 10$ &$ 3.1$  & ${+7.0~}$&${~-6.8}$ & $ 3.3$  & &$ 3.7$&&${-3.7~}$&${~+4.0}$ & ${+0.3~}$&${~-0.7}$ & ${+0.3~}$&${~-0.4}$ & ${-0.2~}$&${~+0.2}$ & &$ 0.8$&&$ 0.4$  & $0.95$  & $ 2.3$  & $1.02$   \\ 
34 &$1.29 \cdot 10$ &$ 6.3$  & ${+9.5~}$&${~-9.6}$ & $ 6.1$  & &$ 5.4$&&${-3.9~}$&${~+3.8}$ & ${+0.6~}$&${~-0.6}$ & ${+0.3~}$&${~-0.2}$ & ${+0.2~}$&${~+0.3}$ & &$ 1.6$&&$ 0.9$  & $0.95$  & $ 2.0$  & $1.03$   \\ 
35 &$1.55$ &$19.6$  & ${+15.9~}$&${~-16.1}$ & $12.6$  & &$ 1.9$&&${-7.8~}$&${~+7.1}$ & ${+1.2~}$&${~-0.4}$ & ${+1.0~}$&${~+0.4}$ & ${-0.5~}$&${~+1.6}$ & &$ 5.0$&&$ 1.7$  & $0.96$  & $ 1.3$  & $1.02$   \\ 
36 &$2.32 \cdot 10^{-1}$ &$40.1$  & ${+64.3~}$&${~-63.6}$ & $41.7$  & &$41.0$&&${-6.1~}$&${~+5.6}$ & ${+6.5~}$&${~+0.9}$ & ${-0.4~}$&${~+3.2}$ & ${+6.5~}$&${~-0.3}$ & &$23.7$&&$ 4.6$  & $0.95$  & $ 1.2$  & $1.07$   \\ 
37 &$3.88 \cdot 10^{2}$ &$ 2.9$  & ${+5.0~}$&${~-4.7}$ & $ 0.8$  & &$ 0.3$&&${+0.2~}$&${~-0.7}$ & ${-3.5~}$&${~+3.9}$ & ${+0.3~}$&${~-0.1}$ & ${-0.1~}$&${~-0.0}$ & &$ 1.2$&&$ 0.4$  & $0.89$  & $ 1.9$  & $1.03$   \\ 
38 &$1.96 \cdot 10^{2}$ &$ 2.4$  & ${+9.3~}$&${~-9.1}$ & $ 3.8$  & &$ 4.7$&&${-5.8~}$&${~+6.0}$ & ${+2.6~}$&${~-2.3}$ & ${+0.2~}$&${~-0.1}$ & ${+0.2~}$&${~+0.0}$ & &$ 0.8$&&$ 0.1$  & $0.94$  & $ 1.8$  & $1.02$   \\ 
39 &$5.96 \cdot 10$ &$ 3.2$  & ${+5.5~}$&${~-4.7}$ & $ 1.3$  & &$ 2.4$&&${-2.5~}$&${~+3.8}$ & ${+0.5~}$&${~-0.1}$ & ${+0.2~}$&${~-0.1}$ & ${+0.3~}$&${~+0.0}$ & &$ 0.8$&&$ 0.3$  & $0.96$  & $ 2.1$  & $1.01$   \\ 
40 &$1.26 \cdot 10$ &$ 6.3$  & ${+8.5~}$&${~-8.0}$ & $ 5.0$  & &$ 4.2$&&${-3.2~}$&${~+4.2}$ & ${+0.1~}$&${~-0.3}$ & ${-0.2~}$&${~+0.1}$ & ${+0.2~}$&${~-0.4}$ & &$ 1.4$&&$ 1.5$  & $0.96$  & $ 1.8$  & $1.03$   \\ 
41 &$2.27$ &$13.7$  & ${+6.4~}$&${~-8.1}$ & $-3.4$  & &$-0.0$&&${-5.3~}$&${~+2.5}$ & ${+0.1~}$&${~-0.5}$ & ${+0.3~}$&${~-0.8}$ & ${-1.3~}$&${~-0.2}$ & &$ 3.6$&&$ 1.5$  & $0.96$  & $ 1.3$  & $1.02$   \\ 
42 &$3.72 \cdot 10^{-1}$ &$25.0$  & ${+15.9~}$&${~-15.1}$ & $ 6.5$  & &$ 8.9$&&${-2.3~}$&${~+6.4}$ & ${+0.5~}$&${~+2.8}$ & ${+0.3~}$&${~-2.3}$ & ${-3.9~}$&${~-1.1}$ & &$ 7.6$&&$ 4.2$  & $0.95$  & $ 1.2$  & $1.03$   \\ 
43 &$2.76 \cdot 10^{2}$ &$ 3.9$  & ${+8.5~}$&${~-8.4}$ & $ 0.1$  & &$-6.7$&&${+0.6~}$&${~-0.3}$ & ${-3.9~}$&${~+4.1}$ & ${+0.3~}$&${~-0.1}$ & ${+0.4~}$&${~+0.4}$ & &$ 1.4$&&$ 0.6$  & $0.89$  & $ 1.3$  & $1.03$   \\ 
44 &$1.22 \cdot 10^{2}$ &$ 3.7$  & ${+8.9~}$&${~-8.7}$ & $ 1.4$  & &$-5.3$&&${-6.0~}$&${~+6.3}$ & ${-0.1~}$&${~-0.3}$ & ${+0.4~}$&${~+0.1}$ & ${-0.4~}$&${~+0.6}$ & &$ 1.2$&&$ 0.4$  & $0.95$  & $ 1.2$  & $1.02$   \\ 
45 &$3.84 \cdot 10$ &$ 4.9$  & ${+6.5~}$&${~-6.5}$ & $ 1.1$  & &$ 1.8$&&${-5.2~}$&${~+5.2}$ & ${-1.0~}$&${~+1.0}$ & ${+0.3~}$&${~-0.2}$ & ${-0.5~}$&${~+0.3}$ & &$ 1.3$&&$ 0.4$  & $0.96$  & $ 1.8$  & $1.00$   \\ 
46 &$9.37$ &$ 8.5$  & ${+5.2~}$&${~-5.7}$ & $ 0.2$  & &$-1.9$&&${-3.9~}$&${~+3.1}$ & ${-1.0~}$&${~+1.1}$ & ${+0.2~}$&${~-0.4}$ & ${-0.2~}$&${~+0.4}$ & &$ 2.0$&&$ 0.7$  & $0.96$  & $ 2.0$  & $1.01$   \\ 
47 &$9.99 \cdot 10^{-1}$ &$31.3$  & ${+29.2~}$&${~-27.3}$ & $18.4$  & &$14.9$&&${-8.0~}$&${~+12.8}$ & ${-0.0~}$&${~+0.0}$ & ${-0.1~}$&${~-1.1}$ & ${-0.1~}$&${~+2.6}$ & &$10.4$&&$ 2.9$  & $0.96$  & $ 0.9$  & $1.03$   \\ 
48 &$1.70 \cdot 10^{-1}$ &$54.7$  & ${+28.7~}$&${~-24.5}$ & $-2.5$  & &$13.2$&&${-6.0~}$&${~+13.6}$ & ${+7.1~}$&${~-2.4}$ & ${+2.6~}$&${~+0.7}$ & ${+1.4~}$&${~+5.1}$ & &$18.8$&&$ 3.5$  & $0.95$  & $ 1.8$  & $0.99$   \\ 
\hline
\hline
\end{tabular}

    \caption{
      Double-differential inclusive jet cross sections measured as a function of \Qsq\ and \ptjet. 
      The bin labels are defined in table~\ref{tab:binlabels}. 
      The data points are statistically correlated and the correlations are displayed in figure~\ref{figCorrelations}.
      The experimental uncertainties quoted are defined in section~\ref{sec:exp_unc}. 
      The total systematic uncertainty, {\DSys{\csdsub}}, sums all systematic uncertainties in quadrature, including the uncertainty due to the LAr noise of $\DLAr{\csdsub} = \unit[0.5]{\%}$, the total normalisation uncertainty of $\DNorm{\csdsub} =\unit[2.5]{\%}$ and the uncorrelated uncertainty of $1\,\%$.  
      The correction factors on the theoretical cross sections $\cHad$, together with their uncertainties $\DHad{}$ are listed on the right.
      The radiative correction factors \cRad\ are already included in the quoted cross sections.
  }
  \label{tab:IncJet}
\end{table}

\begin{table}
  \scriptsize
  \tiny
  \setlength\tabcolsep{4pt} 
  \renewcommand{\arraystretch}{1.2} 
  \center
\begin{tabular}{ccrr@{\hskip0pt}rrc@{\hskip0pt}r@{\hskip0pt}c@{\hskip0pt}r@{\hskip0pt}rr@{\hskip0pt}rr@{\hskip0pt}rr@{\hskip0pt}rr@{\hskip0pt}r@{\hskip0pt}r@{\hskip0pt}c|rr|r}
\multicolumn{24}{c}{ \textbf{Dijet cross sections in bins of \begin{boldmath}$\Qsq$ and $\meanptdi$\end{boldmath}} } \\
\hline
\hline
Bin & \multicolumn{1}{c}{\CS} & \multicolumn{1}{c}{\DStat{\csdsub}} & \multicolumn{2}{c}{\DSys{\csdsub} [\%]} & \multicolumn{1}{c}{\DMod{\csdsub}} & \multicolumn{3}{c}{\DModRW{\csdsub}} & \multicolumn{2}{c}{\DJES{\csdsub} [\%]} & \multicolumn{2}{c}{\DHFS{\csdsub} [\%]} & \multicolumn{2}{c}{\DEe{\csdsub} [\%]} & \multicolumn{2}{c}{\DThe{\csdsub} [\%]} & \multicolumn{3}{c}{\DMCSt{\csdsub}} & \DRad & \cHad & \DHad & \cRad  \\
label & \multicolumn{1}{c}{[pb]} & \multicolumn{1}{c}{[\%]} & \multicolumn{1}{c}{plus} & \multicolumn{1}{c}{minus} & [\%] & &[\%] & & up & down & up & down & up & down & up & down  & & [\%] & &[\%] &  & [\%] &  \\
\hline
1 &$2.99 \cdot 10^{2}$ &$ 3.3$  & ${+14.2~}$&${~-14.0}$ & $ 7.2$  & &$ 6.1$&&${+5.4~}$&${~-5.7}$ & ${-7.9~}$&${~+8.2}$ & ${+0.8~}$&${~-0.2}$ & ${-0.3~}$&${~+1.0}$ & &$ 2.1$&&$ 0.6$  & $0.86$  & $ 5.0$  & $1.02$   \\ 
2 &$1.85 \cdot 10^{2}$ &$ 2.0$  & ${+7.4~}$&${~-7.3}$ & $ 5.0$  & &$ 3.9$&&${-2.0~}$&${~+2.0}$ & ${+0.2~}$&${~+0.5}$ & ${-0.1~}$&${~-0.0}$ & ${+0.5~}$&${~-0.5}$ & &$ 1.0$&&$ 1.1$  & $0.90$  & $ 3.9$  & $1.02$   \\ 
3 &$3.97 \cdot 10$ &$ 2.8$  & ${+7.5~}$&${~-7.4}$ & $ 4.4$  & &$ 3.3$&&${-3.8~}$&${~+3.9}$ & ${+0.6~}$&${~-0.8}$ & ${-0.2~}$&${~-0.1}$ & ${+0.3~}$&${~-0.3}$ & &$ 1.1$&&$ 1.0$  & $0.93$  & $ 2.8$  & $1.02$   \\ 
4 &$6.65$ &$ 5.9$  & ${+7.0~}$&${~-5.8}$ & $ 2.1$  & &$ 1.0$&&${-3.7~}$&${~+5.4}$ & ${+0.4~}$&${~-0.5}$ & ${+0.1~}$&${~-0.4}$ & ${-0.2~}$&${~+0.1}$ & &$ 2.1$&&$ 1.3$  & $0.95$  & $ 2.3$  & $1.01$   \\ 
5 &$1.14$ &$12.7$  & ${+9.1~}$&${~-8.4}$ & $-4.9$  & &$-2.1$&&${-1.6~}$&${~+4.0}$ & ${+1.8~}$&${~-1.9}$ & ${-0.9~}$&${~+0.5}$ & ${+0.2~}$&${~+0.1}$ & &$ 5.0$&&$ 2.0$  & $0.94$  & $ 1.2$  & $1.04$   \\ 
6 &$2.14 \cdot 10^{-1}$ &$21.6$  & ${+19.2~}$&${~-21.2}$ & $11.9$  & &$10.6$&&${-9.1~}$&${~-0.2}$ & ${+1.0~}$&${~-1.5}$ & ${-0.2~}$&${~+1.7}$ & ${+0.8~}$&${~+0.3}$ & &$ 9.4$&&$ 3.8$  & $0.97$  & $ 3.0$  & $1.04$   \\ 
7 &$2.55 \cdot 10^{2}$ &$ 3.0$  & ${+8.8~}$&${~-8.4}$ & $-1.2$  & &$-1.0$&&${+4.0~}$&${~-4.0}$ & ${-6.5~}$&${~+6.9}$ & ${+0.4~}$&${~+0.3}$ & ${+1.1~}$&${~-0.8}$ & &$ 1.6$&&$ 0.6$  & $0.87$  & $ 4.3$  & $1.01$   \\ 
8 &$1.49 \cdot 10^{2}$ &$ 2.1$  & ${+4.1~}$&${~-4.1}$ & $ 1.0$  & &$ 0.7$&&${-1.8~}$&${~+2.2}$ & ${-1.6~}$&${~+1.1}$ & ${+0.9~}$&${~-0.8}$ & ${+0.0~}$&${~+0.1}$ & &$ 0.9$&&$ 0.3$  & $0.90$  & $ 3.5$  & $1.02$   \\ 
9 &$3.20 \cdot 10$ &$ 3.0$  & ${+8.1~}$&${~-8.2}$ & $ 4.6$  & &$ 4.4$&&${-4.0~}$&${~+3.9}$ & ${+0.1~}$&${~-0.4}$ & ${+0.7~}$&${~-0.5}$ & ${+0.3~}$&${~-0.3}$ & &$ 1.0$&&$ 0.8$  & $0.94$  & $ 2.8$  & $1.02$   \\ 
10 &$5.05$ &$ 6.7$  & ${+10.2~}$&${~-10.5}$ & $ 8.2$  & &$ 2.9$&&${-4.6~}$&${~+3.8}$ & ${+1.3~}$&${~-1.2}$ & ${+0.0~}$&${~+0.1}$ & ${+0.9~}$&${~-0.7}$ & &$ 2.0$&&$ 0.9$  & $0.95$  & $ 2.0$  & $1.02$   \\ 
11 &$9.01 \cdot 10^{-1}$ &$14.5$  & ${+8.8~}$&${~-10.1}$ & $ 3.6$  & &$-1.2$&&${-5.4~}$&${~+2.7}$ & ${+1.1~}$&${~-1.4}$ & ${-0.3~}$&${~-0.7}$ & ${-1.3~}$&${~+0.9}$ & &$ 6.4$&&$ 2.5$  & $0.95$  & $ 1.7$  & $1.05$   \\ 
12 &$6.44 \cdot 10^{-2}$ &$73.1$  & ${+50.7~}$&${~-49.7}$ & $-27.6$  & &$-21.2$&&${-11.5~}$&${~+16.2}$ & ${-2.3~}$&${~-4.2}$ & ${-6.1~}$&${~-5.6}$ & ${-2.4~}$&${~-4.2}$ & &$32.1$&&$ 4.3$  & $0.96$  & $ 2.3$  & $1.05$   \\ 
13 &$2.37 \cdot 10^{2}$ &$ 3.0$  & ${+9.2~}$&${~-8.5}$ & $ 2.0$  & &$-3.6$&&${+4.1~}$&${~-3.4}$ & ${-5.6~}$&${~+6.4}$ & ${+0.4~}$&${~-1.3}$ & ${-0.6~}$&${~+0.5}$ & &$ 1.5$&&$ 0.5$  & $0.89$  & $ 3.6$  & $1.02$   \\ 
14 &$1.48 \cdot 10^{2}$ &$ 2.1$  & ${+4.4~}$&${~-3.9}$ & $ 1.7$  & &$ 0.6$&&${-1.8~}$&${~+2.4}$ & ${-0.6~}$&${~+1.1}$ & ${+0.4~}$&${~-0.4}$ & ${+0.1~}$&${~+0.2}$ & &$ 0.8$&&$ 0.5$  & $0.91$  & $ 3.2$  & $1.02$   \\ 
15 &$3.62 \cdot 10$ &$ 2.7$  & ${+7.0~}$&${~-6.7}$ & $ 3.6$  & &$ 3.1$&&${-3.5~}$&${~+4.0}$ & ${-0.0~}$&${~+0.6}$ & ${+0.6~}$&${~-0.4}$ & ${-0.1~}$&${~+0.4}$ & &$ 0.8$&&$ 0.9$  & $0.94$  & $ 2.5$  & $1.03$   \\ 
16 &$6.82$ &$ 5.2$  & ${+6.6~}$&${~-6.5}$ & $ 3.9$  & &$ 0.6$&&${-4.0~}$&${~+4.0}$ & ${+0.7~}$&${~+0.1}$ & ${+0.3~}$&${~-0.5}$ & ${+0.0~}$&${~+0.1}$ & &$ 1.4$&&$ 0.8$  & $0.96$  & $ 2.2$  & $1.05$   \\ 
17 &$1.07$ &$13.5$  & ${+7.4~}$&${~-5.8}$ & $-0.1$  & &$ 0.7$&&${-2.9~}$&${~+4.9}$ & ${+1.9~}$&${~-0.1}$ & ${+1.0~}$&${~+0.3}$ & ${+1.1~}$&${~+1.0}$ & &$ 3.7$&&$ 1.9$  & $0.96$  & $ 1.8$  & $1.04$   \\ 
18 &$1.24 \cdot 10^{-1}$ &$36.5$  & ${+19.9~}$&${~-19.5}$ & $ 5.8$  & &$ 8.7$&&${-9.2~}$&${~+10.3}$ & ${-1.7~}$&${~-0.4}$ & ${+0.9~}$&${~-0.6}$ & ${-1.8~}$&${~+0.7}$ & &$10.4$&&$ 7.9$  & $0.96$  & $ 3.0$  & $0.98$   \\ 
19 &$1.69 \cdot 10^{2}$ &$ 3.8$  & ${+8.7~}$&${~-9.5}$ & $ 0.5$  & &$ 1.8$&&${+3.9~}$&${~-4.8}$ & ${-7.2~}$&${~+6.8}$ & ${+0.5~}$&${~-0.5}$ & ${+0.5~}$&${~-0.3}$ & &$ 1.8$&&$ 0.3$  & $0.89$  & $ 2.7$  & $1.03$   \\ 
20 &$1.14 \cdot 10^{2}$ &$ 2.4$  & ${+6.0~}$&${~-6.3}$ & $ 4.1$  & &$ 3.0$&&${-1.7~}$&${~+1.0}$ & ${-1.7~}$&${~+0.8}$ & ${+0.5~}$&${~-0.6}$ & ${-0.1~}$&${~-0.2}$ & &$ 0.8$&&$ 0.4$  & $0.92$  & $ 2.7$  & $1.02$   \\ 
21 &$2.72 \cdot 10$ &$ 3.1$  & ${+9.0~}$&${~-9.4}$ & $ 5.9$  & &$ 5.3$&&${-4.1~}$&${~+3.0}$ & ${-0.2~}$&${~-0.6}$ & ${+0.1~}$&${~-0.4}$ & ${-0.1~}$&${~-0.5}$ & &$ 0.8$&&$ 0.7$  & $0.94$  & $ 2.6$  & $1.03$   \\ 
22 &$4.84$ &$ 6.7$  & ${+10.5~}$&${~-10.4}$ & $ 6.7$  & &$ 5.8$&&${-3.4~}$&${~+3.9}$ & ${+0.3~}$&${~-0.5}$ & ${-0.1~}$&${~+0.0}$ & ${+0.1~}$&${~-0.6}$ & &$ 1.8$&&$ 2.5$  & $0.95$  & $ 1.8$  & $1.03$   \\ 
23 &$7.86 \cdot 10^{-1}$ &$15.2$  & ${+11.9~}$&${~-9.6}$ & $ 5.4$  & &$ 4.8$&&${-2.0~}$&${~+7.2}$ & ${+0.2~}$&${~+0.2}$ & ${+1.2~}$&${~-0.4}$ & ${+0.5~}$&${~+1.4}$ & &$ 4.3$&&$ 3.1$  & $0.96$  & $ 1.5$  & $1.05$   \\ 
24 &$1.24 \cdot 10^{-1}$ &$41.4$  & ${+30.5~}$&${~-30.2}$ & $12.5$  & &$14.2$&&${-6.3~}$&${~+4.5}$ & ${+1.6~}$&${~-0.2}$ & ${-0.2~}$&${~-0.1}$ & ${+6.2~}$&${~-1.9}$ & &$21.4$&&$ 7.1$  & $0.94$  & $-0.3$  & $1.07$   \\ 
25 &$1.44 \cdot 10^{2}$ &$ 4.2$  & ${+8.9~}$&${~-7.0}$ & $ 1.8$  & &$-1.4$&&${+4.1~}$&${~-3.7}$ & ${-4.4~}$&${~+6.8}$ & ${+0.5~}$&${~-0.5}$ & ${+0.5~}$&${~+0.8}$ & &$ 1.7$&&$ 0.3$  & $0.90$  & $ 2.1$  & $1.03$   \\ 
26 &$1.04 \cdot 10^{2}$ &$ 2.4$  & ${+4.0~}$&${~-3.5}$ & $ 1.0$  & &$ 1.2$&&${-0.9~}$&${~+1.5}$ & ${-0.9~}$&${~+1.6}$ & ${+0.3~}$&${~-0.2}$ & ${+0.1~}$&${~+0.3}$ & &$ 0.7$&&$ 0.3$  & $0.93$  & $ 2.2$  & $1.03$   \\ 
27 &$2.65 \cdot 10$ &$ 3.1$  & ${+7.0~}$&${~-6.9}$ & $ 4.0$  & &$ 3.5$&&${-3.2~}$&${~+3.4}$ & ${+0.1~}$&${~+0.1}$ & ${+0.5~}$&${~-0.3}$ & ${-0.2~}$&${~+0.2}$ & &$ 0.8$&&$ 0.6$  & $0.95$  & $ 2.4$  & $1.02$   \\ 
28 &$4.82$ &$ 6.5$  & ${+7.9~}$&${~-8.4}$ & $ 4.8$  & &$ 2.6$&&${-5.4~}$&${~+4.5}$ & ${+0.7~}$&${~-0.1}$ & ${+0.6~}$&${~-0.1}$ & ${+0.5~}$&${~+0.4}$ & &$ 1.6$&&$ 0.9$  & $0.96$  & $ 2.0$  & $1.03$   \\ 
29 &$8.38 \cdot 10^{-1}$ &$13.5$  & ${+12.6~}$&${~-12.4}$ & $ 8.0$  & &$ 5.5$&&${-4.2~}$&${~+4.8}$ & ${+1.1~}$&${~+0.7}$ & ${+0.2~}$&${~-0.6}$ & ${-0.3~}$&${~+0.2}$ & &$ 3.7$&&$ 4.4$  & $0.95$  & $ 2.5$  & $1.03$   \\ 
30 &$1.65 \cdot 10^{-1}$ &$27.8$  & ${+20.5~}$&${~-18.7}$ & $10.4$  & &$ 7.5$&&${-3.8~}$&${~+9.0}$ & ${-2.3~}$&${~+3.2}$ & ${+1.9~}$&${~-2.3}$ & ${-1.0~}$&${~-0.6}$ & &$ 8.7$&&$ 8.4$  & $0.95$  & $ 1.1$  & $0.96$   \\ 
31 &$1.16 \cdot 10^{2}$ &$ 5.4$  & ${+8.9~}$&${~-8.3}$ & $ 0.9$  & &$-0.3$&&${+3.8~}$&${~-3.8}$ & ${-6.3~}$&${~+7.0}$ & ${-0.4~}$&${~+0.1}$ & ${+1.0~}$&${~-1.0}$ & &$ 2.4$&&$ 0.5$  & $0.90$  & $ 1.5$  & $1.03$   \\ 
32 &$9.39 \cdot 10$ &$ 2.7$  & ${+5.6~}$&${~-5.5}$ & $ 3.6$  & &$ 2.6$&&${-0.8~}$&${~+1.1}$ & ${-1.2~}$&${~+1.5}$ & ${+0.1~}$&${~-0.3}$ & ${+0.2~}$&${~-0.3}$ & &$ 0.9$&&$ 0.2$  & $0.94$  & $ 1.8$  & $1.03$   \\ 
33 &$2.67 \cdot 10$ &$ 3.1$  & ${+5.8~}$&${~-5.9}$ & $ 2.3$  & &$ 3.3$&&${-3.2~}$&${~+2.9}$ & ${+0.1~}$&${~+0.3}$ & ${+0.2~}$&${~-0.3}$ & ${-0.1~}$&${~-0.1}$ & &$ 0.7$&&$ 0.4$  & $0.95$  & $ 2.0$  & $1.03$   \\ 
34 &$5.11$ &$ 6.2$  & ${+8.9~}$&${~-9.0}$ & $ 5.6$  & &$ 4.8$&&${-3.8~}$&${~+3.6}$ & ${+0.8~}$&${~-0.4}$ & ${+0.3~}$&${~-0.1}$ & ${+0.4~}$&${~-0.1}$ & &$ 1.5$&&$ 1.0$  & $0.96$  & $ 1.6$  & $1.03$   \\ 
35 &$8.37 \cdot 10^{-1}$ &$14.4$  & ${+10.5~}$&${~-8.1}$ & $-2.7$  & &$-2.1$&&${-4.7~}$&${~+8.0}$ & ${+0.3~}$&${~-0.9}$ & ${+1.1~}$&${~+0.2}$ & ${-0.3~}$&${~+1.8}$ & &$ 4.3$&&$ 2.3$  & $0.96$  & $ 1.8$  & $1.03$   \\ 
36 &$1.26 \cdot 10^{-1}$ &$36.0$  & ${+29.9~}$&${~-29.4}$ & $15.5$  & &$17.9$&&${-1.1~}$&${~+5.3}$ & ${+2.9~}$&${~-2.9}$ & ${-2.2~}$&${~+0.8}$ & ${+3.3~}$&${~-2.3}$ & &$15.3$&&$ 6.7$  & $0.96$  & $ 3.0$  & $1.04$   \\ 
37 &$1.04 \cdot 10^{2}$ &$ 5.2$  & ${+7.6~}$&${~-6.9}$ & $ 1.6$  & &$-0.4$&&${+3.0~}$&${~-3.0}$ & ${-4.9~}$&${~+5.9}$ & ${+0.4~}$&${~-0.4}$ & ${+0.0~}$&${~+0.0}$ & &$ 1.8$&&$ 0.5$  & $0.90$  & $ 1.0$  & $1.04$   \\ 
38 &$8.76 \cdot 10$ &$ 2.5$  & ${+5.3~}$&${~-5.2}$ & $ 3.2$  & &$ 2.3$&&${-1.4~}$&${~+1.6}$ & ${-1.2~}$&${~+1.3}$ & ${+0.2~}$&${~-0.3}$ & ${+0.1~}$&${~+0.2}$ & &$ 0.7$&&$ 0.3$  & $0.94$  & $ 1.3$  & $1.03$   \\ 
39 &$2.51 \cdot 10$ &$ 3.1$  & ${+6.2~}$&${~-6.1}$ & $ 2.8$  & &$ 3.1$&&${-3.5~}$&${~+3.6}$ & ${+0.0~}$&${~+0.3}$ & ${+0.4~}$&${~-0.3}$ & ${+0.3~}$&${~+0.2}$ & &$ 0.7$&&$ 0.4$  & $0.96$  & $ 1.8$  & $1.02$   \\ 
40 &$5.09$ &$ 6.7$  & ${+13.2~}$&${~-13.3}$ & $ 8.6$  & &$ 8.1$&&${-5.0~}$&${~+4.6}$ & ${+0.2~}$&${~-0.5}$ & ${+0.1~}$&${~-0.2}$ & ${-0.1~}$&${~-0.1}$ & &$ 2.0$&&$ 1.0$  & $0.96$  & $ 1.8$  & $1.03$   \\ 
41 &$9.44 \cdot 10^{-1}$ &$13.8$  & ${+8.0~}$&${~-9.7}$ & $ 1.4$  & &$ 3.5$&&${-7.2~}$&${~+4.6}$ & ${+1.0~}$&${~+0.1}$ & ${-0.6~}$&${~-0.8}$ & ${-1.2~}$&${~-0.1}$ & &$ 3.6$&&$ 2.5$  & $0.95$  & $ 2.5$  & $1.04$   \\ 
42 &$1.09 \cdot 10^{-1}$ &$35.2$  & ${+29.4~}$&${~-30.1}$ & $ 9.5$  & &$21.1$&&${-7.9~}$&${~+8.9}$ & ${+1.2~}$&${~+1.2}$ & ${+0.4~}$&${~-2.1}$ & ${-7.8~}$&${~-2.4}$ & &$12.7$&&$ 8.4$  & $0.97$  & $-0.5$  & $0.99$   \\ 
43 &$7.04 \cdot 10$ &$ 5.9$  & ${+8.9~}$&${~-8.4}$ & $ 3.9$  & &$-3.5$&&${+1.5~}$&${~-2.2}$ & ${-4.8~}$&${~+6.0}$ & ${-0.1~}$&${~-0.3}$ & ${-1.4~}$&${~+0.2}$ & &$ 2.0$&&$ 1.1$  & $0.89$  & $ 0.5$  & $1.05$   \\ 
44 &$5.59 \cdot 10$ &$ 3.7$  & ${+7.2~}$&${~-7.4}$ & $ 5.5$  & &$ 1.2$&&${-2.1~}$&${~+1.7}$ & ${-3.0~}$&${~+2.8}$ & ${+0.2~}$&${~-0.4}$ & ${-1.0~}$&${~+0.5}$ & &$ 1.0$&&$ 0.4$  & $0.95$  & $ 0.7$  & $1.03$   \\ 
45 &$1.81 \cdot 10$ &$ 4.4$  & ${+6.5~}$&${~-7.2}$ & $ 4.3$  & &$ 2.3$&&${-3.8~}$&${~+3.1}$ & ${-2.1~}$&${~+0.4}$ & ${-0.1~}$&${~-0.3}$ & ${-0.7~}$&${~-0.2}$ & &$ 1.2$&&$ 0.5$  & $0.97$  & $ 1.4$  & $1.01$   \\ 
46 &$3.41$ &$ 9.5$  & ${+9.2~}$&${~-8.2}$ & $ 4.3$  & &$ 1.9$&&${-5.1~}$&${~+6.6}$ & ${-0.9~}$&${~+1.5}$ & ${+0.5~}$&${~-0.6}$ & ${-1.8~}$&${~+1.1}$ & &$ 2.4$&&$ 1.4$  & $0.96$  & $ 2.2$  & $1.02$   \\ 
47 &$4.63 \cdot 10^{-1}$ &$26.9$  & ${+15.5~}$&${~-10.7}$ & $ 4.1$  & &$-1.9$&&${-3.4~}$&${~+9.3}$ & ${+1.7~}$&${~+1.0}$ & ${+1.1~}$&${~+2.3}$ & ${+6.8~}$&${~+1.8}$ & &$ 7.9$&&$ 2.9$  & $0.96$  & $ 1.1$  & $1.05$   \\ 
48 &$9.99 \cdot 10^{-2}$ &$45.8$  & ${+25.9~}$&${~-24.4}$ & $ 2.3$  & &$16.9$&&${-7.7~}$&${~+11.7}$ & ${+1.8~}$&${~+0.8}$ & ${-1.4~}$&${~-1.1}$ & ${+0.8~}$&${~+0.9}$ & &$14.4$&&$ 4.9$  & $0.96$  & $ 2.2$  & $0.97$   \\ 
\hline
\hline
\end{tabular}

    \caption{
      Double-differential dijet cross sections measured as a function of \Qsq\ and \meanptdi. 
      For an explanation of the column headings, see table~\ref{tab:IncJet}.
      The LAr noise uncertainty for the dijet cross sections is $\DLAr{\csdsub} = \unit[0.6]{\%}$.
  }
  \label{tab:Dijet}
\end{table}

\begin{table}
  \scriptsize
  \tiny
  \setlength\tabcolsep{3pt} 
  \center
\begin{tabular}{ccrr@{\hskip0pt}rrc@{\hskip0pt}r@{\hskip0pt}c@{\hskip0pt}r@{\hskip0pt}rr@{\hskip0pt}rr@{\hskip0pt}rr@{\hskip0pt}rr@{\hskip0pt}r@{\hskip0pt}r@{\hskip0pt}c|rr|r}
\multicolumn{24}{c}{ \textbf{Trijet cross sections in bins of \begin{boldmath}$\Qsq$ and $\meanpttri$\end{boldmath}} } \\
\hline
\hline
Bin & \multicolumn{1}{c}{\CS} & \multicolumn{1}{c}{\DStat{\csdsub}} & \multicolumn{2}{c}{\DSys{\csdsub} [\%]} & \multicolumn{1}{c}{\DMod{\csdsub}} & \multicolumn{3}{c}{\DModRW{\csdsub}} & \multicolumn{2}{c}{\DJES{\csdsub} [\%]} & \multicolumn{2}{c}{\DHFS{\csdsub} [\%]} & \multicolumn{2}{c}{\DEe{\csdsub} [\%]} & \multicolumn{2}{c}{\DThe{\csdsub} [\%]} & \multicolumn{3}{c}{\DMCSt{\csdsub}} & \DRad & \cHad & \DHad & \cRad  \\
label & \multicolumn{1}{c}{[pb]} & \multicolumn{1}{c}{[\%]} & \multicolumn{1}{c}{plus} & \multicolumn{1}{c}{minus} & [\%] & &[\%] & & up & down & up & down & up & down & up & down  & & [\%] & &[\%] &  & [\%] &  \\
\hline
1 &$4.77 \cdot 10$ &$ 7.1$  & ${+30.4~}$&${~-30.0}$ & $21.5$  & &$17.7$&&${+5.4~}$&${~-5.3}$ & ${-8.8~}$&${~+10.0}$ & ${+0.8~}$&${~+0.1}$ & ${-0.1~}$&${~-0.3}$ & &$ 3.6$&&$ 0.7$  & $0.74$  & $ 7.4$  & $1.02$   \\ 
2 &$2.38 \cdot 10$ &$ 5.4$  & ${+15.5~}$&${~-15.2}$ & $ 9.3$  & &$11.4$&&${-0.5~}$&${~+2.0}$ & ${-0.2~}$&${~+2.6}$ & ${+0.4~}$&${~-0.2}$ & ${-0.5~}$&${~-0.1}$ & &$ 2.3$&&$ 0.6$  & $0.79$  & $ 4.9$  & $1.02$   \\ 
3 &$5.51$ &$ 7.4$  & ${+7.3~}$&${~-6.8}$ & $ 2.3$  & &$ 3.8$&&${-3.0~}$&${~+3.9}$ & ${+0.6~}$&${~+0.7}$ & ${+0.4~}$&${~-0.4}$ & ${+0.1~}$&${~+0.6}$ & &$ 2.7$&&$ 1.3$  & $0.84$  & $ 2.6$  & $1.02$   \\ 
4 &$4.43 \cdot 10^{-1}$ &$19.8$  & ${+12.2~}$&${~-11.4}$ & $ 3.2$  & &$ 5.6$&&${-2.3~}$&${~+2.7}$ & ${+3.2~}$&${~-0.7}$ & ${+0.7~}$&${~+0.8}$ & ${+0.7~}$&${~+2.6}$ & &$ 6.5$&&$ 5.8$  & $0.84$  & $ 0.3$  & $1.04$   \\ 
5 &$4.16 \cdot 10$ &$ 6.3$  & ${+18.8~}$&${~-18.8}$ & $12.4$  & &$10.3$&&${+5.8~}$&${~-6.2}$ & ${-6.1~}$&${~+6.5}$ & ${+0.8~}$&${~-1.0}$ & ${+0.4~}$&${~+0.8}$ & &$ 2.9$&&$ 0.5$  & $0.74$  & $ 7.5$  & $1.03$   \\ 
6 &$1.77 \cdot 10$ &$ 5.8$  & ${+14.1~}$&${~-14.1}$ & $ 8.8$  & &$10.0$&&${-2.1~}$&${~+1.3}$ & ${-2.2~}$&${~+2.5}$ & ${+0.4~}$&${~-0.7}$ & ${-0.0~}$&${~+0.3}$ & &$ 2.2$&&$ 0.6$  & $0.79$  & $ 4.7$  & $1.04$   \\ 
7 &$3.90$ &$ 9.1$  & ${+8.0~}$&${~-7.0}$ & $ 3.5$  & &$ 2.4$&&${-3.4~}$&${~+5.2}$ & ${-0.4~}$&${~-0.6}$ & ${+0.1~}$&${~-0.8}$ & ${-0.7~}$&${~-0.6}$ & &$ 3.0$&&$ 1.2$  & $0.84$  & $ 2.1$  & $1.03$   \\ 
8 &$2.80 \cdot 10^{-1}$ &$27.5$  & ${+19.9~}$&${~-19.5}$ & $-14.0$  & &$-4.5$&&${+2.4~}$&${~+6.2}$ & ${-0.2~}$&${~-3.1}$ & ${+0.8~}$&${~-2.3}$ & ${-2.8~}$&${~+0.8}$ & &$10.8$&&$ 3.9$  & $0.84$  & $ 0.7$  & $1.08$   \\ 
9 &$4.17 \cdot 10$ &$ 5.4$  & ${+16.8~}$&${~-16.6}$ & $ 9.5$  & &$ 9.2$&&${+4.4~}$&${~-4.0}$ & ${-8.4~}$&${~+8.7}$ & ${+0.7~}$&${~-1.2}$ & ${-0.8~}$&${~-0.3}$ & &$ 2.4$&&$ 0.5$  & $0.73$  & $ 7.3$  & $1.03$   \\ 
10 &$2.18 \cdot 10$ &$ 4.8$  & ${+14.8~}$&${~-14.4}$ & $ 7.5$  & &$11.5$&&${-1.9~}$&${~+2.6}$ & ${-1.9~}$&${~+3.5}$ & ${+1.0~}$&${~-0.3}$ & ${-0.0~}$&${~+0.3}$ & &$ 1.7$&&$ 0.6$  & $0.79$  & $ 4.7$  & $1.05$   \\ 
11 &$5.03$ &$ 6.9$  & ${+9.6~}$&${~-10.0}$ & $ 6.3$  & &$ 5.6$&&${-3.8~}$&${~+2.8}$ & ${-0.2~}$&${~+0.4}$ & ${+0.6~}$&${~-1.0}$ & ${-0.2~}$&${~+0.2}$ & &$ 2.1$&&$ 1.1$  & $0.84$  & $ 2.7$  & $1.04$   \\ 
12 &$3.95 \cdot 10^{-1}$ &$19.6$  & ${+11.0~}$&${~-11.6}$ & $ 3.4$  & &$-0.4$&&${-7.7~}$&${~+7.2}$ & ${+0.9~}$&${~-2.2}$ & ${-1.5~}$&${~+1.5}$ & ${+1.6~}$&${~-1.9}$ & &$ 5.8$&&$ 3.5$  & $0.86$  & $ 0.2$  & $1.03$   \\ 
13 &$2.56 \cdot 10$ &$ 8.2$  & ${+19.4~}$&${~-18.2}$ & $ 9.6$  & &$12.9$&&${+6.0~}$&${~-4.4}$ & ${-6.0~}$&${~+7.9}$ & ${+0.9~}$&${~+0.1}$ & ${+1.6~}$&${~+0.8}$ & &$ 3.0$&&$ 0.6$  & $0.73$  & $ 7.3$  & $1.04$   \\ 
14 &$1.53 \cdot 10$ &$ 6.2$  & ${+11.3~}$&${~-11.3}$ & $ 5.7$  & &$ 8.6$&&${-1.1~}$&${~+1.7}$ & ${-2.6~}$&${~+2.1}$ & ${-0.1~}$&${~-0.5}$ & ${-0.4~}$&${~-0.2}$ & &$ 1.9$&&$ 0.8$  & $0.79$  & $ 5.2$  & $1.04$   \\ 
15 &$4.46$ &$ 7.4$  & ${+8.7~}$&${~-8.6}$ & $ 5.3$  & &$ 4.6$&&${-3.2~}$&${~+3.5}$ & ${+0.3~}$&${~+0.2}$ & ${+0.0~}$&${~+0.4}$ & ${+0.5~}$&${~-0.4}$ & &$ 2.1$&&$ 1.3$  & $0.83$  & $ 2.8$  & $1.07$   \\ 
16 &$2.48 \cdot 10^{-1}$ &$30.1$  & ${+22.0~}$&${~-26.9}$ & $16.1$  & &$ 8.4$&&${-16.4~}$&${~+6.3}$ & ${-2.2~}$&${~+0.3}$ & ${+2.7~}$&${~-3.3}$ & ${-1.5~}$&${~+2.1}$ & &$ 9.3$&&$ 3.6$  & $0.86$  & $ 0.9$  & $1.03$   \\ 
17 &$2.05 \cdot 10$ &$ 9.5$  & ${+15.9~}$&${~-12.4}$ & $ 5.1$  & &$ 8.0$&&${+6.8~}$&${~-4.0}$ & ${-5.2~}$&${~+9.8}$ & ${+0.7~}$&${~-0.7}$ & ${-0.7~}$&${~+0.5}$ & &$ 3.2$&&$ 0.8$  & $0.73$  & $ 7.5$  & $1.04$   \\ 
18 &$1.30 \cdot 10$ &$ 7.2$  & ${+10.1~}$&${~-9.9}$ & $ 5.4$  & &$ 7.1$&&${-1.2~}$&${~+0.7}$ & ${-1.9~}$&${~+2.9}$ & ${+0.5~}$&${~-0.3}$ & ${+0.3~}$&${~+0.0}$ & &$ 2.1$&&$ 0.6$  & $0.79$  & $ 5.4$  & $1.04$   \\ 
19 &$3.76$ &$ 8.6$  & ${+6.9~}$&${~-6.8}$ & $ 3.7$  & &$ 2.4$&&${-3.0~}$&${~+2.8}$ & ${-0.1~}$&${~+1.8}$ & ${+0.6~}$&${~-0.7}$ & ${-0.3~}$&${~+0.3}$ & &$ 2.3$&&$ 1.8$  & $0.84$  & $ 2.8$  & $1.04$   \\ 
20 &$2.57 \cdot 10^{-1}$ &$27.8$  & ${+14.6~}$&${~-14.3}$ & $-0.4$  & &$-4.9$&&${-7.2~}$&${~+7.6}$ & ${-1.0~}$&${~+1.6}$ & ${-0.0~}$&${~-0.7}$ & ${-1.0~}$&${~+0.5}$ & &$ 7.4$&&$ 8.2$  & $0.86$  & $ 1.7$  & $1.03$   \\ 
21 &$2.15 \cdot 10$ &$ 8.7$  & ${+13.7~}$&${~-10.7}$ & $ 2.2$  & &$ 6.5$&&${+5.7~}$&${~-3.2}$ & ${-6.4~}$&${~+9.5}$ & ${+0.8~}$&${~+0.1}$ & ${+0.8~}$&${~-0.2}$ & &$ 2.8$&&$ 0.6$  & $0.73$  & $ 7.2$  & $1.04$   \\ 
22 &$1.39 \cdot 10$ &$ 6.2$  & ${+13.5~}$&${~-13.6}$ & $ 7.6$  & &$10.3$&&${-1.6~}$&${~+1.1}$ & ${-2.8~}$&${~+2.2}$ & ${+0.4~}$&${~-0.4}$ & ${-0.1~}$&${~-0.0}$ & &$ 1.6$&&$ 1.0$  & $0.79$  & $ 5.2$  & $1.04$   \\ 
23 &$3.32$ &$ 9.1$  & ${+12.3~}$&${~-11.5}$ & $ 8.1$  & &$ 6.5$&&${-2.6~}$&${~+4.6}$ & ${-0.3~}$&${~+1.6}$ & ${+1.0~}$&${~-0.3}$ & ${+0.4~}$&${~+1.0}$ & &$ 2.3$&&$ 1.9$  & $0.83$  & $ 2.5$  & $1.05$   \\ 
24 &$1.32 \cdot 10^{-1}$ &$50.8$  & ${+32.7~}$&${~-26.4}$ & $14.4$  & &$ 6.7$&&${-13.9~}$&${~+20.9}$ & ${+9.8~}$&${~-0.2}$ & ${+3.0~}$&${~+1.1}$ & ${+3.1~}$&${~+6.2}$ & &$14.8$&&$ 3.6$  & $0.85$  & $ 0.8$  & $1.05$   \\ 
25 &$1.73 \cdot 10$ &$10.4$  & ${+8.6~}$&${~-9.3}$ & $-1.0$  & &$ 1.6$&&${+3.9~}$&${~-4.4}$ & ${-6.7~}$&${~+6.1}$ & ${+0.2~}$&${~-0.4}$ & ${-1.2~}$&${~+0.7}$ & &$ 3.1$&&$ 0.5$  & $0.73$  & $ 6.8$  & $1.04$   \\ 
26 &$1.32 \cdot 10$ &$ 6.7$  & ${+4.5~}$&${~-4.4}$ & $ 0.2$  & &$ 1.1$&&${-0.7~}$&${~+0.9}$ & ${-2.0~}$&${~+2.3}$ & ${+0.2~}$&${~-0.5}$ & ${-0.8~}$&${~+0.4}$ & &$ 1.7$&&$ 1.3$  & $0.78$  & $ 5.6$  & $1.04$   \\ 
27 &$3.75$ &$ 8.6$  & ${+9.9~}$&${~-10.9}$ & $ 6.7$  & &$ 5.7$&&${-4.9~}$&${~+2.2}$ & ${-1.2~}$&${~-0.5}$ & ${-0.6~}$&${~-0.3}$ & ${-0.6~}$&${~-0.7}$ & &$ 2.1$&&$ 1.2$  & $0.84$  & $ 3.6$  & $1.04$   \\ 
28 &$2.95 \cdot 10^{-1}$ &$26.7$  & ${+10.8~}$&${~-12.2}$ & $-2.1$  & &$-5.1$&&${-5.7~}$&${~+3.3}$ & ${-1.4~}$&${~-0.1}$ & ${+0.3~}$&${~-0.1}$ & ${-2.7~}$&${~+0.2}$ & &$ 7.4$&&$ 4.0$  & $0.83$  & $ 1.5$  & $1.08$   \\ 
29 &$1.34 \cdot 10$ &$11.0$  & ${+9.4~}$&${~-10.7}$ & $-0.5$  & &$-4.6$&&${+2.7~}$&${~-2.5}$ & ${-8.2~}$&${~+6.4}$ & ${-0.1~}$&${~-0.6}$ & ${-1.5~}$&${~+1.4}$ & &$ 3.0$&&$ 0.7$  & $0.73$  & $ 6.1$  & $1.04$   \\ 
30 &$9.07$ &$ 8.8$  & ${+6.5~}$&${~-6.0}$ & $ 3.0$  & &$ 0.5$&&${-2.1~}$&${~+2.7}$ & ${-2.8~}$&${~+3.2}$ & ${-0.1~}$&${~+0.1}$ & ${-0.1~}$&${~+0.5}$ & &$ 2.2$&&$ 1.4$  & $0.78$  & $ 4.8$  & $1.03$   \\ 
31 &$2.77$ &$12.0$  & ${+7.9~}$&${~-5.7}$ & $ 2.6$  & &$ 1.7$&&${-1.8~}$&${~+4.8}$ & ${+0.0~}$&${~+2.7}$ & ${+0.3~}$&${~-0.2}$ & ${-0.0~}$&${~+1.8}$ & &$ 2.9$&&$ 1.6$  & $0.83$  & $ 3.2$  & $1.04$   \\ 
32 &$3.13 \cdot 10^{-1}$ &$27.5$  & ${+14.6~}$&${~-15.3}$ & $-3.0$  & &$ 5.1$&&${-5.1~}$&${~-0.2}$ & ${+1.7~}$&${~+0.5}$ & ${-1.5~}$&${~+0.6}$ & ${+3.0~}$&${~-2.0}$ & &$ 9.8$&&$ 8.4$  & $0.84$  & $ 0.8$  & $1.06$   \\ 
\hline
\hline
\end{tabular}

    \caption{
      Double-differential trijet cross sections measured as a function of \Qsq\ and \meanpttri. 
      For an explanation of the column headings, see table~\ref{tab:IncJet}.
      The LAr noise uncertainty for the trijet cross sections is $\DLAr{\csdsub} = \unit[0.9]{\%}$.
  }
  \label{tab:Trijet}
\renewcommand{\arraystretch}{1.3} 

\end{table}

\begin{table}
  \scriptsize
  \tiny
  \setlength\tabcolsep{3pt} 
  \center
\begin{tabular}{ccrr@{\hskip0pt}rrc@{\hskip0pt}r@{\hskip0pt}c@{\hskip0pt}r@{\hskip0pt}rr@{\hskip0pt}rr@{\hskip0pt}rr@{\hskip0pt}rr@{\hskip0pt}r@{\hskip0pt}r@{\hskip0pt}c|rr|r}
\multicolumn{24}{c}{ \textbf{Normalised inclusive jet cross sections in bins of \begin{boldmath}$\Qsq$ and $\ptjet$\end{boldmath}} } \\
\hline
\hline
Bin & \multicolumn{1}{c}{$\sigma/\sigma_{\rm NC}$} & \multicolumn{1}{c}{\DStat{\csdsub}} & \multicolumn{2}{c}{\DSys{\csdsub} [\%]} & \multicolumn{1}{c}{\DMod{\csdsub}} & \multicolumn{3}{c}{\DModRW{\csdsub}} &  \multicolumn{2}{c}{\DJES{\csdsub} [\%]} &  \multicolumn{2}{c}{\DHFS{\csdsub} [\%]} & \multicolumn{2}{c}{\DEe{\csdsub} [\%]} & \multicolumn{2}{c}{\DThe{\csdsub} [\%]} & \multicolumn{3}{c}{\DMCSt{\csdsub}} & \DRad & \cHad & \DHad & \cRad  \\
label & \multicolumn{1}{c}{} & \multicolumn{1}{c}{[\%]} & \multicolumn{1}{c}{plus} & \multicolumn{1}{c}{minus} &  [\%] & & [\%] & & up & down & up & down & up & down & up & down  & & [\%] & & [\%] &  & [\%] &  \\
\hline
1 &$1.10 \cdot 10^{-1}$ &$ 2.0$  & ${+4.6~}$&${~-3.4}$ & $ 0.5$  & &$ 0.9$&&${+1.0~}$&${~-1.1}$ & ${-2.3~}$&${~+3.8}$ & ${+0.2~}$&${~+0.6}$ & ${+0.5~}$&${~+0.1}$ & &$ 1.3$&&$ 0.4$  & $0.86$  & $ 6.1$  & $1.01$   \\ 
2 &$3.23 \cdot 10^{-2}$ &$ 2.2$  & ${+12.6~}$&${~-12.1}$ & $ 2.8$  & &$ 2.3$&&${-8.4~}$&${~+9.5}$ & ${+7.1~}$&${~-7.6}$ & ${-0.4~}$&${~+0.9}$ & ${+0.6~}$&${~-0.3}$ & &$ 1.4$&&$ 0.8$  & $0.90$  & $ 4.5$  & $1.02$   \\ 
3 &$7.83 \cdot 10^{-3}$ &$ 2.8$  & ${+8.2~}$&${~-8.2}$ & $ 4.8$  & &$ 3.2$&&${-5.1~}$&${~+5.3}$ & ${+0.2~}$&${~-1.5}$ & ${-0.5~}$&${~+0.8}$ & ${+0.3~}$&${~-0.3}$ & &$ 1.4$&&$ 0.9$  & $0.93$  & $ 3.0$  & $1.02$   \\ 
4 &$1.33 \cdot 10^{-3}$ &$ 5.7$  & ${+9.0~}$&${~-8.9}$ & $ 4.1$  & &$ 4.8$&&${-5.3~}$&${~+5.5}$ & ${+1.2~}$&${~-1.4}$ & ${-0.1~}$&${~+0.3}$ & ${-0.3~}$&${~+0.4}$ & &$ 2.5$&&$ 0.9$  & $0.95$  & $ 2.2$  & $1.02$   \\ 
5 &$1.59 \cdot 10^{-4}$ &$16.4$  & ${+13.7~}$&${~-14.6}$ & $ 6.4$  & &$ 5.9$&&${-8.7~}$&${~+7.9}$ & ${+1.2~}$&${~-3.9}$ & ${-0.4~}$&${~+1.6}$ & ${+0.7~}$&${~-0.3}$ & &$ 6.4$&&$ 1.5$  & $0.95$  & $ 0.9$  & $1.03$   \\ 
6 &$2.41 \cdot 10^{-5}$ &$31.1$  & ${+22.5~}$&${~-24.0}$ & $12.0$  & &$-1.1$&&${-13.9~}$&${~+10.5}$ & ${+3.2~}$&${~-3.6}$ & ${+0.5~}$&${~+0.9}$ & ${-1.2~}$&${~+4.0}$ & &$14.6$&&$ 2.8$  & $0.95$  & $ 0.2$  & $1.05$   \\ 
7 &$1.23 \cdot 10^{-1}$ &$ 2.0$  & ${+8.9~}$&${~-8.7}$ & $ 6.2$  & &$ 4.6$&&${+0.2~}$&${~-0.9}$ & ${-3.4~}$&${~+3.9}$ & ${-0.3~}$&${~+0.2}$ & ${+0.5~}$&${~-0.4}$ & &$ 1.2$&&$ 0.4$  & $0.87$  & $ 5.4$  & $1.02$   \\ 
8 &$4.09 \cdot 10^{-2}$ &$ 2.2$  & ${+11.3~}$&${~-10.7}$ & $ 5.6$  & &$ 2.4$&&${-7.1~}$&${~+8.4}$ & ${+4.0~}$&${~-4.9}$ & ${-0.2~}$&${~+0.1}$ & ${+0.5~}$&${~-0.3}$ & &$ 1.1$&&$ 0.3$  & $0.90$  & $ 4.2$  & $1.02$   \\ 
9 &$9.63 \cdot 10^{-3}$ &$ 3.0$  & ${+6.7~}$&${~-5.9}$ & $ 4.4$  & &$ 1.1$&&${-3.2~}$&${~+4.5}$ & ${-0.4~}$&${~-0.5}$ & ${-0.3~}$&${~-0.1}$ & ${-0.1~}$&${~-0.2}$ & &$ 1.2$&&$ 0.7$  & $0.93$  & $ 2.8$  & $1.02$   \\ 
10 &$1.44 \cdot 10^{-3}$ &$ 6.9$  & ${+12.6~}$&${~-12.6}$ & $ 9.0$  & &$ 6.1$&&${-5.3~}$&${~+5.7}$ & ${+0.7~}$&${~-1.9}$ & ${-1.3~}$&${~+0.9}$ & ${+0.4~}$&${~-0.8}$ & &$ 2.3$&&$ 0.6$  & $0.95$  & $ 1.9$  & $1.03$   \\ 
11 &$1.77 \cdot 10^{-4}$ &$21.2$  & ${+15.9~}$&${~-11.7}$ & $-2.8$  & &$ 2.8$&&${-7.8~}$&${~+12.5}$ & ${+0.1~}$&${~+0.8}$ & ${+1.9~}$&${~-2.1}$ & ${+0.0~}$&${~+4.4}$ & &$ 7.4$&&$ 1.3$  & $0.95$  & $ 0.9$  & $1.04$   \\ 
12 &$2.84 \cdot 10^{-5}$ &$34.8$  & ${+25.5~}$&${~-26.1}$ & $17.3$  & &$ 4.2$&&${-8.6~}$&${~+11.3}$ & ${+2.5~}$&${~-7.0}$ & ${-5.1~}$&${~+2.5}$ & ${+1.8~}$&${~-4.8}$ & &$13.5$&&$ 2.8$  & $0.96$  & $ 2.0$  & $1.03$   \\ 
13 &$1.36 \cdot 10^{-1}$ &$ 1.9$  & ${+4.1~}$&${~-4.1}$ & $ 0.6$  & &$ 2.3$&&${+0.9~}$&${~-1.1}$ & ${-2.6~}$&${~+2.7}$ & ${-0.2~}$&${~+0.1}$ & ${-0.1~}$&${~+0.0}$ & &$ 1.0$&&$ 0.2$  & $0.88$  & $ 4.7$  & $1.02$   \\ 
14 &$4.59 \cdot 10^{-2}$ &$ 1.9$  & ${+11.6~}$&${~-12.0}$ & $ 4.0$  & &$ 6.3$&&${-7.1~}$&${~+7.3}$ & ${+4.6~}$&${~-5.9}$ & ${-0.5~}$&${~+0.2}$ & ${+0.0~}$&${~-0.4}$ & &$ 0.9$&&$ 0.6$  & $0.91$  & $ 3.8$  & $1.02$   \\ 
15 &$1.16 \cdot 10^{-2}$ &$ 2.7$  & ${+7.6~}$&${~-7.4}$ & $ 4.5$  & &$ 3.4$&&${-4.1~}$&${~+4.6}$ & ${+0.6~}$&${~-1.6}$ & ${-0.2~}$&${~+0.4}$ & ${+0.2~}$&${~-0.2}$ & &$ 1.0$&&$ 0.8$  & $0.93$  & $ 2.8$  & $1.03$   \\ 
16 &$1.91 \cdot 10^{-3}$ &$ 5.9$  & ${+7.9~}$&${~-7.2}$ & $ 4.7$  & &$ 2.5$&&${-4.2~}$&${~+5.3}$ & ${+0.9~}$&${~-1.3}$ & ${-0.1~}$&${~+0.0}$ & ${+0.1~}$&${~-0.4}$ & &$ 1.7$&&$ 0.8$  & $0.95$  & $ 2.0$  & $1.04$   \\ 
17 &$3.06 \cdot 10^{-4}$ &$16.0$  & ${+12.2~}$&${~-13.0}$ & $ 8.4$  & &$ 6.0$&&${-4.7~}$&${~+2.9}$ & ${+1.3~}$&${~-2.3}$ & ${-1.3~}$&${~+1.5}$ & ${+0.8~}$&${~-1.3}$ & &$ 5.3$&&$ 1.3$  & $0.96$  & $ 1.6$  & $1.05$   \\ 
18 &$5.74 \cdot 10^{-5}$ &$22.1$  & ${+18.4~}$&${~-17.6}$ & $12.1$  & &$ 7.3$&&${-4.9~}$&${~+7.0}$ & ${+2.0~}$&${~+1.3}$ & ${+1.2~}$&${~-0.2}$ & ${+0.0~}$&${~+1.4}$ & &$ 7.3$&&$ 5.1$  & $0.96$  & $ 0.6$  & $1.00$   \\ 
19 &$1.53 \cdot 10^{-1}$ &$ 2.2$  & ${+3.8~}$&${~-4.5}$ & $ 1.6$  & &$-0.5$&&${+0.4~}$&${~-0.7}$ & ${-3.4~}$&${~+2.6}$ & ${-0.6~}$&${~+0.5}$ & ${+1.2~}$&${~-1.0}$ & &$ 1.2$&&$ 0.2$  & $0.89$  & $ 4.0$  & $1.02$   \\ 
20 &$5.75 \cdot 10^{-2}$ &$ 2.2$  & ${+8.9~}$&${~-9.2}$ & $ 3.1$  & &$ 2.5$&&${-6.8~}$&${~+6.9}$ & ${+3.4~}$&${~-4.5}$ & ${-0.0~}$&${~+0.4}$ & ${+1.1~}$&${~-0.8}$ & &$ 0.9$&&$ 0.5$  & $0.92$  & $ 3.4$  & $1.02$   \\ 
21 &$1.48 \cdot 10^{-2}$ &$ 3.0$  & ${+5.1~}$&${~-4.6}$ & $ 1.3$  & &$ 0.3$&&${-3.8~}$&${~+4.5}$ & ${+0.4~}$&${~-1.2}$ & ${-0.5~}$&${~+0.3}$ & ${+0.9~}$&${~-0.8}$ & &$ 1.0$&&$ 0.5$  & $0.94$  & $ 2.7$  & $1.03$   \\ 
22 &$2.76 \cdot 10^{-3}$ &$ 6.1$  & ${+7.8~}$&${~-7.4}$ & $ 4.4$  & &$ 3.8$&&${-3.3~}$&${~+4.0}$ & ${+0.4~}$&${~-0.8}$ & ${-0.3~}$&${~+0.6}$ & ${+0.7~}$&${~-0.3}$ & &$ 1.7$&&$ 2.3$  & $0.95$  & $ 1.7$  & $1.03$   \\ 
23 &$3.67 \cdot 10^{-4}$ &$17.9$  & ${+17.8~}$&${~-17.7}$ & $12.3$  & &$ 8.5$&&${-7.4~}$&${~+7.4}$ & ${+0.2~}$&${~+0.7}$ & ${+0.1~}$&${~+0.9}$ & ${+2.4~}$&${~-1.8}$ & &$ 5.0$&&$ 2.5$  & $0.96$  & $ 1.5$  & $1.05$   \\ 
24 &$5.27 \cdot 10^{-5}$ &$36.3$  & ${+17.0~}$&${~-16.4}$ & $-3.2$  & &$-4.3$&&${-8.8~}$&${~+9.1}$ & ${+2.2~}$&${~-1.3}$ & ${-1.4~}$&${~-1.1}$ & ${+4.6~}$&${~-3.3}$ & &$11.7$&&$ 3.3$  & $0.94$  & $ 0.5$  & $1.06$   \\ 
25 &$1.61 \cdot 10^{-1}$ &$ 2.6$  & ${+6.3~}$&${~-6.4}$ & $ 3.9$  & &$ 3.0$&&${+0.8~}$&${~-1.1}$ & ${-3.5~}$&${~+3.5}$ & ${+0.0~}$&${~+0.1}$ & ${+0.5~}$&${~-0.2}$ & &$ 1.1$&&$ 0.3$  & $0.89$  & $ 3.3$  & $1.02$   \\ 
26 &$6.61 \cdot 10^{-2}$ &$ 2.4$  & ${+9.6~}$&${~-9.1}$ & $ 2.3$  & &$ 4.7$&&${-5.9~}$&${~+6.8}$ & ${+4.0~}$&${~-4.3}$ & ${-0.5~}$&${~+0.1}$ & ${-0.2~}$&${~-0.2}$ & &$ 1.0$&&$ 0.3$  & $0.93$  & $ 2.9$  & $1.03$   \\ 
27 &$1.82 \cdot 10^{-2}$ &$ 3.3$  & ${+8.5~}$&${~-8.1}$ & $ 4.6$  & &$ 5.3$&&${-3.3~}$&${~+4.2}$ & ${+1.2~}$&${~-1.6}$ & ${-0.1~}$&${~+0.2}$ & ${+0.1~}$&${~-0.2}$ & &$ 1.0$&&$ 0.7$  & $0.94$  & $ 2.8$  & $1.02$   \\ 
28 &$3.03 \cdot 10^{-3}$ &$ 7.7$  & ${+9.4~}$&${~-8.3}$ & $ 3.5$  & &$ 5.6$&&${-3.9~}$&${~+5.7}$ & ${+2.3~}$&${~-1.8}$ & ${-0.0~}$&${~+0.3}$ & ${+0.6~}$&${~-0.1}$ & &$ 2.1$&&$ 0.7$  & $0.95$  & $ 2.1$  & $1.03$   \\ 
29 &$5.20 \cdot 10^{-4}$ &$16.0$  & ${+11.9~}$&${~-9.2}$ & $ 4.0$  & &$ 5.0$&&${-3.4~}$&${~+8.2}$ & ${+2.2~}$&${~-2.2}$ & ${-0.4~}$&${~-0.2}$ & ${-0.5~}$&${~+1.5}$ & &$ 4.8$&&$ 1.8$  & $0.96$  & $ 0.9$  & $1.01$   \\ 
30 &$1.17 \cdot 10^{-4}$ &$24.9$  & ${+15.9~}$&${~-16.0}$ & $-9.4$  & &$-6.1$&&${-3.5~}$&${~+3.7}$ & ${-0.2~}$&${~-0.4}$ & ${+0.5~}$&${~-0.8}$ & ${-2.2~}$&${~-0.5}$ & &$ 8.5$&&$ 6.2$  & $0.95$  & $ 1.4$  & $1.00$   \\ 
31 &$1.93 \cdot 10^{-1}$ &$ 2.5$  & ${+5.0~}$&${~-4.9}$ & $-0.3$  & &$-3.5$&&${+0.9~}$&${~-0.8}$ & ${-2.6~}$&${~+2.9}$ & ${-0.4~}$&${~+0.3}$ & ${+0.7~}$&${~-0.8}$ & &$ 1.0$&&$ 0.3$  & $0.89$  & $ 2.6$  & $1.03$   \\ 
32 &$8.74 \cdot 10^{-2}$ &$ 2.3$  & ${+7.3~}$&${~-7.6}$ & $ 1.6$  & &$ 1.0$&&${-6.2~}$&${~+6.3}$ & ${+2.8~}$&${~-3.6}$ & ${-0.3~}$&${~+0.1}$ & ${+0.6~}$&${~-0.7}$ & &$ 0.8$&&$ 0.1$  & $0.94$  & $ 2.3$  & $1.02$   \\ 
33 &$2.36 \cdot 10^{-2}$ &$ 3.1$  & ${+7.3~}$&${~-7.2}$ & $ 3.1$  & &$ 4.9$&&${-3.7~}$&${~+4.0}$ & ${+0.6~}$&${~-1.1}$ & ${-0.1~}$&${~+0.1}$ & ${+0.1~}$&${~-0.1}$ & &$ 0.8$&&$ 0.4$  & $0.95$  & $ 2.3$  & $1.02$   \\ 
34 &$4.62 \cdot 10^{-3}$ &$ 6.3$  & ${+10.1~}$&${~-10.1}$ & $ 5.9$  & &$ 6.7$&&${-3.9~}$&${~+3.8}$ & ${+1.0~}$&${~-1.0}$ & ${-0.1~}$&${~+0.3}$ & ${+0.5~}$&${~-0.0}$ & &$ 1.6$&&$ 0.9$  & $0.95$  & $ 2.0$  & $1.03$   \\ 
35 &$5.56 \cdot 10^{-4}$ &$19.6$  & ${+16.0~}$&${~-16.2}$ & $12.4$  & &$ 3.8$&&${-8.0~}$&${~+7.2}$ & ${+1.7~}$&${~-1.0}$ & ${+0.3~}$&${~+1.1}$ & ${-0.0~}$&${~+1.1}$ & &$ 5.1$&&$ 1.7$  & $0.96$  & $ 1.3$  & $1.02$   \\ 
36 &$8.35 \cdot 10^{-5}$ &$40.1$  & ${+66.5~}$&${~-65.6}$ & $41.7$  & &$43.9$&&${-6.2~}$&${~+5.6}$ & ${+7.3~}$&${~+0.1}$ & ${-1.2~}$&${~+4.1}$ & ${+7.2~}$&${~-0.9}$ & &$23.9$&&$ 4.5$  & $0.95$  & $ 1.2$  & $1.07$   \\ 
37 &$1.88 \cdot 10^{-1}$ &$ 2.9$  & ${+3.6~}$&${~-3.4}$ & $ 0.4$  & &$ 0.9$&&${+0.3~}$&${~-0.8}$ & ${-2.4~}$&${~+2.8}$ & ${-0.2~}$&${~+0.3}$ & ${+0.5~}$&${~-0.7}$ & &$ 1.2$&&$ 0.4$  & $0.89$  & $ 1.9$  & $1.03$   \\ 
38 &$9.50 \cdot 10^{-2}$ &$ 2.4$  & ${+9.5~}$&${~-9.3}$ & $ 3.5$  & &$ 5.2$&&${-5.8~}$&${~+5.9}$ & ${+3.6~}$&${~-3.3}$ & ${-0.2~}$&${~+0.3}$ & ${+0.8~}$&${~-0.6}$ & &$ 0.8$&&$ 0.1$  & $0.94$  & $ 1.8$  & $1.02$   \\ 
39 &$2.89 \cdot 10^{-2}$ &$ 3.2$  & ${+5.4~}$&${~-4.4}$ & $ 0.9$  & &$ 2.9$&&${-2.4~}$&${~+3.7}$ & ${+1.5~}$&${~-1.0}$ & ${-0.2~}$&${~+0.3}$ & ${+0.9~}$&${~-0.6}$ & &$ 0.8$&&$ 0.3$  & $0.96$  & $ 2.1$  & $1.01$   \\ 
40 &$6.12 \cdot 10^{-3}$ &$ 6.3$  & ${+8.3~}$&${~-7.9}$ & $ 4.7$  & &$ 4.7$&&${-3.2~}$&${~+4.2}$ & ${+1.0~}$&${~-1.3}$ & ${-0.7~}$&${~+0.5}$ & ${+0.8~}$&${~-1.1}$ & &$ 1.4$&&$ 1.5$  & $0.96$  & $ 1.8$  & $1.03$   \\ 
41 &$1.10 \cdot 10^{-3}$ &$13.7$  & ${+6.3~}$&${~-7.9}$ & $-3.8$  & &$ 0.4$&&${-5.3~}$&${~+2.5}$ & ${+1.0~}$&${~-1.4}$ & ${-0.1~}$&${~-0.3}$ & ${-0.7~}$&${~-0.8}$ & &$ 3.7$&&$ 1.5$  & $0.96$  & $ 1.3$  & $1.02$   \\ 
42 &$1.80 \cdot 10^{-4}$ &$25.0$  & ${+15.9~}$&${~-15.1}$ & $ 6.2$  & &$ 9.5$&&${-2.3~}$&${~+6.3}$ & ${+1.5~}$&${~+1.8}$ & ${-0.2~}$&${~-1.9}$ & ${-3.3~}$&${~-1.8}$ & &$ 7.7$&&$ 4.2$  & $0.95$  & $ 1.2$  & $1.03$   \\ 
43 &$2.36 \cdot 10^{-1}$ &$ 3.9$  & ${+6.7~}$&${~-6.4}$ & $-0.7$  & &$-5.6$&&${+1.0~}$&${~-0.7}$ & ${-2.1~}$&${~+2.3}$ & ${-0.1~}$&${~+0.4}$ & ${+1.5~}$&${~-0.7}$ & &$ 1.5$&&$ 0.6$  & $0.89$  & $ 1.3$  & $1.03$   \\ 
44 &$1.04 \cdot 10^{-1}$ &$ 3.7$  & ${+7.8~}$&${~-7.6}$ & $ 0.6$  & &$-4.0$&&${-5.6~}$&${~+6.0}$ & ${+2.0~}$&${~-2.4}$ & ${-0.0~}$&${~+0.6}$ & ${+0.8~}$&${~-0.6}$ & &$ 1.3$&&$ 0.4$  & $0.95$  & $ 1.2$  & $1.02$   \\ 
45 &$3.27 \cdot 10^{-2}$ &$ 4.9$  & ${+6.3~}$&${~-6.3}$ & $ 0.2$  & &$ 3.3$&&${-4.8~}$&${~+4.8}$ & ${+1.1~}$&${~-1.1}$ & ${-0.1~}$&${~+0.3}$ & ${+0.8~}$&${~-0.9}$ & &$ 1.3$&&$ 0.4$  & $0.96$  & $ 1.8$  & $1.00$   \\ 
46 &$8.00 \cdot 10^{-3}$ &$ 8.5$  & ${+4.0~}$&${~-4.6}$ & $-0.6$  & &$-0.7$&&${-3.6~}$&${~+2.7}$ & ${+0.9~}$&${~-0.8}$ & ${-0.2~}$&${~+0.0}$ & ${+0.9~}$&${~-0.7}$ & &$ 2.0$&&$ 0.7$  & $0.96$  & $ 2.0$  & $1.01$   \\ 
47 &$8.53 \cdot 10^{-4}$ &$31.3$  & ${+29.9~}$&${~-28.3}$ & $17.4$  & &$17.5$&&${-7.5~}$&${~+12.3}$ & ${+3.3~}$&${~-3.3}$ & ${-0.9~}$&${~-0.3}$ & ${+1.9~}$&${~+0.7}$ & &$10.6$&&$ 2.8$  & $0.96$  & $ 0.9$  & $1.03$   \\ 
48 &$1.45 \cdot 10^{-4}$ &$54.7$  & ${+31.1~}$&${~-27.2}$ & $-4.1$  & &$16.1$&&${-5.4~}$&${~+13.0}$ & ${+11.0~}$&${~-6.2}$ & ${+1.8~}$&${~+1.6}$ & ${+3.7~}$&${~+3.0}$ & &$19.2$&&$ 3.5$  & $0.95$  & $ 1.8$  & $0.99$   \\ 
\hline
\hline
\end{tabular}

    \caption{
      Double-differential normalised inclusive jet cross sections measured as a function of \Qsq\ and \ptjet. 
      For an explanation of the column headings, see table~\ref{tab:IncJet}.
      The residual normalisation uncertainty is $\DNorm{\csdsub} =\unit[0.8]{\%}$.
  }
  \label{tab:NormIncJet}
\end{table}

\begin{table}
  \scriptsize
  \tiny
  \setlength\tabcolsep{3pt} 
  \center
\begin{tabular}{ccrr@{\hskip0pt}rrc@{\hskip0pt}r@{\hskip0pt}c@{\hskip0pt}r@{\hskip0pt}rr@{\hskip0pt}rr@{\hskip0pt}rr@{\hskip0pt}rr@{\hskip0pt}r@{\hskip0pt}r@{\hskip0pt}c|rr|r}
\multicolumn{24}{c}{ \textbf{Normalised dijet cross sections in bins of \begin{boldmath}$\Qsq$ and $\meanptdi$\end{boldmath}} } \\
\hline
\hline
Bin & \multicolumn{1}{c}{$\sigma/\sigma_{\rm NC}$} & \multicolumn{1}{c}{\DStat{\csdsub}} & \multicolumn{2}{c}{\DSys{\csdsub} [\%]} & \multicolumn{1}{c}{\DMod{\csdsub}} & \multicolumn{3}{c}{\DModRW{\csdsub}} &  \multicolumn{2}{c}{\DJES{\csdsub} [\%]} &  \multicolumn{2}{c}{\DHFS{\csdsub} [\%]} & \multicolumn{2}{c}{\DEe{\csdsub} [\%]} & \multicolumn{2}{c}{\DThe{\csdsub} [\%]} & \multicolumn{3}{c}{\DMCSt{\csdsub}} & \DRad & \cHad & \DHad & \cRad  \\
label & \multicolumn{1}{c}{} & \multicolumn{1}{c}{[\%]} & \multicolumn{1}{c}{plus} & \multicolumn{1}{c}{minus} &  [\%] & & [\%] & & up & down & up & down & up & down & up & down  & & [\%] & & [\%] &  & [\%] &  \\
\hline
1 &$2.21 \cdot 10^{-2}$ &$ 3.3$  & ${+13.7~}$&${~-13.6}$ & $ 7.1$  & &$ 6.0$&&${+5.5~}$&${~-5.8}$ & ${-7.6~}$&${~+7.9}$ & ${+0.2~}$&${~+0.5}$ & ${-0.4~}$&${~+1.1}$ & &$ 2.2$&&$ 0.6$  & $0.86$  & $ 5.0$  & $1.02$   \\ 
2 &$1.36 \cdot 10^{-2}$ &$ 2.0$  & ${+6.9~}$&${~-6.9}$ & $ 4.9$  & &$ 3.8$&&${-2.1~}$&${~+2.1}$ & ${+0.7~}$&${~+0.0}$ & ${-0.8~}$&${~+0.6}$ & ${+0.4~}$&${~-0.5}$ & &$ 1.0$&&$ 1.1$  & $0.90$  & $ 3.9$  & $1.02$   \\ 
3 &$2.93 \cdot 10^{-3}$ &$ 2.8$  & ${+7.1~}$&${~-7.1}$ & $ 4.3$  & &$ 3.2$&&${-3.9~}$&${~+4.1}$ & ${+1.1~}$&${~-1.4}$ & ${-0.9~}$&${~+0.6}$ & ${+0.2~}$&${~-0.3}$ & &$ 1.1$&&$ 1.0$  & $0.93$  & $ 2.8$  & $1.02$   \\ 
4 &$4.90 \cdot 10^{-4}$ &$ 5.9$  & ${+6.7~}$&${~-5.4}$ & $ 2.0$  & &$ 0.8$&&${-3.9~}$&${~+5.6}$ & ${+0.9~}$&${~-1.0}$ & ${-0.5~}$&${~+0.2}$ & ${-0.3~}$&${~+0.1}$ & &$ 2.2$&&$ 1.3$  & $0.95$  & $ 2.3$  & $1.01$   \\ 
5 &$8.38 \cdot 10^{-5}$ &$12.7$  & ${+9.4~}$&${~-8.7}$ & $-5.2$  & &$-2.4$&&${-1.7~}$&${~+4.1}$ & ${+2.3~}$&${~-2.4}$ & ${-1.5~}$&${~+1.1}$ & ${+0.2~}$&${~+0.1}$ & &$ 5.1$&&$ 1.9$  & $0.94$  & $ 1.2$  & $1.04$   \\ 
6 &$1.58 \cdot 10^{-5}$ &$21.6$  & ${+19.4~}$&${~-21.5}$ & $12.1$  & &$10.7$&&${-9.4~}$&${~-0.1}$ & ${+1.3~}$&${~-1.9}$ & ${-0.7~}$&${~+2.2}$ & ${+0.8~}$&${~+0.4}$ & &$ 9.6$&&$ 3.8$  & $0.97$  & $ 3.0$  & $1.04$   \\ 
7 &$2.90 \cdot 10^{-2}$ &$ 3.0$  & ${+8.4~}$&${~-7.9}$ & $-1.7$  & &$-1.1$&&${+4.0~}$&${~-4.1}$ & ${-6.0~}$&${~+6.4}$ & ${-0.3~}$&${~+1.1}$ & ${+1.4~}$&${~-1.0}$ & &$ 1.6$&&$ 0.6$  & $0.87$  & $ 4.3$  & $1.01$   \\ 
8 &$1.69 \cdot 10^{-2}$ &$ 2.1$  & ${+3.0~}$&${~-2.9}$ & $ 0.5$  & &$ 0.6$&&${-1.9~}$&${~+2.2}$ & ${-1.0~}$&${~+0.5}$ & ${+0.1~}$&${~+0.1}$ & ${+0.2~}$&${~-0.1}$ & &$ 0.9$&&$ 0.3$  & $0.90$  & $ 3.5$  & $1.02$   \\ 
9 &$3.64 \cdot 10^{-3}$ &$ 3.0$  & ${+7.6~}$&${~-7.6}$ & $ 4.1$  & &$ 4.4$&&${-4.2~}$&${~+4.0}$ & ${+0.8~}$&${~-1.1}$ & ${-0.2~}$&${~+0.5}$ & ${+0.5~}$&${~-0.5}$ & &$ 1.0$&&$ 0.8$  & $0.94$  & $ 2.8$  & $1.02$   \\ 
10 &$5.73 \cdot 10^{-4}$ &$ 6.7$  & ${+9.9~}$&${~-10.2}$ & $ 7.8$  & &$ 2.9$&&${-4.7~}$&${~+3.9}$ & ${+2.1~}$&${~-1.9}$ & ${-0.9~}$&${~+1.1}$ & ${+1.2~}$&${~-1.0}$ & &$ 2.0$&&$ 0.9$  & $0.95$  & $ 2.0$  & $1.02$   \\ 
11 &$1.02 \cdot 10^{-4}$ &$14.5$  & ${+8.6~}$&${~-10.0}$ & $ 3.1$  & &$-1.4$&&${-5.6~}$&${~+2.8}$ & ${+1.8~}$&${~-2.0}$ & ${-1.1~}$&${~+0.1}$ & ${-1.1~}$&${~+0.7}$ & &$ 6.5$&&$ 2.5$  & $0.95$  & $ 1.7$  & $1.05$   \\ 
12 &$7.31 \cdot 10^{-6}$ &$73.1$  & ${+52.3~}$&${~-52.0}$ & $-29.5$  & &$-21.9$&&${-11.8~}$&${~+16.6}$ & ${-0.9~}$&${~-5.8}$ & ${-8.1~}$&${~-3.7}$ & ${-1.9~}$&${~-4.7}$ & &$32.8$&&$ 4.2$  & $0.96$  & $ 2.3$  & $1.05$   \\ 
13 &$3.04 \cdot 10^{-2}$ &$ 3.0$  & ${+8.3~}$&${~-7.5}$ & $ 1.5$  & &$-3.4$&&${+4.1~}$&${~-3.5}$ & ${-5.1~}$&${~+5.9}$ & ${-0.3~}$&${~-0.6}$ & ${-0.3~}$&${~+0.2}$ & &$ 1.6$&&$ 0.4$  & $0.89$  & $ 3.6$  & $1.02$   \\ 
14 &$1.89 \cdot 10^{-2}$ &$ 2.1$  & ${+3.4~}$&${~-3.0}$ & $ 1.2$  & &$ 0.9$&&${-1.9~}$&${~+2.5}$ & ${+0.1~}$&${~+0.5}$ & ${-0.4~}$&${~+0.3}$ & ${+0.4~}$&${~-0.2}$ & &$ 0.8$&&$ 0.5$  & $0.91$  & $ 3.2$  & $1.02$   \\ 
15 &$4.64 \cdot 10^{-3}$ &$ 2.7$  & ${+6.5~}$&${~-6.2}$ & $ 3.1$  & &$ 3.5$&&${-3.6~}$&${~+4.1}$ & ${+0.6~}$&${~-0.1}$ & ${-0.2~}$&${~+0.3}$ & ${+0.2~}$&${~+0.1}$ & &$ 0.8$&&$ 0.9$  & $0.94$  & $ 2.5$  & $1.03$   \\ 
16 &$8.75 \cdot 10^{-4}$ &$ 5.2$  & ${+6.1~}$&${~-5.9}$ & $ 3.5$  & &$ 0.9$&&${-4.1~}$&${~+4.1}$ & ${+1.3~}$&${~-0.6}$ & ${-0.4~}$&${~+0.2}$ & ${+0.3~}$&${~-0.2}$ & &$ 1.4$&&$ 0.8$  & $0.96$  & $ 2.2$  & $1.05$   \\ 
17 &$1.37 \cdot 10^{-4}$ &$13.5$  & ${+7.5~}$&${~-5.5}$ & $-0.7$  & &$ 1.0$&&${-3.0~}$&${~+5.0}$ & ${+2.6~}$&${~-0.8}$ & ${+0.3~}$&${~+1.0}$ & ${+1.4~}$&${~+0.7}$ & &$ 3.8$&&$ 1.8$  & $0.96$  & $ 1.8$  & $1.04$   \\ 
18 &$1.58 \cdot 10^{-5}$ &$36.5$  & ${+20.1~}$&${~-19.6}$ & $ 5.1$  & &$ 9.3$&&${-9.4~}$&${~+10.6}$ & ${-0.8~}$&${~-1.3}$ & ${-0.2~}$&${~+0.4}$ & ${-1.4~}$&${~+0.2}$ & &$10.7$&&$ 7.8$  & $0.96$  & $ 3.0$  & $0.98$   \\ 
19 &$3.50 \cdot 10^{-2}$ &$ 3.8$  & ${+8.2~}$&${~-9.0}$ & $-0.0$  & &$ 2.3$&&${+3.9~}$&${~-4.9}$ & ${-6.8~}$&${~+6.4}$ & ${-0.1~}$&${~+0.2}$ & ${+0.9~}$&${~-0.6}$ & &$ 1.8$&&$ 0.3$  & $0.89$  & $ 2.7$  & $1.03$   \\ 
20 &$2.36 \cdot 10^{-2}$ &$ 2.4$  & ${+5.4~}$&${~-5.7}$ & $ 3.7$  & &$ 3.5$&&${-1.7~}$&${~+1.1}$ & ${-1.1~}$&${~+0.2}$ & ${-0.1~}$&${~+0.1}$ & ${+0.3~}$&${~-0.6}$ & &$ 0.8$&&$ 0.4$  & $0.92$  & $ 2.7$  & $1.02$   \\ 
21 &$5.64 \cdot 10^{-3}$ &$ 3.1$  & ${+8.8~}$&${~-9.4}$ & $ 5.4$  & &$ 5.9$&&${-4.2~}$&${~+3.1}$ & ${+0.4~}$&${~-1.3}$ & ${-0.7~}$&${~+0.3}$ & ${+0.3~}$&${~-0.9}$ & &$ 0.9$&&$ 0.7$  & $0.94$  & $ 2.6$  & $1.03$   \\ 
22 &$1.00 \cdot 10^{-3}$ &$ 6.7$  & ${+10.5~}$&${~-10.3}$ & $ 6.3$  & &$ 6.4$&&${-3.5~}$&${~+4.1}$ & ${+1.0~}$&${~-1.2}$ & ${-0.8~}$&${~+0.8}$ & ${+0.6~}$&${~-1.0}$ & &$ 1.8$&&$ 2.5$  & $0.95$  & $ 1.8$  & $1.03$   \\ 
23 &$1.63 \cdot 10^{-4}$ &$15.2$  & ${+11.8~}$&${~-9.5}$ & $ 4.9$  & &$ 5.4$&&${-2.1~}$&${~+7.3}$ & ${+0.8~}$&${~-0.5}$ & ${+0.5~}$&${~+0.4}$ & ${+1.0~}$&${~+1.0}$ & &$ 4.4$&&$ 3.1$  & $0.96$  & $ 1.5$  & $1.05$   \\ 
24 &$2.56 \cdot 10^{-5}$ &$41.4$  & ${+31.2~}$&${~-30.8}$ & $12.1$  & &$15.1$&&${-6.4~}$&${~+4.6}$ & ${+2.4~}$&${~-1.1}$ & ${-1.0~}$&${~+0.8}$ & ${+6.8~}$&${~-2.4}$ & &$21.9$&&$ 7.0$  & $0.94$  & $-0.3$  & $1.07$   \\ 
25 &$4.13 \cdot 10^{-2}$ &$ 4.2$  & ${+8.1~}$&${~-6.1}$ & $ 1.4$  & &$-0.7$&&${+4.1~}$&${~-3.8}$ & ${-3.9~}$&${~+6.4}$ & ${-0.0~}$&${~-0.0}$ & ${+0.8~}$&${~+0.5}$ & &$ 1.7$&&$ 0.3$  & $0.90$  & $ 2.1$  & $1.03$   \\ 
26 &$2.99 \cdot 10^{-2}$ &$ 2.4$  & ${+3.2~}$&${~-2.8}$ & $ 0.6$  & &$ 1.9$&&${-1.0~}$&${~+1.5}$ & ${-0.4~}$&${~+1.2}$ & ${-0.1~}$&${~+0.3}$ & ${+0.3~}$&${~-0.1}$ & &$ 0.7$&&$ 0.3$  & $0.93$  & $ 2.2$  & $1.03$   \\ 
27 &$7.62 \cdot 10^{-3}$ &$ 3.1$  & ${+6.9~}$&${~-6.7}$ & $ 3.6$  & &$ 4.3$&&${-3.2~}$&${~+3.5}$ & ${+0.6~}$&${~-0.4}$ & ${-0.1~}$&${~+0.2}$ & ${+0.1~}$&${~-0.2}$ & &$ 0.8$&&$ 0.6$  & $0.95$  & $ 2.4$  & $1.02$   \\ 
28 &$1.39 \cdot 10^{-3}$ &$ 6.5$  & ${+7.7~}$&${~-8.1}$ & $ 4.4$  & &$ 3.4$&&${-5.4~}$&${~+4.5}$ & ${+1.2~}$&${~-0.7}$ & ${+0.1~}$&${~+0.4}$ & ${+0.8~}$&${~-0.0}$ & &$ 1.6$&&$ 0.9$  & $0.96$  & $ 2.0$  & $1.03$   \\ 
29 &$2.41 \cdot 10^{-4}$ &$13.5$  & ${+12.6~}$&${~-12.3}$ & $ 7.6$  & &$ 6.3$&&${-4.3~}$&${~+4.8}$ & ${+1.6~}$&${~+0.2}$ & ${-0.4~}$&${~-0.1}$ & ${-0.0~}$&${~-0.2}$ & &$ 3.8$&&$ 4.3$  & $0.95$  & $ 2.5$  & $1.03$   \\ 
30 &$4.73 \cdot 10^{-5}$ &$27.8$  & ${+20.4~}$&${~-18.6}$ & $10.1$  & &$ 8.2$&&${-3.8~}$&${~+9.1}$ & ${-1.9~}$&${~+2.8}$ & ${+1.5~}$&${~-1.9}$ & ${-0.8~}$&${~-0.9}$ & &$ 8.8$&&$ 8.3$  & $0.95$  & $ 1.1$  & $0.96$   \\ 
31 &$4.19 \cdot 10^{-2}$ &$ 5.4$  & ${+8.4~}$&${~-7.9}$ & $ 0.6$  & &$ 1.1$&&${+3.8~}$&${~-3.8}$ & ${-6.0~}$&${~+6.6}$ & ${-0.9~}$&${~+0.6}$ & ${+1.3~}$&${~-1.4}$ & &$ 2.4$&&$ 0.5$  & $0.90$  & $ 1.5$  & $1.03$   \\ 
32 &$3.38 \cdot 10^{-2}$ &$ 2.7$  & ${+5.7~}$&${~-5.6}$ & $ 3.4$  & &$ 3.9$&&${-0.8~}$&${~+1.1}$ & ${-0.8~}$&${~+1.1}$ & ${-0.3~}$&${~+0.1}$ & ${+0.5~}$&${~-0.6}$ & &$ 0.9$&&$ 0.2$  & $0.94$  & $ 1.8$  & $1.03$   \\ 
33 &$9.62 \cdot 10^{-3}$ &$ 3.1$  & ${+6.1~}$&${~-6.3}$ & $ 2.1$  & &$ 4.6$&&${-3.2~}$&${~+3.0}$ & ${+0.5~}$&${~-0.1}$ & ${-0.3~}$&${~+0.2}$ & ${+0.3~}$&${~-0.5}$ & &$ 0.7$&&$ 0.4$  & $0.95$  & $ 2.0$  & $1.03$   \\ 
34 &$1.84 \cdot 10^{-3}$ &$ 6.2$  & ${+9.4~}$&${~-9.5}$ & $ 5.4$  & &$ 6.2$&&${-3.9~}$&${~+3.7}$ & ${+1.2~}$&${~-0.8}$ & ${-0.2~}$&${~+0.4}$ & ${+0.7~}$&${~-0.5}$ & &$ 1.5$&&$ 1.0$  & $0.96$  & $ 1.6$  & $1.03$   \\ 
35 &$3.01 \cdot 10^{-4}$ &$14.4$  & ${+10.2~}$&${~-7.7}$ & $-2.9$  & &$-0.8$&&${-4.7~}$&${~+8.1}$ & ${+0.7~}$&${~-1.3}$ & ${+0.6~}$&${~+0.7}$ & ${+0.1~}$&${~+1.4}$ & &$ 4.4$&&$ 2.3$  & $0.96$  & $ 1.8$  & $1.03$   \\ 
36 &$4.54 \cdot 10^{-5}$ &$36.0$  & ${+31.1~}$&${~-30.6}$ & $15.4$  & &$19.7$&&${-1.1~}$&${~+5.4}$ & ${+3.4~}$&${~-3.5}$ & ${-2.8~}$&${~+1.4}$ & ${+3.8~}$&${~-2.7}$ & &$15.5$&&$ 6.7$  & $0.96$  & $ 3.0$  & $1.04$   \\ 
37 &$5.02 \cdot 10^{-2}$ &$ 5.2$  & ${+6.4~}$&${~-5.8}$ & $ 1.3$  & &$ 0.1$&&${+3.1~}$&${~-3.1}$ & ${-4.0~}$&${~+4.9}$ & ${-0.0~}$&${~+0.1}$ & ${+0.6~}$&${~-0.6}$ & &$ 1.8$&&$ 0.5$  & $0.90$  & $ 1.0$  & $1.04$   \\ 
38 &$4.24 \cdot 10^{-2}$ &$ 2.5$  & ${+4.6~}$&${~-4.5}$ & $ 2.9$  & &$ 2.8$&&${-1.3~}$&${~+1.5}$ & ${-0.3~}$&${~+0.4}$ & ${-0.2~}$&${~+0.1}$ & ${+0.6~}$&${~-0.3}$ & &$ 0.7$&&$ 0.3$  & $0.94$  & $ 1.3$  & $1.03$   \\ 
39 &$1.21 \cdot 10^{-2}$ &$ 3.1$  & ${+6.0~}$&${~-5.8}$ & $ 2.5$  & &$ 3.6$&&${-3.4~}$&${~+3.5}$ & ${+1.0~}$&${~-0.7}$ & ${-0.1~}$&${~+0.2}$ & ${+0.9~}$&${~-0.4}$ & &$ 0.7$&&$ 0.4$  & $0.96$  & $ 1.8$  & $1.02$   \\ 
40 &$2.47 \cdot 10^{-3}$ &$ 6.7$  & ${+13.2~}$&${~-13.4}$ & $ 8.3$  & &$ 8.7$&&${-4.9~}$&${~+4.6}$ & ${+1.2~}$&${~-1.6}$ & ${-0.4~}$&${~+0.3}$ & ${+0.5~}$&${~-0.7}$ & &$ 2.1$&&$ 1.0$  & $0.96$  & $ 1.8$  & $1.03$   \\ 
41 &$4.57 \cdot 10^{-4}$ &$13.8$  & ${+7.9~}$&${~-9.6}$ & $ 1.1$  & &$ 3.9$&&${-7.2~}$&${~+4.6}$ & ${+1.9~}$&${~-0.8}$ & ${-1.1~}$&${~-0.4}$ & ${-0.7~}$&${~-0.7}$ & &$ 3.7$&&$ 2.5$  & $0.95$  & $ 2.5$  & $1.04$   \\ 
42 &$5.27 \cdot 10^{-5}$ &$35.2$  & ${+30.0~}$&${~-30.3}$ & $ 9.1$  & &$22.0$&&${-7.8~}$&${~+8.8}$ & ${+2.5~}$&${~-0.1}$ & ${-0.2~}$&${~-1.5}$ & ${-7.1~}$&${~-3.2}$ & &$12.8$&&$ 8.3$  & $0.97$  & $-0.5$  & $0.99$   \\ 
43 &$6.00 \cdot 10^{-2}$ &$ 5.9$  & ${+6.8~}$&${~-6.4}$ & $ 3.3$  & &$-2.4$&&${+1.9~}$&${~-2.6}$ & ${-3.1~}$&${~+4.4}$ & ${-0.5~}$&${~+0.1}$ & ${-0.4~}$&${~-0.8}$ & &$ 2.0$&&$ 1.1$  & $0.89$  & $ 0.5$  & $1.05$   \\ 
44 &$4.77 \cdot 10^{-2}$ &$ 3.7$  & ${+6.0~}$&${~-6.2}$ & $ 4.9$  & &$ 2.5$&&${-1.8~}$&${~+1.3}$ & ${-1.1~}$&${~+1.0}$ & ${-0.2~}$&${~+0.1}$ & ${+0.1~}$&${~-0.7}$ & &$ 1.1$&&$ 0.4$  & $0.95$  & $ 0.7$  & $1.03$   \\ 
45 &$1.55 \cdot 10^{-2}$ &$ 4.4$  & ${+6.1~}$&${~-6.8}$ & $ 3.5$  & &$ 3.7$&&${-3.5~}$&${~+2.7}$ & ${-0.2~}$&${~-1.5}$ & ${-0.5~}$&${~+0.2}$ & ${+0.4~}$&${~-1.3}$ & &$ 1.2$&&$ 0.5$  & $0.97$  & $ 1.4$  & $1.01$   \\ 
46 &$2.91 \cdot 10^{-3}$ &$ 9.5$  & ${+8.6~}$&${~-7.5}$ & $ 3.4$  & &$ 3.6$&&${-4.7~}$&${~+6.2}$ & ${+1.3~}$&${~-0.8}$ & ${+0.0~}$&${~-0.0}$ & ${-0.5~}$&${~-0.2}$ & &$ 2.4$&&$ 1.3$  & $0.96$  & $ 2.2$  & $1.02$   \\ 
47 &$3.95 \cdot 10^{-4}$ &$26.9$  & ${+16.4~}$&${~-9.8}$ & $ 3.0$  & &$ 0.1$&&${-2.8~}$&${~+8.8}$ & ${+4.6~}$&${~-1.9}$ & ${+0.5~}$&${~+3.0}$ & ${+8.7~}$&${~+0.1}$ & &$ 8.1$&&$ 2.8$  & $0.96$  & $ 1.1$  & $1.05$   \\ 
48 &$8.53 \cdot 10^{-5}$ &$45.8$  & ${+27.5~}$&${~-25.8}$ & $ 1.3$  & &$18.9$&&${-7.4~}$&${~+11.4}$ & ${+4.3~}$&${~-1.6}$ & ${-2.0~}$&${~-0.5}$ & ${+2.3~}$&${~-0.5}$ & &$14.7$&&$ 4.9$  & $0.96$  & $ 2.2$  & $0.97$   \\ 
\hline
\hline
\end{tabular}

    \caption{
      Double-differential normalised dijet cross sections measured as a function of \Qsq\ and \meanptdi. 
      For an explanation of the column headings, see table~\ref{tab:IncJet}.
      The residual normalisation uncertainty is $\DNorm{\csdsub} =\unit[0.8]{\%}$
      and the LAr noise uncertainty is $\DLAr{\csdsub} = \unit[0.6]{\%}$.
  }
  \label{tab:NormDijet}
\end{table}

\begin{table}
  \scriptsize
  \tiny
  \setlength\tabcolsep{3pt} 
  \center
\begin{tabular}{ccrr@{\hskip0pt}rrc@{\hskip0pt}r@{\hskip0pt}c@{\hskip0pt}r@{\hskip0pt}rr@{\hskip0pt}rr@{\hskip0pt}rr@{\hskip0pt}rr@{\hskip0pt}r@{\hskip0pt}r@{\hskip0pt}c|rr|r}
\multicolumn{24}{c}{ \textbf{Normalised trijet cross sections in bins of \begin{boldmath}$\Qsq$ and $\meanpttri$\end{boldmath}} } \\
\hline
\hline
Bin & \multicolumn{1}{c}{$\sigma/\sigma_{\rm NC}$} & \multicolumn{1}{c}{\DStat{\csdsub}} & \multicolumn{2}{c}{\DSys{\csdsub} [\%]} & \multicolumn{1}{c}{\DMod{\csdsub}} & \multicolumn{3}{c}{\DModRW{\csdsub}} &  \multicolumn{2}{c}{\DJES{\csdsub} [\%]} &  \multicolumn{2}{c}{\DHFS{\csdsub} [\%]} & \multicolumn{2}{c}{\DEe{\csdsub} [\%]} & \multicolumn{2}{c}{\DThe{\csdsub} [\%]} & \multicolumn{3}{c}{\DMCSt{\csdsub}} & \DRad & \cHad & \DHad & \cRad  \\
label & \multicolumn{1}{c}{} & \multicolumn{1}{c}{[\%]} & \multicolumn{1}{c}{plus} & \multicolumn{1}{c}{minus} &  [\%] & & [\%] & & up & down & up & down & up & down & up & down  & & [\%] & & [\%] &  & [\%] &  \\
\hline
1 &$3.52 \cdot 10^{-3}$ &$ 7.1$  & ${+30.6~}$&${~-30.2}$ & $21.8$  & &$17.8$&&${+5.5~}$&${~-5.4}$ & ${-8.5~}$&${~+9.6}$ & ${-0.0~}$&${~+0.9}$ & ${-0.2~}$&${~-0.2}$ & &$ 3.7$&&$ 0.7$  & $0.74$  & $ 7.4$  & $1.02$   \\ 
2 &$1.76 \cdot 10^{-3}$ &$ 5.4$  & ${+15.3~}$&${~-15.0}$ & $ 9.3$  & &$11.4$&&${-0.6~}$&${~+2.1}$ & ${+0.3~}$&${~+2.1}$ & ${-0.3~}$&${~+0.5}$ & ${-0.6~}$&${~-0.1}$ & &$ 2.4$&&$ 0.6$  & $0.79$  & $ 4.9$  & $1.02$   \\ 
3 &$4.06 \cdot 10^{-4}$ &$ 7.4$  & ${+6.9~}$&${~-6.3}$ & $ 2.1$  & &$ 3.7$&&${-3.1~}$&${~+4.1}$ & ${+1.1~}$&${~+0.2}$ & ${-0.3~}$&${~+0.2}$ & ${-0.0~}$&${~+0.7}$ & &$ 2.8$&&$ 1.3$  & $0.84$  & $ 2.6$  & $1.02$   \\ 
4 &$3.27 \cdot 10^{-5}$ &$19.8$  & ${+12.2~}$&${~-11.2}$ & $ 3.1$  & &$ 5.6$&&${-2.5~}$&${~+2.8}$ & ${+3.7~}$&${~-1.1}$ & ${+0.2~}$&${~+1.4}$ & ${+0.7~}$&${~+2.7}$ & &$ 6.7$&&$ 5.7$  & $0.84$  & $ 0.3$  & $1.04$   \\ 
5 &$4.72 \cdot 10^{-3}$ &$ 6.3$  & ${+18.4~}$&${~-18.3}$ & $12.1$  & &$10.4$&&${+5.9~}$&${~-6.2}$ & ${-5.5~}$&${~+6.0}$ & ${-0.1~}$&${~-0.0}$ & ${+0.6~}$&${~+0.6}$ & &$ 3.0$&&$ 0.5$  & $0.74$  & $ 7.5$  & $1.03$   \\ 
6 &$2.00 \cdot 10^{-3}$ &$ 5.9$  & ${+13.5~}$&${~-13.6}$ & $ 8.3$  & &$10.0$&&${-2.2~}$&${~+1.4}$ & ${-1.4~}$&${~+1.8}$ & ${-0.6~}$&${~+0.3}$ & ${+0.2~}$&${~+0.0}$ & &$ 2.3$&&$ 0.6$  & $0.79$  & $ 4.7$  & $1.04$   \\ 
7 &$4.42 \cdot 10^{-4}$ &$ 9.1$  & ${+7.4~}$&${~-6.5}$ & $ 2.9$  & &$ 2.3$&&${-3.5~}$&${~+5.3}$ & ${+0.4~}$&${~-1.4}$ & ${-0.9~}$&${~+0.3}$ & ${-0.4~}$&${~-0.8}$ & &$ 3.0$&&$ 1.2$  & $0.84$  & $ 2.1$  & $1.03$   \\ 
8 &$3.18 \cdot 10^{-5}$ &$27.5$  & ${+20.7~}$&${~-20.3}$ & $-15.0$  & &$-4.8$&&${+2.4~}$&${~+6.4}$ & ${+0.7~}$&${~-4.0}$ & ${-0.2~}$&${~-1.2}$ & ${-2.5~}$&${~+0.6}$ & &$11.0$&&$ 3.9$  & $0.84$  & $ 0.7$  & $1.08$   \\ 
9 &$5.35 \cdot 10^{-3}$ &$ 5.4$  & ${+16.5~}$&${~-16.3}$ & $ 9.2$  & &$ 9.7$&&${+4.5~}$&${~-4.0}$ & ${-7.9~}$&${~+8.2}$ & ${-0.1~}$&${~-0.4}$ & ${-0.5~}$&${~-0.6}$ & &$ 2.4$&&$ 0.5$  & $0.73$  & $ 7.3$  & $1.03$   \\ 
10 &$2.80 \cdot 10^{-3}$ &$ 4.8$  & ${+14.7~}$&${~-14.4}$ & $ 7.1$  & &$12.0$&&${-2.0~}$&${~+2.7}$ & ${-1.3~}$&${~+2.9}$ & ${+0.2~}$&${~+0.4}$ & ${+0.2~}$&${~-0.1}$ & &$ 1.7$&&$ 0.6$  & $0.79$  & $ 4.7$  & $1.05$   \\ 
11 &$6.45 \cdot 10^{-4}$ &$ 6.9$  & ${+9.3~}$&${~-9.7}$ & $ 5.8$  & &$ 6.0$&&${-3.9~}$&${~+2.9}$ & ${+0.5~}$&${~-0.3}$ & ${-0.3~}$&${~-0.2}$ & ${+0.1~}$&${~-0.2}$ & &$ 2.1$&&$ 1.1$  & $0.84$  & $ 2.7$  & $1.04$   \\ 
12 &$5.07 \cdot 10^{-5}$ &$19.6$  & ${+11.1~}$&${~-11.8}$ & $ 2.8$  & &$-0.1$&&${-7.9~}$&${~+7.4}$ & ${+1.6~}$&${~-3.0}$ & ${-2.4~}$&${~+2.4}$ & ${+2.0~}$&${~-2.4}$ & &$ 5.9$&&$ 3.5$  & $0.86$  & $ 0.2$  & $1.03$   \\ 
13 &$5.29 \cdot 10^{-3}$ &$ 8.2$  & ${+19.4~}$&${~-18.2}$ & $ 9.1$  & &$13.7$&&${+6.1~}$&${~-4.5}$ & ${-5.4~}$&${~+7.3}$ & ${+0.1~}$&${~+0.9}$ & ${+2.1~}$&${~+0.4}$ & &$ 3.1$&&$ 0.6$  & $0.73$  & $ 7.3$  & $1.04$   \\ 
14 &$3.16 \cdot 10^{-3}$ &$ 6.2$  & ${+11.2~}$&${~-11.3}$ & $ 5.2$  & &$ 9.3$&&${-1.2~}$&${~+1.8}$ & ${-2.0~}$&${~+1.5}$ & ${-0.8~}$&${~+0.2}$ & ${+0.0~}$&${~-0.6}$ & &$ 2.0$&&$ 0.8$  & $0.79$  & $ 5.2$  & $1.04$   \\ 
15 &$9.23 \cdot 10^{-4}$ &$ 7.4$  & ${+8.6~}$&${~-8.4}$ & $ 4.9$  & &$ 5.1$&&${-3.3~}$&${~+3.6}$ & ${+0.9~}$&${~-0.4}$ & ${-0.6~}$&${~+1.1}$ & ${+0.9~}$&${~-0.8}$ & &$ 2.1$&&$ 1.3$  & $0.83$  & $ 2.8$  & $1.07$   \\ 
16 &$5.13 \cdot 10^{-5}$ &$30.1$  & ${+21.9~}$&${~-26.9}$ & $15.6$  & &$ 9.2$&&${-16.8~}$&${~+6.4}$ & ${-1.4~}$&${~-0.6}$ & ${+1.8~}$&${~-2.3}$ & ${-1.0~}$&${~+1.6}$ & &$ 9.5$&&$ 3.5$  & $0.86$  & $ 0.9$  & $1.03$   \\ 
17 &$5.90 \cdot 10^{-3}$ &$ 9.5$  & ${+15.7~}$&${~-12.3}$ & $ 4.6$  & &$ 8.9$&&${+6.8~}$&${~-4.0}$ & ${-4.6~}$&${~+9.2}$ & ${+0.1~}$&${~-0.1}$ & ${-0.3~}$&${~+0.0}$ & &$ 3.2$&&$ 0.8$  & $0.73$  & $ 7.5$  & $1.04$   \\ 
18 &$3.73 \cdot 10^{-3}$ &$ 7.2$  & ${+10.1~}$&${~-9.9}$ & $ 4.9$  & &$ 8.0$&&${-1.3~}$&${~+0.8}$ & ${-1.4~}$&${~+2.3}$ & ${-0.1~}$&${~+0.3}$ & ${+0.6~}$&${~-0.4}$ & &$ 2.2$&&$ 0.6$  & $0.79$  & $ 5.4$  & $1.04$   \\ 
19 &$1.08 \cdot 10^{-3}$ &$ 8.6$  & ${+6.4~}$&${~-6.4}$ & $ 3.3$  & &$ 3.2$&&${-3.1~}$&${~+2.8}$ & ${+0.4~}$&${~+1.2}$ & ${-0.0~}$&${~-0.2}$ & ${+0.0~}$&${~-0.1}$ & &$ 2.3$&&$ 1.7$  & $0.84$  & $ 2.8$  & $1.04$   \\ 
20 &$7.38 \cdot 10^{-5}$ &$27.8$  & ${+14.1~}$&${~-13.9}$ & $-1.0$  & &$-3.9$&&${-7.3~}$&${~+7.7}$ & ${-0.4~}$&${~+1.0}$ & ${-0.7~}$&${~-0.0}$ & ${-0.6~}$&${~-0.0}$ & &$ 7.5$&&$ 8.2$  & $0.86$  & $ 1.7$  & $1.03$   \\ 
21 &$7.74 \cdot 10^{-3}$ &$ 8.7$  & ${+14.0~}$&${~-11.2}$ & $ 1.9$  & &$ 8.0$&&${+5.7~}$&${~-3.2}$ & ${-6.1~}$&${~+9.1}$ & ${+0.4~}$&${~+0.6}$ & ${+1.2~}$&${~-0.5}$ & &$ 2.8$&&$ 0.6$  & $0.73$  & $ 7.2$  & $1.04$   \\ 
22 &$4.98 \cdot 10^{-3}$ &$ 6.2$  & ${+14.3~}$&${~-14.4}$ & $ 7.4$  & &$11.7$&&${-1.6~}$&${~+1.1}$ & ${-2.5~}$&${~+1.8}$ & ${-0.0~}$&${~+0.1}$ & ${+0.2~}$&${~-0.4}$ & &$ 1.7$&&$ 1.0$  & $0.79$  & $ 5.2$  & $1.04$   \\ 
23 &$1.19 \cdot 10^{-3}$ &$ 9.1$  & ${+12.8~}$&${~-12.2}$ & $ 7.9$  & &$ 8.2$&&${-2.7~}$&${~+4.7}$ & ${+0.2~}$&${~+1.1}$ & ${+0.5~}$&${~+0.3}$ & ${+0.9~}$&${~+0.7}$ & &$ 2.3$&&$ 1.9$  & $0.83$  & $ 2.5$  & $1.05$   \\ 
24 &$4.74 \cdot 10^{-5}$ &$50.8$  & ${+33.8~}$&${~-27.6}$ & $13.9$  & &$10.4$&&${-14.1~}$&${~+21.2}$ & ${+10.9~}$&${~-1.3}$ & ${+1.8~}$&${~+2.4}$ & ${+4.1~}$&${~+5.4}$ & &$15.0$&&$ 3.6$  & $0.85$  & $ 0.8$  & $1.05$   \\ 
25 &$8.40 \cdot 10^{-3}$ &$10.4$  & ${+7.8~}$&${~-8.5}$ & $-1.4$  & &$ 2.2$&&${+4.1~}$&${~-4.5}$ & ${-5.7~}$&${~+5.0}$ & ${-0.3~}$&${~+0.0}$ & ${-0.6~}$&${~+0.0}$ & &$ 3.1$&&$ 0.5$  & $0.73$  & $ 6.8$  & $1.04$   \\ 
26 &$6.40 \cdot 10^{-3}$ &$ 6.7$  & ${+3.4~}$&${~-3.3}$ & $-0.1$  & &$ 1.6$&&${-0.6~}$&${~+0.8}$ & ${-1.1~}$&${~+1.3}$ & ${-0.2~}$&${~-0.0}$ & ${-0.3~}$&${~-0.2}$ & &$ 1.7$&&$ 1.3$  & $0.78$  & $ 5.6$  & $1.04$   \\ 
27 &$1.81 \cdot 10^{-3}$ &$ 8.6$  & ${+9.7~}$&${~-10.9}$ & $ 6.4$  & &$ 6.3$&&${-4.9~}$&${~+2.1}$ & ${-0.2~}$&${~-1.6}$ & ${-1.1~}$&${~+0.1}$ & ${+0.1~}$&${~-1.3}$ & &$ 2.1$&&$ 1.2$  & $0.84$  & $ 3.6$  & $1.04$   \\ 
28 &$1.43 \cdot 10^{-4}$ &$26.7$  & ${+10.5~}$&${~-11.7}$ & $-2.5$  & &$-4.5$&&${-5.6~}$&${~+3.2}$ & ${-0.3~}$&${~-1.2}$ & ${-0.3~}$&${~+0.4}$ & ${-2.0~}$&${~-0.5}$ & &$ 7.5$&&$ 3.9$  & $0.83$  & $ 1.5$  & $1.08$   \\ 
29 &$1.14 \cdot 10^{-2}$ &$11.0$  & ${+7.6~}$&${~-8.8}$ & $-1.3$  & &$-3.4$&&${+3.2~}$&${~-3.0}$ & ${-6.5~}$&${~+4.7}$ & ${-0.5~}$&${~-0.2}$ & ${-0.4~}$&${~+0.3}$ & &$ 3.0$&&$ 0.7$  & $0.73$  & $ 6.1$  & $1.04$   \\ 
30 &$7.74 \cdot 10^{-3}$ &$ 8.8$  & ${+5.2~}$&${~-4.8}$ & $ 2.3$  & &$ 1.8$&&${-1.8~}$&${~+2.3}$ & ${-1.0~}$&${~+1.4}$ & ${-0.5~}$&${~+0.6}$ & ${+1.0~}$&${~-0.6}$ & &$ 2.2$&&$ 1.4$  & $0.78$  & $ 4.8$  & $1.03$   \\ 
31 &$2.36 \cdot 10^{-3}$ &$12.0$  & ${+7.2~}$&${~-5.4}$ & $ 1.8$  & &$ 3.1$&&${-1.4~}$&${~+4.5}$ & ${+2.0~}$&${~+0.7}$ & ${-0.1~}$&${~+0.3}$ & ${+1.2~}$&${~+0.6}$ & &$ 3.0$&&$ 1.6$  & $0.83$  & $ 3.2$  & $1.04$   \\ 
32 &$2.67 \cdot 10^{-4}$ &$27.5$  & ${+15.8~}$&${~-16.1}$ & $-3.8$  & &$ 6.4$&&${-4.9~}$&${~-0.6}$ & ${+3.4~}$&${~-1.2}$ & ${-1.9~}$&${~+1.0}$ & ${+4.1~}$&${~-3.0}$ & &$10.0$&&$ 8.3$  & $0.84$  & $ 0.8$  & $1.06$   \\ 
\hline
\hline
\end{tabular}

    \caption{
      Double-differential normalised trijet cross sections measured as a function of \Qsq\ and \meanpttri. 
      For an explanation of the column headings, see table~\ref{tab:IncJet}.
      The residual normalisation uncertainty is $\DNorm{\csdsub} =\unit[0.8]{\%}$
      and the LAr noise uncertainty is $\DLAr{\csdsub} = \unit[0.9]{\%}$.
  }
  \label{tab:NormTrijet}
\end{table}

\clearpage
\begin{table}
  \scriptsize
  \center
\begin{tabular}{c c r r r c c c c c | c c c}
\multicolumn{13}{c}{ \textbf{Inclusive jet cross sections at high \begin{boldmath}\Qsq\ for 5<\ptjet<7\,\GeV\end{boldmath}}} \\
\hline
\Qsq-range & \multicolumn{1}{c}{\CS} & \multicolumn{1}{c}{\DStat{\csdsub}} & \multicolumn{1}{c}{\DSys{\csdsub}} & \multicolumn{1}{c}{\DMod{\csdsub}} & \DJES{\csdsub} & \DHFS{\csdsub} & \DEe{\csdsub} & \DThe{\csdsub} & \DID{\csdsub} & \cHad & \DHad & \cEW \\
 $[\GeVsq]$ & \multicolumn{1}{c}{[pb]} & \multicolumn{1}{c}{[\%]} & \multicolumn{1}{c}{[\%]} & \multicolumn{1}{c}{[\%]} & [\%] & [\%] & [\%] & [\%] & [\%] &  & [\%] &  \\
\hline
 150--200  & $8.85\trenn 10^{1}$ & 3.7~ & 4.4~ & $+2.0$~ & $^{~-0.85}_{~+1.07}$ & $^{~+1.97}_{~-2.12}$ & $^{~-1.09}_{~+1.20}$ & $^{~-0.36}_{~+0.36}$ & $^{~+0.48}_{~-0.48}$ & 0.90 & 2.5 & 1.00 \\
 200--270  & $7.19\trenn 10^{1}$ & 3.9~ & 4.2~ & $+1.6$~ & $^{~-0.89}_{~+0.91}$ & $^{~+1.72}_{~-2.08}$ & $^{~-1.23}_{~+1.02}$ & $^{~-0.42}_{~+0.41}$ & $^{~+0.49}_{~-0.49}$ & 0.90 & 2.5 & 1.00 \\
 270--400  & $6.59\trenn 10^{1}$ & 4.0~ & 3.8~ & $-1.3$~ & $^{~-0.65}_{~+0.66}$ & $^{~+1.42}_{~-1.45}$ & $^{~-0.86}_{~+1.12}$ & $^{~-0.26}_{~+0.32}$ & $^{~+0.47}_{~-0.47}$ & 0.90 & 2.2 & 1.00 \\
 400--700  & $4.68\trenn 10^{1}$ & 5.6~ & 3.6~ & $+0.6$~ & $^{~-0.32}_{~+0.48}$ & $^{~+1.24}_{~-1.16}$ & $^{~-1.35}_{~+1.21}$ & $^{~-0.16}_{~+0.22}$ & $^{~+0.40}_{~-0.40}$ & 0.90 & 2.2 & 1.00 \\
 700--5000 & $4.04\trenn 10^{1}$ & 6.4~ & 3.5~ & $+1.2$~ & $^{~-0.29}_{~+0.42}$ & $^{~+0.71}_{~-0.61}$ & $^{~+0.15}_{~+0.10}$ & $^{~-0.33}_{~+0.34}$ & $^{~+1.12}_{~-1.12}$ & 0.91 & 2.0 & 1.02 \\
 5000--15\,000& $9.13\trenn 10^{-1}$ & 79.2~&17.4~ &$+16.7$~ & $^{~-0.74}_{~-0.21}$ & $^{~+0.84}_{~-0.49}$ & $^{~+3.47}_{~+1.20}$ & $^{~-0.78}_{~-0.61}$ & $^{~+2.03}_{~-2.03}$ & 0.91 & 2.0 & 1.11 \\
\hline

\end{tabular}
    \caption{
      Inclusive jet cross sections for $5<\ptjet<7\,\GeV$ measured as a function of \Qsq\ in the range $150<\Qsq<15\,000\,\GeVsq$ and $0.2<y<0.7$.
      The total systematic uncertainty, $\DSys{\csdsub}$, sums all the quoted systematic uncertainties in quadrature, including in addition the uncertainty due to the LAr noise of $\DLAr{\csdsub} = \unit[0.5]{\%}$ and the total normalisation uncertainty of $\DNorm{\csdsub} =\unit[2.5]{\%}$. The labels \DEe\ and \DThe\  denote the uncertainty on the scattered electron energy and azimuthal angle, respectively. The label \DID\ denotes the uncertainty on the electron identification as defined in reference~\cite{H1Multijets}. The label \cEW\ denotes the multiplicative corrections for electroweak effects.
      The column labels \cHad\ and \DHad\ denote the multiplicative hadronisation correction factors and their uncertainties, respectively.
      The cross section values and uncertainties have been determined in the scope of the analysis of reference~\cite{H1Multijets}.
    }
  \label{tab:IncJetHQ}
\end{table}

\begin{table}
  \scriptsize
  \center
  \begin{tabular}{c c r r r c c c c | c c }
    \multicolumn{11}{c}{ \textbf{Normalised inclusive jet cross sections at \begin{boldmath}high-\Qsq\ for 5<\ptjet<7\,\GeV\end{boldmath}}} \\
    \hline
     \Qsq-range & \multicolumn{1}{c}{\CSN} & \multicolumn{1}{c}{\DStat{\csdsubn}} & \multicolumn{1}{c}{\DSys{\csdsubn}} & \multicolumn{1}{c}{\DMod{\csdsubn}} & \DJES{\csdsubn} & \DHFS{\csdsubn} & \DEe{\csdsubn} & \DThe{\csdsubn} & \cHad & \DHad  \\
     $[\GeVsq]$ & \multicolumn{1}{c}{} & \multicolumn{1}{c}{[\%]} & \multicolumn{1}{c}{[\%]} & \multicolumn{1}{c}{[\%]} & [\%] & [\%] & [\%] & [\%] &  & [\%]  \\
     \hline
     150--200   & $2.05\trenn 10^{-1}$ &  3.7~ &  ~2.7 & $ 1.9$~ & $^{~-1.20}_{~+ 1.44}$ & $^{~+1.30}_{~-1.40}$ & $^{~-1.08}_{~+1.24}$ & $^{~+0.13}_{~-0.12}$ & 0.90 & 2.5 \\
     200--270   & $2.27\trenn 10^{-1}$ &  3.9~ &  ~2.5 & $ 1.2$~ & $^{~-1.29}_{~+ 1.31}$ & $^{~+1.15}_{~-1.48}$ & $^{~-1.27}_{~+1.01}$ & $^{~+0.05}_{~-0.08}$ & 0.90 & 2.5 \\
     270--400   & $2.39\trenn 10^{-1}$ &  3.9~ &  ~2.6 & $-2.1$~ & $^{~-1.12}_{~+ 1.15}$ & $^{~+1.05}_{~-1.07}$ & $^{~-0.86}_{~+1.14}$ & $^{~+0.11}_{~-0.06}$ & 0.90 & 2.2 \\
     400--700   & $2.14\trenn 10^{-1}$ &  5.6~ &  ~2.1 & $ 1.3$~ & $^{~-0.90}_{~+ 1.07}$ & $^{~+1.02}_{~-0.93}$ & $^{~-1.31}_{~+1.17}$ & $^{~+0.05}_{~-0.01}$ & 0.90 & 2.2 \\
     700--5000  & $2.25\trenn 10^{-1}$ &  6.4~ &  ~1.3 & $-0.5$~ & $^{~-0.79}_{~+ 0.93}$ & $^{~+0.64}_{~-0.54}$ & $^{~-0.01}_{~+0.26}$ & $^{~+0.14}_{~-0.11}$ & 0.91 & 2.0 \\
     5000--15\,000 & $1.07\trenn 10^{-1}$ & 79.1~ & ~15.1 & $-14.6$~& $^{~-1.30}_{~+ 0.33}$ & $^{~+0.81}_{~-0.48}$ & $^{~+2.94}_{~+1.77}$ & $^{~-0.48}_{~-0.92}$ & 0.91 & 2.0 \\
     \hline
   \end{tabular}
  \caption{
    Normalised inclusive jet cross sections  for $5<\ptjet<7\,\GeV$ measured  as a function of \Qsq\ in the range $150<\Qsq<15\,000\,\GeVsq$ and $0.2<y<0.7$.
    The normalisation uncertainty $\DNorm{}$ equals zero for this measurement.
    The total systematic uncertainty, $\DSys{\csdsub}$, includes the quoted uncertainties and the uncertainty on the LAr noise $\DLAr{\csdsub} = \unit[0.5]{\%}$.
    Other details are given in the caption of table~\ref{tab:IncJetHQ}.
  }
  \label{tab:NormIncJetHQ}
\end{table}

\begin{table}
  \scriptsize
  \center
  \begin{tabular}{c | l | l}
\multicolumn{3}{c}{\bf \begin{boldmath}\end{boldmath}} \\
\hline
\mur{} [GeV] & \multicolumn{1}{c|}{\asmz{}} & \multicolumn{1}{c}{$\alpha_{\rm s}(\mu_r)$} \\
\hline
 5.29 & $0.1257\,(10)_{\text{exp}}\,({}^{+113}_{-105})_{\text{th}}$ & $0.2357\,(37)_{\text{exp}}\,({}^{+456}_{-354})_{\text{th}}$ \\
 7.21 & $0.1252\,(10)_{\text{exp}}\,({}^{+104}_{-93})_{\text{th}}$ & $0.2131\,(31)_{\text{exp}}\,({}^{+335}_{-263})_{\text{th}}$ \\
 9.38 & $0.1218\,(13)_{\text{exp}}\,({}^{+79}_{-66})_{\text{th}}$ & $0.1899\,(32)_{\text{exp}}\,({}^{+204}_{-159})_{\text{th}}$ \\
 12.2 & $0.1186\,(8)_{\text{exp}}\,({}^{+64}_{-51})_{\text{th}}$ & $0.1713\,(17)_{\text{exp}}\,({}^{+141}_{-107})_{\text{th}}$ \\
 14.2 & $0.1170\,(14)_{\text{exp}}\,({}^{+59}_{-51})_{\text{th}}$ & $0.1625\,(27)_{\text{exp}}\,({}^{+119}_{-97})_{\text{th}}$ \\
 17.3 & $0.1179\,(10)_{\text{exp}}\,({}^{+53}_{-45})_{\text{th}}$ & $0.1576\,(19)_{\text{exp}}\,({}^{+98}_{-81})_{\text{th}}$ \\
 20.7 & $0.1146\,(24)_{\text{exp}}\,({}^{+69}_{-52})_{\text{th}}$ & $0.1466\,(40)_{\text{exp}}\,({}^{+117}_{-84})_{\text{th}}$ \\
 25.4 & $0.1142\,(30)_{\text{exp}}\,({}^{+68}_{-55})_{\text{th}}$ & $0.1405\,(46)_{\text{exp}}\,({}^{+106}_{-82})_{\text{th}}$ \\
 34.6 & $0.1155\,(11)_{\text{exp}}\,({}^{+39}_{-25})_{\text{th}}$ & $0.1348\,(15)_{\text{exp}}\,({}^{+53}_{-34})_{\text{th}}$ \\
 54.8 & $0.1129\,(39)_{\text{exp}}\,({}^{+51}_{-27})_{\text{th}}$ & $0.1218\,(46)_{\text{exp}}\,({}^{+60}_{-32})_{\text{th}}$ \\
\hline
  \end{tabular}
  \caption{The strong coupling extracted from the normalized inclusive jet, dijet and trijet data
    at NLO as a function of the renormalisation scale \mur{}. For each
    \mur{} the values of the strong coupling $\alpha_{\rm s}(\mu_r)$
    and the equivalent values \asmz{} are given with experimental
    (exp) and theoretical (th) uncertainties.
  }
  \label{tab:AlphasRunning}
\end{table}

\clearpage


\newpage

\newpage

\end{document}